\documentclass[11pt]{article}
\pdfoutput=1
\usepackage{jcapmod}
\usepackage{shorthand}

\usepackage{mathtools}
\usepackage{booktabs}
\usepackage[english]{babel}
\usepackage{amsmath,amssymb,amsbsy,amstext, amsthm, simplewick, amsfonts}
\usepackage{hyperref}
\usepackage{graphicx}
\usepackage[small]{caption}
\usepackage{siunitx}
\usepackage{upgreek}
\usepackage{framed}
\usepackage{wrapfig}
\usepackage{multirow}
\usepackage{bbm}
\usepackage[svgnames,dvipsnames,x11names]{xcolor}
\usepackage[utf8x]{inputenc}
\usepackage{selinput}
\usepackage{bm}
\usepackage{float}
\usepackage{geometry}
\usepackage{hyperref}
\usepackage{yfonts}
\usepackage{caption}
\usepackage{subcaption}
\usepackage{sidecap}
\usepackage{longtable}
\usepackage{anyfontsize}
\usepackage{dsfont}

\usepackage[framemethod=default]{mdframed}
\newmdenv[skipabove=7pt,
skipbelow=7pt,
rightline=false,
leftline=false,
topline=false,
bottomline=false,
backgroundcolor=gray!10,
linecolor=gray,
innerleftmargin=5pt,
innerrightmargin=5pt,
innertopmargin=5pt,
innerbottommargin=5pt,
leftmargin=0cm,
rightmargin=0cm,
linewidth=4pt]{eBox}
\newmdenv[skipabove=7pt,
skipbelow=7pt,
rightline=false,
leftline=false,
topline=false,
bottomline=false,
backgroundcolor=gray!10,
linecolor=gray,
innerleftmargin=5pt,
innerrightmargin=5pt,
innertopmargin=-5pt,
innerbottommargin=5pt,
leftmargin=0cm,
rightmargin=0cm,
linewidth=4pt]{eBox2}

\definecolor{blue3}{RGB}{31, 119, 180}
\definecolor{red3}{RGB}{	214, 39, 40}
\definecolor{orange3}{RGB}{255, 127, 14}
\definecolor{green3}{RGB}{44, 160, 44}
\definecolor{repBlue}{RGB}{31, 119, 180}
\definecolor{repRed}{RGB}{	214, 39, 40}
\definecolor{repGreen}{RGB}{44, 160, 44}

\renewcommand{\(}{\left(}
\renewcommand{\)}{\right)}
\renewcommand{\[}{\left[}
\renewcommand{\]}{\right]}

\def\be{\begin{equation}}
\def\ee{\end{equation}}
\newcommand{\bea}{\begin{eqnarray}}
\newcommand{\eea}{\end{eqnarray}}

\usepackage{colortbl}
\definecolor{lightgreen}{cmyk}{0.2, 0, 0.2, 0.2}
\definecolor{lightgray}{cmyk}{0.1,0.2,0,0.1}
\definecolor{lightgray2}{cmyk}{0.1,0.1,0,0.1}

\makeatletter
\newlength{\apb@width}
\newcommand{\autoparbox}[2][c]{\settowidth{\apb@width}{#2}\parbox[#1]{\apb@width}{#2}}

\makeatother


\def\beq{\begin{equation}}
\def\eeq{\end{equation}}

\def\s{ \chi}

\allowdisplaybreaks[1]
\begin{document}

\newgeometry{top=2cm, bottom=2cm, left=2.9cm, right=2.9cm}

\begin{titlepage}
\setcounter{page}{1} \baselineskip=15.5pt 
\thispagestyle{empty}

\begin{center}
{\fontsize{20}{18} \bf On the IR divergences in de Sitter space: \\ [12pt]
\fontsize{17}{18} \it 
loops, resummation and the semi-classical wavefunction}\\ 
\end{center}

\vskip 20pt

\begin{center}
\noindent
{\fontsize{12}{18}\selectfont Sebasti\'an C\'espedes$^1$, Anne-Christine Davis$^2$ and Dong-Gang Wang$^2$}
\end{center}

\vskip 20pt

\begin{center}
  \vskip8pt
   {$^1$ \fontsize{12}{18}\it Department of Physics,  Imperial College, London, SW7 2AZ, UK
}

  \vskip8pt
{$^2$ \fontsize{12}{18}\it Department of Applied Mathematics and Theoretical Physics,\\ University of Cambridge,
Wilberforce Road, Cambridge, CB3 0WA, UK}

\end{center}

%
%

%=========================================
\vspace{0.4cm}
 \begin{center}{\bf Abstract} 
 \end{center}
 \noindent
In this  paper, we revisit the infrared (IR) divergences  in de Sitter (dS) space using the wavefunction method, and explicitly explore how the resummation of higher-order loops leads to the stochastic formalism.
In light of recent developments of the cosmological bootstrap, we track the behaviour of these nontrivial IR effects from perturbation theory to the non-perturbative regime.
Specifically, we first examine the perturbative computation of wavefunction coefficients, and show that
there is a clear distinction between classical components from tree-level diagrams and quantum ones from loop processes.
Cosmological correlators at loop level receive contributions from tree-level wavefunction coefficients, which we dub classical loops.
This distinction significantly simplifies the analysis of loop-level IR divergences, as we find the leading contributions always come from these classical loops. 
Then we compare with correlators from the perturbative stochastic computation, and find the results there are essentially the ones from classical loops, while quantum loops are only present as subleading corrections.
This demonstrates that the leading IR effects are contained in the semi-classical wavefunction which is a  resummation of all the tree-level diagrams.
With this insight, we go beyond perturbation theory and present a new derivation of the stochastic formalism using the saddle-point approximation. We show that the  Fokker-Planck equation follows as a consequence of two effects:
the drift from the Schr\"odinger equation that describes the bulk time evolution, and the diffusion from the Polchinski's equation which corresponds to the exact renormalization group flow of the coarse-grained theory on the boundary. {Our analysis highlights the precise and simple link between the stochastic formalism and the semi-classical wavefunction.}

\noindent

\end{titlepage}

\newpage

\restoregeometry
\setcounter{tocdepth}{3}
\setcounter{page}{1}
\tableofcontents

\newpage

%=======================================
\section{Introduction}
%=======================================

In the past several decades, great efforts have been made in understanding the infrared (IR) divergences in de Sitter (dS) space. 
At the conceptual level, this issue  has  important implications, such as for the radiative stability of dS spacetime, the measure problem of eternal inflation, and also the holographic description for our Universe.
Furthermore, the IR effects in dS are of phenomenological interest for inflationary cosmology. There, light scalars are supposed to be responsible for the quantum origin of cosmic structures, and correlation functions with nontrivial IR behaviour at the end of inflation may become primary targets that can be tested in upcoming cosmological observations.

\vskip4pt
In our current understanding, these IR divergences arise in the perturbation theory  of interacting light scalars.
For quantum field theories in a fixed dS background, it is widely recognized that 
correlators of these fields may exhibit secular growth on superhorizon scales  \cite{Ford:1984hs, Antoniadis:1985pj, Tsamis:1994ca, Tsamis:1996qm, Polyakov:2007mm,Polyakov:2009nq,Polyakov:2012uc, Burgess:2009bs, Seery:2010kh, Burgess:2010dd, Giddings:2010nc,Giddings:2010ui, Hu:2018nxy}.
These divergences already show up at tree level, and become more and more singular for higher-order loops, which in due course invalidate the perturbative analysis. Therefore in order to check whether final results are IR-safe or not, in principle one needs to go beyond perturbation theory and perform a resummation of all the higher-order loop diagrams, though in practice this seems to be out of reach.
Meanwhile, in parallel with the field-theoretic computations, a non-perturbative approach called the stochastic formalism was proposed  by Starobinsky et al., which applies the Fokker-Planck equation to describe the probability distribution of long wavelength perturbations during inflation~\cite{Starobinsky:1986fx,Starobinsky:1994bd}.
{For scalar field theories with potentials bounded from below}, it has been shown that ultimately, an equilibrium state will be reached for long modes on superhorizon scales. Thus final correlation functions remain IR-finite.  Since then, the stochastic formalism has been studied as a solution of the IR divergences in dS using various methods, see \cite{Garbrecht:2013coa,Garbrecht:2014dca,Burgess:2014eoa,Burgess:2015ajz,Moss:2016uix,Gorbenko:2019rza,Baumgart:2019clc,Mirbabayi:2019qtx,Mirbabayi:2020vyt,Pinol:2020cdp,Cohen:2020php,Cohen:2021fzf,Cohen:2021jbo,Cable:2022uwd,Cable:2023gdz,Honda:2023unh,Palma:2023idj} for recent examples.

\vskip4pt
In spite of the extensive analysis {and various derivations in the literature}, some basic questions remain about the stochastic formalism. The first one is about the explicit connection with the field-theoretic approaches.
While the equilibrium state emerges as a solution of the Fokker-Planck equation in the non-perturbative regime, {it is interesting to see how this description is connected to the IR divergences in field-theoretical computations. Recently it has been shown that the original stochastic formalism 
is the leading approximation of strong IR effects in de Sitter space \cite{Gorbenko:2019rza,Baumgart:2019clc}. Next, we can go one step further and ask, what does the stochastic formalism actually resum, or which part of the field-theoretical computation is being resummed by the
original Fokker-Planck equation?}
\footnote{{This is a different question from the derivation of the stochastic formalism.} In the literature, the pertubative stochastic computation was found to match 1-loop corrections from in-in formalism \cite{Tsamis:2005hd,Tokuda:2017fdh}, while more recently Ref.~\cite{Baumgart:2019clc}  showed that the leading divergences from higher-loop diagrams agree with the stochastic results at the corresponding order. We refer the reader to the end of Section \ref{sec:resum} for detailed discussions.}
Since the computation becomes more and more complicated as we increase the order of  loops, at first sight it looks hopeless to provide an exact answer to this question.
Meanwhile, the stochastic formalism is essentially a coarse-grained description for long wavelength modes where short wavelength ones are being smoothed out.  
This recalls to the similar idea of the Wilsonian effective field theory: integrating out high energy modes leads to a low-energy description which changes with the ultraviolet (UV) cutoff scale as the renormalization (RG) group flow. It is very tempting to see how the story in cosmology can be phrased in terms of this Wilsonian picture.

\vskip4pt
Another interesting question  concerns the situation where we have multiple interacting light scalars in de Sitter space. For studies of cosmic inflation, this corresponds to the multi-field models, where the inflaton is not the only light scalar degree of freedom.
Unlike the inflaton field, the additional light scalars have less constraints and typically lead to IR divergences with various forms~\cite{Achucarro:2016fby}.
Recently, a bootstrap analysis was performed within perturbation theory, and it was shown that the conversion process in multi-field models is actually described by the IR-singular secular growth in the field-theoretic approach \cite{Wang:2022eop}.
This in the end provides a rigorous derivation for the well-known prediction of local non-Gaussianity from multi-field inflation.
Meanwhile a non-perturbative treatment using the stochastic formalism is presented in Ref. \cite{Achucarro:2021pdh}~(See also  \cite{Panagopoulos:2019ail}). 
It remains to be seen how new phenomenologies of primordial non-Gaussianity would appear in the non-perturbative regime of the multi-field system.

\vskip4pt
Motivated by these questions, the purpose of this work is less of making conceptually new discoveries on this extensively-discussed topic, rather than trying to gain a better understanding using the recently developed wavefunction approach for cosmological perturbations.
Our investigation is initiated by  the rapid development of the {\it cosmological bootstrap} in the past several years, where a timeless description of cosmological observables is being constructed by imposing consistency requirements \cite{Arkani-Hamed:2018kmz,Baumann:2019oyu,Baumann:2020dch, Arkani-Hamed:2017fdk,Arkani-Hamed:2018bjr,Benincasa:2018ssx, Sleight:2019mgd, Sleight:2019hfp,Sleight:2020obc, Pajer:2020wxk, Jazayeri:2021fvk,Bonifacio:2021azc, Pimentel:2022fsc, Jazayeri:2022kjy, Goodhew:2020hob, Cespedes:2020xqq, Melville:2021lst, Goodhew:2021oqg, Baumann:2021fxj, Meltzer:2021zin,Hogervorst:2021uvp,DiPietro:2021sjt, Baumann:2022jpr, Goodhew:2022ayb, Qin:2022fbv, Xianyu:2022jwk, Salcedo:2022aal, Wang:2022eop,DuasoPueyo:2023viy}. A central object in the bootstrap analysis is the wavefunction of the Universe $\Psi[\phi({\bf x}), t]$, which is a functional of boundary field fluctuations ~\cite{Maldacena:2002vr,Harlow:2011ke,Anninos:2014lwa}.
For a quantum system, the probability distribution is famously related to the wavefunction   as
\be
P = |\Psi|^2~.
\ee
Meanwhile the Fokker-Planck equation is a description for the evolution of the probability distribution of long wavelength perturbations.
In this sense, for the purpose of building the link between the field-theoretic and  stochastic approaches, the wavefunction of the Universe provides the most obvious object to look at. 
This line of thinking has been explored in Ref.~\cite{Gorbenko:2019rza}, where a systematic derivation of the stochastic formalism is presented using the wavefunction method.

\vskip4pt
In this paper,  we adopt a similar approach to track the behaviour of IR-divergent correlators from the perturbation theory to the non-perturbative regime. 
{Built on previous studies, we further identify the origin of leading IR effects, and find  a precise link between the stochastic formalism and the semi-classical piece of the wavefunction. This new understanding helps us simplify the analysis of IR divergences with  more detailed results through  concrete examples and explicit computations.}
In perturbation theory, we will propose the notion of \textit{classical loops}, and show that for diagrams beyond tree level, these are the leading contributions to loop-level correlators with IR divergences. Then through explicit comparison with the perturbative stochastic computation, we will demonstrate the stochastic formalism as a resummation of these classical loop contributions, while the quantum loops only lead to subleading corrections.

\vskip4pt
Beyond  perturbation theory, we further explore the implications of the semi-classical wavefunction, which now by itself is a resummation of all the tree-level contributions. 
Considering a coarse-grained theory on the boundary, { we apply Polchinski's RG flow techniques to derive a non-perturbative equation for the probability distribution. This  RG-type equation in the end} leads to the diffusion term in the Fokker-Planck equation, while the drift term results from the Schr\"odinger equation which describes the time evolution of the semi-classical wavefunction.

\vskip4pt
The  structure of the paper is organized as follows. In Section \ref{sec:wav}, we first briefly review the wavefunction method for computing cosmological correlation functions in perturbation theory. Here, we specifically make a distinction between classical and quantum effects, and
propose the notion of quantum and classical loops for the loop-level correlators.
After this preparation, we move to look into IR divergences in de Sitter space.
In Section \ref{sec:pert}, we perform the perturbative computations from various types of interacting theories of light scalars. Through the explicit one-loop analysis and generalizations to multi-loop diagrams, we establish that the leading IR-divergent contributions to correlators always originate from classical loops.
In Section \ref{sec:resum}, we explicitly compare the perturbative results from  the wavefunction computation and the stochastic formalism, and identify that the latter approach is resumming classical loops.
In Section \ref{sec:beyond}, we go beyond perturbation theory and re-derive the Fokker-Planck equation by identifying the diffusion as a consequence of the boundary exact renormalization group (RG) equation for the semi-classical wavefunction. 
We conclude in Section \ref{sec:concl} with a summary and outlook.

\vskip4pt
More technical details are presented in the Appendices. In Appendix \ref{app:inin},  we perform the in-in computation for one-loop IR-divergent correlators, which match the results from the wavefunction formalism in Section \ref{sec:1loop}.
In Appendix \ref{app:wave}, we use the coarse-grained wavefunction to reproduce the probability distributions of long wavelength fluctuations within perturbation theory, and show that the contributions from tree-level wavefunction coefficients satisfy the Fokker-Planck equation, which provides another justification for the relation between the stochastic formalism and the semi-classical wavefunction. 
In Appendix \ref{app:2field}, we extend the derivation of the Fokker-Planck equation to a two-field system.
Throughout the paper, we mainly take the convention of natural units $c = \hbar = 1$, with the metric signature $(-, +, +, +)$, except for special cases (as in Section \ref{sec:wav}) where we keep $\hbar$ explicit for the distinction of quantum and classical effects. The bulk fields are denoted by $\Phi$, $\Sigma$, while  $\phi$ and $\chi$ are boundary fluctuations.

%\newpage
\section{The Wavefunction Method: Classical {\it vs} Quantum}
\label{sec:wav}

Before delving into the resolution of the infrared divergence problem in de Sitter space, in this section we first outline the wavefunction of the Universe formalism. Our primary goal here is to derive the Feynman rules required for perturbative calculations  by employing functional quantization in a rigorous manner. We explicitly delineate the clear demarcation between classical and quantum processes within this framework, with both types of processes making contributions to the cosmological correlation functions at the level of loop diagrams.

\vskip6pt

The wavefunction method for computing correlation functions offers significant advantages, both in terms of conceptual clarity and computational efficiency.
For a quantum system, the wavefunction contains all the information about correlations and is naturally a more primitive object to study.
In cosmology, this has been appreciated in the recent advances of the {\it Cosmological Bootstrap} program, where we have a collection of quantum fields on a fixed de Sitter background and the objects of interest are  correlations on the late-time boundary of the spacetime. 
In perturbation theory, it has been shown that the analytical structures of the wavefunction are nicely manifested as consequences of unitarity, locality and spacetime symmetry. Meanwhile, at the practical level, the wavefunction in many cases becomes a much simpler object for computations.
In this section, we review the wavefunction method for computing correlation functions, and in particular we turn our attention to examining the loop-level diagrams of the cosmological correlators, 
highlighting the distinctions between classical and quantum contributions there.

\vskip4pt
In general, for the quantum-mechanical state of a given system $| {\Psi} \rangle$, the wavefunction is defined  in the Heisenberg picture as its inner product with the basis of field eigenstates $|\phi(t_0, {\bf x})\rangle$ at a fixed time $t_0$
\be \label{wavef:def}
\Psi[\phi({\bf x})]\equiv  \langle \phi(t_0, {\bf x}) |{\Psi} \rangle ~.
\ee
Typically we are interested in the interacting vacuum state $| {\Psi} \rangle = |\Omega \rangle$ defined at the infinite past ($t_i=-\infty$), and thus the wavefunctional $\Psi[\phi(x)]$ describes the overlap  between the vacuum state and a given field profile $\phi(x)$ at $t_0$.
Formally, the wavefunction of this scalar field $\phi$ can be presented as the following path integral
\be \label{pathInt}
\Psi[\phi({\bf x})] =\int\limits_{\substack{\hspace{-0.3cm}\Phi(t_0) \,=\,\phi\\ \hspace{-0.5cm}\Phi(-\infty)\,=\,0}}
\mathcal{D}\Phi \exp\({\frac{i}{\hbar} S[\Phi]}\)~,
\ee
where we assume the bulk field $\Phi$ goes to zero (vacuum) in the infinite past, and is given by the profile $\phi({\bf x})$ at the late time $t_0$. 
Then the path integral sums over all the possible field configurations with these two fixed boundary conditions. 
Recall that in QFT textbooks, the path integrals normally have a final state at the infinite future ($t_f\rightarrow+\infty$), the major difference in the wavefunction formalism is the presence of a fixed future boundary.
This description applies to general FLRW spacetimes, including Minkowski and de Sitter spaces.
For the rest of this section, we shall keep the analysis formal and general, which may simply extend to other spacetime background and field components as well. 

\vskip4pt
Another thing to notice is that in \eqref{pathInt} we keep $\hbar$  explicit,
for the convenience of tracking quantum and semi-classical contributions in the wavefunction method. 
As is well-known, in the path integral quantization the distinction between classical and quantum physics can be clearly demonstrated by taking the $\hbar\rightarrow0$ limit. 
There the path integral is dominated by the configuration $\phi(t,{\bf x})$ which has an extremum with $\delta S = 0$. This configuration precisely leads to the Euler-Lagrange equation satisfied by a classical field.
In standard quantum field theory, one common lore is that the processes that can be described by the classical equations of motion give rise to tree-level diagrams, while loop diagrams, which are truly quantum effects, cannot be generated in classical field theories and are only caused by  higher-order (in $\hbar$) corrections to the classical computation.
This difference is formally demonstrated by the Schwinger-Dyson equation where the contact terms with $\hbar$ are responsible for generating loop processes.
In the rest of this section, we will clarify that for cosmology, this distinction between classical and quantum effects is similarly present in the wavefunction method, and the basic object of interest here is the wavefunction coefficients.

\subsection{Classical and Quantum Wavefunction Coefficients}

In perturbation theory, one convenient way to study  the wavefunction is to write it as an expansion in powers of the field fluctuations on the boundary. After the spatial Fourier space transformation, it can usually be expressed as
\begin{align}
\Psi[\phi] =& \exp \[  \frac{1}{2} \int_{{\bf k}_1,{\bf k}_2} 
%\frac{{\rm d}^{3}k_1 {\rm d}^{3}k_2}{(2\pi)^{6}} 
\psi_2({\bf k}_1,{\bf k}_2) ~\phi_{{\bf k}_1}\phi_{{\bf k}_2} + \sum_{n = 3}^{\infty}\frac{1}{n!}\int_{{\bf k}_1,...,{\bf k}_n}
%\frac{{\rm d}^{3}k_1\cdot\cdot\cdot {\rm d}^{3}k_n}{(2\pi)^{3n}}
\, \psi_n({\bf k}_1,...,{\bf k}_n)  ~\phi_{{\bf k}_1}\cdot\cdot\cdot \phi_{{\bf k}_n}\,
%(2\pi)^3  \delta({\bf k}_1+\cdots+{\bf k}_n)\, \psi_n({\bf k}_1,...,{\bf k}_n) 
\]\,,
 \label{Psi:coeff}
\end{align}
where the momentum integrals are given by
\be
\int_{\bf k} \equiv \int\frac{{\rm d}^3k}{(2\pi)^3}~,
\ee
and $\psi_2$, $\psi_3$, ... are the wavefunction coefficients in Fourier space. These are the central objects in our analysis, which encode the statistics of $\phi$ and can also be seen as correlations functions of boundary CFT operators.
In the following we shall derive the Feynman rules for computing $\psi_n$, and focus on the differences in the computations of tree-level and loop-level processes. 
For demonstration, we mainly use the following simple scalar field theory in a general FLRW universe as a concrete example
\be \label{example}
S[\Phi] = \int {\rm d}\eta {\rm d}^3x~a(\eta)^4\[ -\frac{1}{2}\(\partial_\mu\Phi\)^2 - \frac{1}{2} m^2 \Phi^2 -\frac{g}{3!} \Phi^3 \]~,
\ee
where the spacetime metric is the one with conformal time $\eta$: ${\rm d}s^2=a(\eta)^2\[-{\rm d}\eta^2+{\rm d}{\bf x}^2\right]$, and we choose the mass and coupling to be independent of $\hbar$.

\paragraph{Tree diagrams} The derivation of Feynman rules for the tree diagrams has been well presented in the literature. Here we briefly summarize the basic steps, and the readers may refer to \cite{Maldacena:2002vr,Anninos:2014lwa,Goodhew:2020hob,Bauman:2023aj} for more details.
The starting point is the saddle-point approximation, where the action is on-shell
\be \label{saddle}
\Psi[\phi]  \simeq  \exp\({\frac{i}{\hbar} S[\Phi_{\rm cl}]}\)~,
\ee
with $\Phi_{\rm cl}$ being the solution of the classical equation of motion. For the example in \eqref{example}, this is given by
\be \label{eom}
(\square -m^2)\Phi_{\rm cl} = \frac{g}{2} \Phi_{\rm cl}^2~,
\ee
where $\square = g^{\mu\nu} \nabla_\mu \nabla_\nu$ is the d'Alembert operator in the curved spacetime. 
We refer to the saddle-point approximation in \eqref{saddle} as the {\it semi-classical wavefunction}, which will be of central interest in our analysis.
Next we work in the Fourier space aiming to derive the explicit expressions of wavefunction coefficients in \eqref{Psi:coeff}. As we impose the boundary condition $\Phi_{\rm cl}(\eta_0,{\bf k}) = \phi_{\bf k}$, formally the on-shell equation \eqref{eom} can be solved as
\be \label{formalsol}
\Phi_{\rm cl}(\eta, {\bf k}) = \phi_{\bf k} K( k,\eta) + \frac{i}{\hbar} \left.\int {\rm d} \eta' G(k; \eta, \eta') \frac{\delta S_{\rm int}}{\delta \Phi_{\bf k}(\eta')}\right|_{\Phi=\Phi_{\rm cl}}~.
\ee
Here the first term is the part that solves the {\it free} equation of motion, and $K(k, \eta)$ is also known as the bulk-to-boundary propagator given by
\be \label{btod}
\left(\square_k -m^2\right) ~K(k,\eta) = 0~~~~~~{\rm with}~~ \lim_{\eta\rightarrow \eta_0}K(k,\eta)=1 ~~{\rm and}~~ \lim_{\eta\rightarrow -\infty}K(k,\eta)=0~,
\ee 
where $\square_k$ is the spatial Fourier transformation of the d'Alembert operator.
The second term treats the interaction as a source, and introduces the Green's function $G(k; \eta, \eta')$  which satisfies
\be \label{green}
\left(\square_k -m^2\right)~ G(k;\eta,\eta') = \frac{i\hbar}{a^4}\delta(\eta-\eta')~~~~~~{\rm with}~~ \lim_{\eta,\eta'\rightarrow -\infty}G(k;\eta,\eta')  =\lim_{\eta,\eta'\rightarrow \eta_0} G(k;\eta,\eta') =0~.
\ee
In classical theories, the appearance of $\hbar$ is not necessary as in the end \eqref{formalsol} is $\hbar$-independent. Here it has been introduced in such a way  to match the one in a quantum theory as we shall show very soon.
For those familiar with bulk-to-bulk propagators in the in-in formalism, one nontrivial difference to notice is that the wavefunction $G$ propagator has a vanishing boundary condition as $\eta\rightarrow\eta_0$.

\begin{figure} [t]
   \centering
            \includegraphics[width=.8\textwidth]{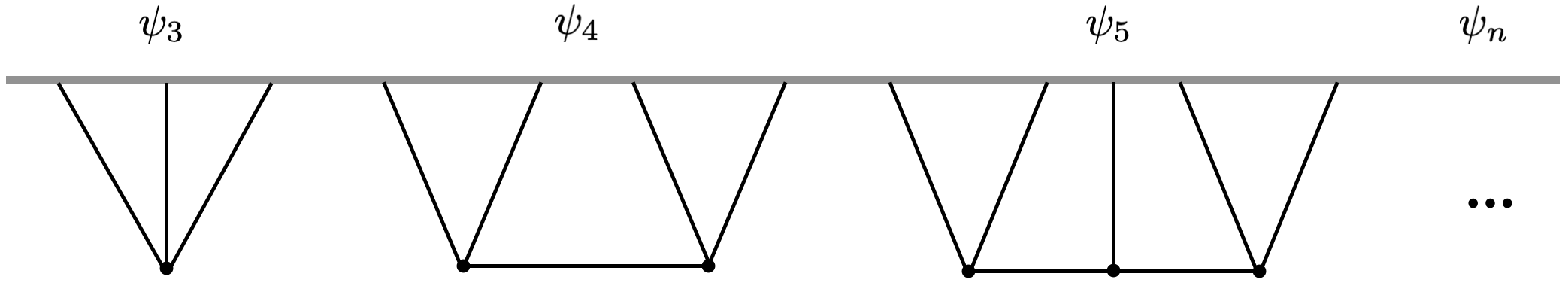}  
   \caption{(Semi-)classical wavefunction coefficients from tree-level diagrams.}
  \label{fig:trees}
\end{figure}

\vskip4pt
Then the tree-level wavefunction coefficients can be derived by substituting the formal solution \eqref{formalsol} into the saddle-point wavefunction expression \eqref{saddle} iteratively. Explicitly, up to $g^2$ order for the example in \eqref{example}, the action can be written into the following form
\begin{small}
\begin{align}
S[\Phi_{\rm cl}] = &  \int_{{\bf k}_1,{\bf k}_2}    \left[\left. \frac{1}{2}a(\eta)^2 \Phi_{{\bf k}_1} \partial_\eta\Phi_{{\bf k}_2} \right|_{\eta=\eta_0}  +
\frac{1}{2}\int {\rm d}\eta 
  a(\eta)^4  \Phi_{{\bf k}_1}  ( \square_k -   m^2) \Phi_{{\bf k}_2} \right]{(2\pi)^3}\delta^{(3)}({\bf k}_1+{\bf k}_2)  \nn\\
 & -\frac{g}{3!} \int_{{\bf k}_1,{\bf k}_2,{\bf k}_3}  \int {\rm d}\eta a(\eta)^4 K(k_1,\eta)K(k_2,\eta)K(k_3,\eta) \phi_{{\bf k}_1}\phi_{{\bf k}_2}\phi_{{\bf k}_3} {(2\pi)^3}\delta^{(3)}({\bf k}_1+{\bf k}_2+{\bf k}_3)\nn\\
& + \frac{ig^2}{4!\hbar} \int_{{\bf k}_1,...,{\bf k}_4}  \int  {\rm d}\eta \int  {\rm d}\eta' a(\eta)^4 a(\eta')^4 \[  K(k_1,\eta)K(k_2,\eta) G(s;\eta,\eta')K(k_3,\eta')K(k_4,\eta')\right.\nn\\
&~~~~~~~~~~~~\left. {(2\pi)^3}\delta^{(3)}({\bf k}_1+{\bf k}_2+{\bf s}){(2\pi)^3}\delta^{(3)}({\bf k}_3+{\bf k}_4-{\bf s})  + \text{$t$- and $u$-channels}\] \phi_{{\bf k}_1}\phi_{{\bf k}_2}\phi_{{\bf k}_3}\phi_{{\bf k}_4} \nn\\
&+ ...~
 \label{Psi:tree}
\end{align}
\end{small}Putting it back into \eqref{saddle} and comparing with the notation in \eqref{Psi:coeff}, we identify the tree-level wavefunction coefficients.
The first line in \eqref{Psi:tree} comes from the free theory: the boundary term yields the two-point wavefunction coefficient  $\psi_2^{\rm free} = i \hbar^{-1} a(\eta_0)^2 \partial_{\eta_0} K(k_1,\eta_0)  {(2\pi)^3}\delta^{(3)}({\bf k}_1+{\bf k}_2) $, while the second term is proportional to the classical equation of motion and thus goes to zero once the on-shell condition is imposed.
The second line gives the 3-point function $\psi_3$ from the contact diagram shown by the first diagram in Figure \ref{fig:trees}; the third line generates the single-exchange 4-point function $\psi_4$ as shown by the second diagram in Figure \ref{fig:trees}.
Continuing this process, we can generate the wavefunction results for all the tree-level diagrams, such as the 5-point function from the double-exchange diagram and so on.

\vskip4pt
From this explicit derivation, it is clear that all the tree diagrams follow from the semi-classical wavefunction \eqref{saddle} and the formal solution of the classical equation \eqref{formalsol}.
Diagrammatically, we can compute these wavefunction coefficients in perturbation theory using the following Feynman rules: {\it i}) assign $ig/\hbar$ to each vertex; {\it ii}) assign a $K$ propagator to each external line and a Green's function $G$ to each internal line; {\it iii}) integrate over the bulk time for all the insertions ${\rm d}\eta ~a(\eta)^4$. 
As there is an $
\hbar^{-1}$ with each insertion and a $\hbar$ for each internal line, all the tree-level $\psi_n$s   have a $\hbar^{-1}$, which is expected from our starting point -- the saddle-point approximation \eqref{saddle}.

\paragraph{Loop diagrams}

Notably, the above derivation cannot generate loops.
For flat spacetime scattering amplitudes, it is well-known that tree-level processes can also arise in classical field theory, while loop diagrams are only present in a quantum theory.
Similar argument is expected in cosmology as well, but has not been explicitly presented in literature yet. To identify the truly quantum effects, we need to go beyond the saddle-point approximation \eqref{saddle}.
 Now we explicitly look into the origin of loops in the wavefunction method.

\vskip4pt
Let's first consider the functional quantization of a free theory in the wavefunction formalism. Without imposing the on-shell condition, the path integral \eqref{pathInt} can be expressed in the following form
\begin{small}
\be
\Psi_0[\phi ] =\int\limits_{\substack{\hspace{-0.3cm}\Phi(t_0) \,=\,\phi\\ \hspace{-0.5cm}\Phi(-\infty)\,=\,0}}
\mathcal{D}\Phi \exp\[{\frac{i}{2\hbar}  \int _{{\bf k},{\bf k}'}   \left(a^2\left.    \phi_{{\bf k}} \partial_\eta \Phi_{{\bf k'}} \right|_{\eta=\eta_0}  +
 \int {\rm d}\eta 
  a(\eta)^4 \Phi_{{\bf k}'}  ( \square_k -   m^2) \Phi_{{\bf k}} \right)}(2\pi)^3 \delta^{(3)}({\bf k}+{\bf k}')\]~,
\ee
\end{small}The main difference with the first line in \eqref{Psi:tree} is that now the second term is no longer zero as the on-shell condition becomes invalid in a quantum theory. While the boundary term is fixed as in  classical theories, the integration over $\Phi$ can be solved as
\be
\Psi_0 [\phi] =    \exp\(\frac{1}{2}  \int_{{\bf k},{\bf k}'}   \psi_2^{\rm free} \phi_{\bf k} \delta\phi_{{\bf k}'} \)    \[{\rm det}\(
 -i(\square_k-m^2 )/2\pi\hbar\)\]^{-1/2}~.
\ee
Furthermore, using the Gaussian integral, we can compute the two-point function of the bulk field $\Phi$
\be \label{b2bw}
\langle \Phi_{\bf k}(\eta) \Phi_{\bf k'}(\eta') \rangle \equiv \frac{\displaystyle \int   \mathcal{D}\Phi \Phi_{\bf k}(\eta) \Phi_{\bf k'}(\eta') \exp\({\frac{i}{\hbar} S_0[\Phi]}\) }{\displaystyle\int   \mathcal{D}\Phi  \exp\({\frac{i}{\hbar} S_0[\Phi]}\) }  = G(k,\eta,\eta') (2\pi)^3 \delta^{(3)}({\bf k}+{\bf k}')~,
\ee
where $G(k,\eta,\eta')$ is the same with the Green's function  \eqref{green} defined in classical theories. 
As  is clear from this new definition, $G$ is also called the {\it bulk-to-bulk} propagator which takes a bulk field from a time $\eta'$ to another time $\eta$. 
Now it also becomes obvious for our choice of putting $\hbar$ in \eqref{green}. In the quantum theory, the $G$ propagators carry an $\hbar$ by definition.
As an aside, because of the boundary condition at $\eta_0$ in \eqref{green}, this bulk-to-bulk propagator differs from the one in the in-in formalism, which will become important for our discussion on the IR divergences later. The bulk-to-boundary propagator $K$ can be introduced in a similar manner by using the conjugate momentum $\Pi = a^2\partial_\eta \Phi$ on the boundary
\be \label{b2dw}
\langle \Pi_{\bf k}(\eta_0) \Phi_{\bf k'}(\eta) \rangle = a(\eta')^2 \partial_{\eta'} \langle \Phi_{\bf k}(\eta') \Phi_{\bf k'}(\eta) \rangle|_{\eta'=\eta_0} = i\hbar K(k,\eta) (2\pi)^3 \delta^{(3)}({\bf k}+{\bf k}') ~,
\ee
where in the second equality we have used the identity $ K(k,\eta) = a(\eta')^2 \partial_{\eta'} G(k,\eta',\eta)|_{\eta'=\eta_0}$.

\vskip4pt
Next, we move to compute wavefunction coefficients in perturbation theory. In the path integral formalism, they are expressed as the following matrix element 
\be \label{psin:def}
\psi_n ({\bf k}_1,...,{\bf k}_n) \equiv \left.\frac{\delta^n \Psi[\phi]}{\delta\phi_{{\bf k}_1}...\delta\phi_{{\bf k}_n}}\right|_{\phi=0} 
= \frac{(i/\hbar)^n}{\Psi[0]}  \int   \mathcal{D}\Phi \Pi_{{\bf k}_1}(\eta_0)... \Pi_{{\bf k}_n}(\eta_0) \exp\(\frac{i}{\hbar} S[\Phi]\)~.
\ee
With the presence of interactions, 
we can perturbatively expand the exponential function in  \eqref{psin:def} into $ \exp \(iS_0/\hbar\)\(1 +i S_{\rm int}/\hbar + ...\right)$ and perform the functional integral to produce Wick contractions. 
This procedure   leads to the Feynman rules for computing  wavefunction coefficients.
The contractions between $\Pi$ and $\Phi$ are given by \eqref{b2dw},
which simply means that the external legs are associated with the bulk-to-boundary propagators.
Then we also find time-ordered correlation functions of the bulk field $\Phi$, whose contractions yield $G$ propagators in \eqref{b2bw}.
 At the tree-level, this derivation is in precise agreement with the Feynman rules in classical theories, as it should be.
 However, one expects nontrivial consequences in a quantum theory, as it also becomes possible to contract bulk fields in such a way that closed loops can be formed using the bulk-to-bulk propagators.

\begin{figure} [t]
   \centering
            \includegraphics[width=.7\textwidth]{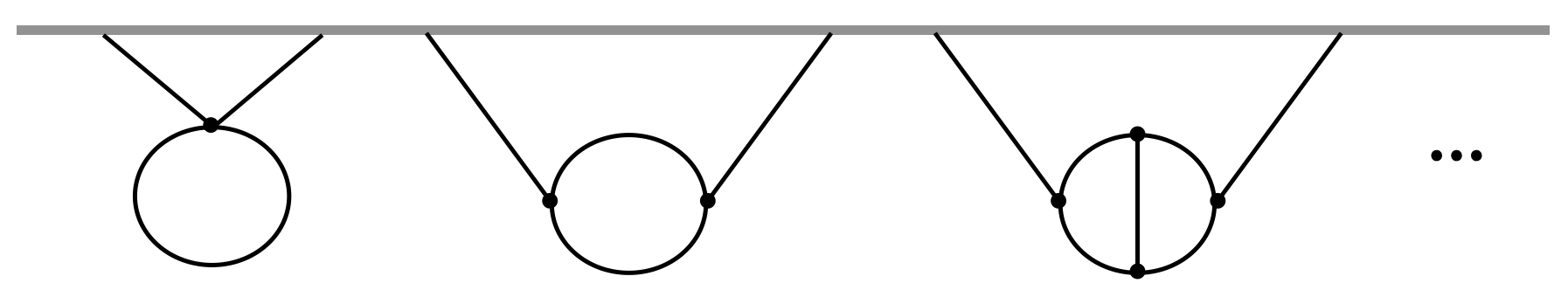}  
   \caption{Quantum wavefunction coefficients from loop-level diagrams.}
  \label{fig:loops}
\end{figure}

 \vskip4pt
 As a concrete example, let's take a look at the leading correction to the two-point wavefunction coefficient in the quantum version of the $\Phi^3$ theory in \eqref{example}.
 At the $g^2$ order, this correction corresponds to a one-loop process, which is explicitly given by
\begin{small}\begin{align}
 \label{psi2:loop}
 \psi_2^{\rm 1-loop}  =&  \frac{g^2}{(3!)^2\hbar^4  \Psi[0]} \int   \mathcal{D}\Phi \Pi_{{\bf k}}(\eta_0)\Pi_{{\bf k}'}(\eta_0) \int_{\rm x,y}  {\rm d}\eta_1  {\rm d}\eta_2 a(\eta_1)^4 a(\eta_2)^4 \Phi(\eta_1,{\bf x})^3 \Phi(\eta_2,{\bf y})^3 \exp\(\frac{i}{\hbar} S_0[\Phi]\) \nn\\
 =& \frac{g^2}{(3!)^2\hbar^4 } \int  {\rm d}\eta_1  {\rm d}\eta_2 a(\eta_1)^4 a(\eta_2)^4 \int_{{\bf p}_1,{\bf p}_2,{\bf p}_3,{\bf q}_1,{\bf q}_2,{\bf q}_3} {(2\pi)^3}\delta^{(3)}({\bf p}_1+{\bf p}_2+{\bf p}_3){(2\pi)^3}\delta^{(3)}({\bf q}_1+{\bf q}_2+{\bf q}_3)
 \nn\\
 & ~~~~~~ {\displaystyle \times 
 \langle \Pi_{{\bf k}}(\eta_0)\Pi_{{\bf k}'}(\eta_0)   \Phi(\eta_1,{\bf p}_1) \Phi(\eta_1,{\bf p}_2) \Phi(\eta_1,{\bf p}_3) \Phi(\eta_2,{\bf q}_1) \Phi(\eta_2,{\bf q}_1) \Phi(\eta_2,{\bf q}_3)  \rangle }~,
\end{align}
\end{small}where in the second line we have performed the Fourier transformation for $\bf x$ and $\bf y$.
Next we can apply the Wick contractions using \eqref{b2bw} and \eqref{b2dw}. In addition to the contraction between the bulk field and the boundary conjugate momentum, there are also two contractions of two $\Phi$ fields  which form a closed loop. In the end the wavefunction coefficient becomes
\be \label{psi2:phi3}
 \psi_2^{\prime\rm 1-loop}  = -\frac{g^2}{2} \int{\rm d}\eta_1  {\rm d}\eta_2  a(\eta_1)^4 a(\eta_2)^4    K (k,\eta_1)  K (k,\eta_2)   \int_{\bf p} G(p,\eta_1,\eta_2) G(|{\bf k}+{\bf p}|,\eta_1,\eta_2)  ~,
\ee
where
the prime means the momentum conserving $(2\pi)^3\delta^{(3)} ({\bf k}+{\bf k}')$ has been stripped. Keeping doing this computation to higher orders in perturbation theory, we find higher-loop corrections to $\psi_2$, whose full expression   is given by 
\be \label{psi2-loops}
\psi_2 ({\bf k},{\bf k}') \equiv \left.\frac{\delta^2 \Psi[\phi]}{\delta\phi_{{\bf k}}\delta\phi_{{\bf k}'}}\right|_{\phi=0}  = \psi_2^{\rm free} + \psi_2^{\rm 1-loop} + \psi_2^{\rm 2-loop} + ...
\ee
As we can see, these corrections to the two-point function cannot arise in the classical computation using saddle-point approximations and on-shell conditions. This is because in a classical theory one cannot contract two bulk fields within a correlator as we have done in \eqref{psi2:loop}, while for quantum theories  this is allowed through contact terms in the Schwinger-Dyson equation \cite{Schwartz:2014sze}.
Thus the loop-level wavefunction coefficients have a truly quantum origin.

\vskip4pt
Now we need to add one more Feynman rule for the computation of wavefunction coefficients in quantum theories: use multiple bulk-to-bulk propagators to form loops and integrate over the loop momentum.
Another interesting way to see the quantum origin of loop diagrams is to notice that the loop expansion in \eqref{psi2-loops} is also an expansion in $\hbar$. 
Since there is an $\hbar$ for each bulk-to-bulk propagator, for wavefunction coefficients the order of $\hbar$ increases with the number of loops $L$ as $\hbar^{L-1}$, while all the tree-level wavefunction coefficients are order $\hbar^{-1}$.

\vskip6pt
In summary, by deriving the Feynman rules from scratch for the wavefunction method,  we identify that classical and quantum effects are associated with tree-level and loop-level wavefunction coefficients respectively.\footnote{Strictly speaking, the tree-level wavefunction coefficients are due to semi-classical effects, as the generation of field fluctuations in cosmology is  quantum. Throughout this paper, we do not distinguish ``classical" from  ``semi-classical", but just keep in mind that we are analyzing quantum field theories in curved spacetime.} This is in analogy with our understanding for  scattering amplitudes in Minkowski spacetime.
For cosmology, the distinction between classical and quantum effects is somehow obscured
in the in-in formalism computation for correlation functions. One explanation  is that, as shown by the definition \eqref{wavef:def},  the wavefunction  is an {\it in-out} object  like the S-matrix in Minkowski spacetime but with a fixed-time future boundary (see \cite{Salcedo:2022aal, Agui-Salcedo:2023wlq, Melville:2023kgd} for elaboration of this perspective). 
Therefore it is more natural to discuss quantum and classical origins in the wavefunction coefficients. While the equal-time cosmological correlators are  intrinsically {\it in-in}, and thus the quantum and classical contributions are in general mixed.
Meanwhile, for cosmological observations we are interested in correlation functions at the end of inflation, which can be related to the late-time measurement in CMB and large-scale structure surveys.
Next, we shall see for loop diagrams how the combination of classical and quantum wavefunction coefficients lead us to these physical observables.

\subsection{Classical Loops in Cosmological Correlators}

In perturbation theory, the way to compute equal-time cosmological correlators of boundary field fluctuations  from the wavefunction of the Universe is to apply the Born rule
\begin{equation} \label{bornrule}
\langle\phi_{{\bf k}_1}\cdots \phi_{{\bf k}_n}\rangle =\frac{ \displaystyle\int{\cal D} \phi\,~\phi_{{\bf k}_1}\cdots \phi_{{\bf k}_n} \left\lvert\Psi[\phi, \eta_0]\right\rvert^2}{\displaystyle\int{\cal D} \phi ~ \left\lvert\Psi[\phi,\eta_0]\right\rvert^2}\,.
\end{equation}
With the wavefunction coefficients, we can do the Taylor expansion and then perform the Gaussian integral over the boundary field $\phi$ and derive the relations with the boundary correlators. 
At the tree-level, these equations are simple algebraic ones. For instance, the two-, three- and four-point correlators of the $\Phi^3$ theory in \eqref{example} are related to the corresponding wavefunction coefficients via
\bea
\langle \phi_{\bf k}\phi_{-{\bf k}}  \rangle' &=& -\frac{1}{2\hskip 1pt{\rm Re}\hskip 1pt\psi_2'(k)} \,, \label{equ:2pt}\\
\langle \phi_{{\bf k}_1}\phi_{{\bf k}_2}  \phi_{{\bf k}_3} \rangle' &=& -\frac{{\rm Re}\hskip 1pt\psi_3'({\bf k}_1,{\bf k}_2,{\bf k}_3)}{4   {\rm Re}\hskip 1pt\psi_2'(k_1) ~ {\rm Re}\hskip 1pt\psi_2'(k_2)~{\rm Re}\hskip 1pt\psi_2'(k_3)}\,.
\label{equ:3pt}\\
\langle \phi_{{\bf k}_1}\phi_{{\bf k}_2}  \phi_{{\bf k}_3}\phi_{{\bf k}_4} \rangle' &=&   \frac{{\rm Re}\hskip 1pt\psi_4'({\bf k}_1,{\bf k}_2,{\bf k}_3,{\bf k}_4) }{8 \prod_{a=1}^4 {\rm Re}\hskip 1pt \psi_2'(k_a)} - \langle \phi_{{\bf k}_1}\phi_{{\bf k}_2}  \phi_{{\bf k}_3}\phi_{{\bf k}_4} \rangle'_{\rm d} \, , \label{equ:4pt}
\eea
where the 
{\it disconnected} part has contributions from a product of two $\psi_3$
\begin{small}
\be \label{disconn}
\langle \phi_{{\bf k}_1}\phi_{{\bf k}_2}  \phi_{{\bf k}_3}\phi_{{\bf k}_4} \rangle'_{\rm d} =
\frac{1}{{8 \prod_{a=1}^4 {\rm Re} \hskip 1pt\psi_2'(k_a)}} \[ \frac{{\rm Re}\hskip 1pt\psi_3' ({\bf k}_1,{\bf k}_2,{\bf s}) \hskip 1pt {\rm Re}\hskip 1pt\psi_3'(-{\bf s},{\bf k}_3,{\bf k}_4) }{{\rm Re}\hskip 1pt\psi'_2(s)}  + \text{$t$- and $u$-channels}\, \] \, .
\ee \end{small}The primes on correlators and $\psi_n$ mean that we stripped the momentum-conserving $\delta$-functions.
Therefore, knowing the classical wavefunction coefficients, such as \eqref{Psi:tree}, we are able to compute  the tree-level correlation functions.

\vskip4pt
For loop diagrams, the relation between correlators and $\psi_n$ becomes more complicated, as both classical and quantum wavefunction coefficients are expected to contribute to the final correlation functions. For the $\Phi^3$ theory, the Born rule leads to the following two-point correlator up to $g^2$ order
\bea
\langle \phi_{{\bf k}_1}\phi_{{\bf k}_2} \rangle
 &=&  \int \mathcal{D}\phi ~\phi_{{\bf k}_1}\phi_{{\bf k}_2} \exp\Big\{\frac{1}{2} \int_{{\bf q}_1,{\bf q}_2} \phi_{{\bf q}_1}\phi_{{\bf q}_2}~2{\rm Re} \[\psi_2 + \psi_2^{\rm 1-loop}\] \nn\\
 && ~~ + \frac{1}{3!} \int_{{\bf q}_1,{\bf q}_2,{\bf q}_3} \phi_{{\bf q}_1}\phi_{{\bf q}_2} \phi_{{\bf q}_3} ~2{\rm Re}\psi_3 + \frac{1}{4!} \int_{{\bf q}_1,...,{\bf q}_4} \phi_{{\bf q}_1}\phi_{{\bf q}_2} \phi_{{\bf q}_3}\phi_{{\bf q}_4}~2{\rm Re}\psi_4
 \Big\}~,
\eea
where $\psi_2$, $\psi_3$ and $\psi_4$ are the classical ones that can be read off from the tree-level computation in \eqref{Psi:tree}, while $\psi_2^{\rm 1-loop}$ is the quantum wavefunction coefficient in \eqref{psi2:loop}.
Expanding the exponential and performing the Gaussian integral over $\phi$, the one-loop correction then can be explicitly decomposed into three contributions
\bea \label{classicaloop}
\langle \phi_{{\bf k}_1}\phi_{{\bf k}_2} \rangle_{\rm 1-loop}   &=&  \frac{{\rm Re}\hskip 1pt \psi^{\rm 1-loop}_{{\bf k}_1{\bf k}_2}}{  2{\rm Re}\hskip 1pt\psi'_2 (k_1) \hskip 1pt {\rm Re}\hskip 1pt\psi'_2 (k_2)} 
 - \frac{1}{ 8{\rm Re}\hskip 1pt\psi'_2 (k_1) \hskip 1pt {\rm Re}\hskip 1pt\psi'_2 (k_2)} 
 \int_{\bf p} \frac{{\rm Re}\hskip 1pt \psi_4({{\bf k}_1,{\bf k}_2},{\bf p},-{\bf p})}{{\rm Re}\hskip 1pt\psi'_2 (p)} \nn\\
 && + \frac{1}{ 8{\rm Re}\hskip 1pt\psi'_2 (k_1) \hskip 1pt {\rm Re}\hskip 1pt\psi'_2 (k_2)} 
 \int_{\bf p} \[\frac{ {\rm Re}\hskip 1pt \psi'_3({\bf k}_1,{\bf p},-{\bf p}-{\bf k}_1) 
 {\rm Re}\hskip 1pt \psi_3({\bf k}_2,-{\bf p},{\bf p}+{\bf k}_1) }{ {\rm Re}\hskip 1pt\psi'_2 (p) {\rm Re}\hskip 1pt\psi'_2 (|{\bf p}+{\bf k}_1|)}\right. \nn\\
 &&  \left. 
 %+ \frac{1}{ 8{\rm Re}\hskip 1pt\psi'_2 (k_1) \hskip 1pt {\rm Re}\hskip 1pt\psi'_2 (k_2)}  \int_{\bf p} 
~~~~~~~~~~~~~~~~~~~~~~~~ + \frac{ {\rm Re}\hskip 1pt \psi'_3({\bf k}_1,{\bf k}_2,-{\bf k}_1-{\bf k}_2) 
 {\rm Re}\hskip 1pt \psi_3({\bf k}_1+{\bf k}_2,{\bf p},-{\bf p}) }{ {\rm Re}\hskip 1pt\psi'_2 (p) {\rm Re}\hskip 1pt\psi'_2 (|{\bf k}_1+{\bf k}_2|)} \]~.
\eea
 From the expression above we can see that the first term has a truly quantum origin, which cannot be present in any classical field theory. However, the in-in correlation functions also contains classical loop contributions from tree-level wavefunction coefficients. For the case here, the two-point correlator receives contributions from the product of two contact $\psi_3$ and the single-exchange $\psi_4$ at the one-loop level. These are the {\it classical loops} in cosmological correlators, which can be computed by solving the classical equation of motion.\footnote{We acknowledge  insightful discussions with Enrico Pajer and Santiago Agui Salcedo to clarify this point.}

\vskip4pt
This analysis applies also to higher-order loops. All in all, in this way we can always distinguish the classical contributions from the quantum ones. In practice, however, at higher order it becomes more complicated to derive the relation between wavefunction coefficients and boundary correlators, as more diagrams would be involved both at the tree level and loop level. For instance in the $\Phi^3$ theory, to compute $\langle \phi \phi \rangle$  at the two-loop order would also require $\psi_4^{\rm 1-loop}$ and triple-exchange $\psi_6$, in addition to the contact $\psi_3$ and single-exchange $\psi_4$.
Generally speaking, the Born rule \eqref{bornrule} tells us that for a given order in perturbation theory, the correlators receive contributions from all the wavefunction coefficients and their products at the same order. The schematic form can be found in \eqref{lloop}, and also see Figure \ref{fig:mloop} for a specific example of higher loop diagrams.
In any case,  the distinction between classical and quantum contributions to loop diagrams can still be helpful.
In the following sections, we shall take this approach to examine the IR divergences caused by interacting massless scalars in de Sitter space. 
As these divergences are associated with long wavelength modes which become semi-classical after horizon-exit, we shall see how the distinction between classical and quantum loops leads to simplifications and interesting insights about the IR divergences in dS, and in particular clarifies the explicit connection with the stochastic formalism.

\vskip4pt
As a concluding remark of this section, we note that in many circumstances for cosmology, the classical loop contributions can dominate over the quantum ones, which means that we only need to  use the classical equation of motion to capture the leading effects in the loop computations. This has been noticed for a long time in the studies of some late-time processes, such as the formation of large-scale structure and induced gravitational waves, though it was not fully clear how the loop computations there were related to the ones in quantum field theory.
Using the wavefunction formalism, here we have performed a systematical analysis to distinguish  the (semi-)classical contributions from the quantum  ones.
For the previous (semi-)classical approximations, it would be interesting to understand better their regime of validity through this approach and  evaluate the effects of quantum corrections.

%=======================================
\section{IR Divergences in the Perturbative Computation}
%=======================================
\label{sec:pert}

Within the wavefunction formalism, in this section we turn our attention to examine the infrared (IR) divergences of de Sitter boundary correlation functions in perturbation theory. 
In particular, after a general review of the subject  in Section \ref{sec:review}, we will first collect some tree-level computation of wavefunction coefficients with nontrivial IR behaviour in Section \ref{sec:tree}. After that, in Section \ref{sec:1loop} we shall examine the one-loop structure of IR-divergent correlators via the wavefunction approach, and identify the distinct contributions from classical and quantum loops in three types of theories. 
In Section \ref{sec:multiloop}, we extend the one-loop analysis to the general IR-divergent correlators in higher-order loop diagrams, and find that contributions from classical loops are always dominant, which significantly simplifies the identification of the leading logarithmic behaviour in perturbation theory.

\subsection{Loops and IR Divergences}
\label{sec:review}

In perturbative computations, correlation functions in de Sitter space may keep growing on superhorizon scales. This typically happens  with very light scalars whose amplitudes freeze after horizon exit and thus can cumulatively source the growth of correlators. 
As a result, these correlators become singular towards the future boundary of dS with $\eta_0\rightarrow0$, which signals the breakdown of perturbation theory. This is famously known as the IR divergences in dS and has been extensively discussed in the literature.
The appearance of IR divergences is a property of spacetimes with boundaries~\cite{Symanzik:1981wd}, and is very similar in AdS (see for instance~\cite{Skenderis:2002wp}).  In the case of de Sitter the fact that the boundary is spacelike has further complications.

\vskip4pt
Meanwhile, as physical observables, cosmological correlators are supposed to be IR-safe and stable with respect to higher order radiative corrections. Thus one may expect these IR divergences  become finite once we go beyond the perturbative regime.\footnote{In this paper, we restrict our analysis to quantum field theories in a fixed dS background, and thus neglect the backreaction effects with dynamical gravity which might become important for the stability of the dS spacetime \cite{Polyakov:2007mm,Polyakov:2009nq,Polyakov:2012uc,Brandenberger:2002sk,Brandenberger:2018fdd,Comeau:2023euf}}
It has been identified as early in~\cite{Starobinsky:1986fx} that the theory is finite because it reaches an equilibrium state.  The idea behind this approach is that modes can be split between the long wavelength classical solution and the short wavelength quantum noise.  Heuristically this corresponds to a stochastic Langevin equation,  which can be translated into a non-perturbative  Fokker-Planck equation.  By use of these equations one can compute correlation functions which are now finite,  and in particular, time independent at late times.
From a field-theoretic point of view, this approach is expected to resum all the divergent contributions, especially the higher order loops. 
To understand better the resummation, we first taker a closer look at the perturbative computations.
Here let's briefly summarize the state of the art on the topic, especially from the boundary perspective of the bootstrap approach.

\vskip4pt
The IR divergences in dS already show up in tree-level processes. As we will show in more detail in Section \ref{sec:tree}, these IR divergences are the {\it secular} terms with logarithmic growth $\log(-k_\Sigma\eta_0)$ that come from the bulk time integration for each vertex, with $k_\Sigma$ normally being the sum of energies for one vertex. 
In a realistic scenario, cosmic inflation must end, thus the conformal time $\eta_0$ at the end of inflation provides a natural IR cutoff to regularize the logarithmic singularity. In this case, perturbative computations remain valid as long as the logarithmic function multiplied by the coupling is smaller than one.
Recently, a systematical investigation on the IR behaviour of both contact and exchange diagrams has been presented in Ref. \cite{Wang:2022eop}  by using the cosmological bootstrap (see also \cite{Bzowski:2013sza, Bzowski:2015pba,Pajer:2016ieg, Bzowski:2018fql, Bzowski:2019kwd, Cespedes:2020xqq, Bzowski:2022rlz,Bzowski:2023nef} for relevant works in this direction).
There it has been shown that the secular growth of late-time correlators is described by the boundary conformal field theory with anomalies, and the IR-divergent correlators from massless exchange diagrams lead to local-type non-Gaussianities from multi-field inflation with full singularity structure.
Furthermore, it has  been noticed that in the wavefunction approach, the disconnected contributions to exchange correlators, such as the one shown in \eqref{equ:4pt} and  \eqref{disconn},  typically contain leading IR divergences, while the connected part may be less singular \cite{Wang:2022eop}.

\vskip4pt
Beyond tree-level diagrams, higher order IR divergences are expected from loops.
In addition to secular  divergences from time integrals, there is another type of IR divergences that arise from loop momentum integration. 
This has long been noticed in the in-in computation \cite{Weinberg:2005vy,vanderMeulen:2007ah,Senatore:2009cf,Senatore:2012nq,Pimentel:2012tw}. Typically we may encounter the following loop integral
\be
\int_{\bf p} \frac{1}{p^3}= 
 \frac{1}{2\pi^2} \int_{1/L_{\rm IR}}^{\Lambda}   \frac{dp}{p}   =  \frac{1}{2\pi^2}  \log\(\Lambda L_{\rm IR} \)~,
\ee
where we have introduced two cutoff scales for regularization. 
The UV cutoff $\Lambda$ becomes essential when we screen out the short wavelength modes to study the coarse-grained perturbations at large scales. We will elaborate on this in Sections \ref{sec:resum} and \ref{sec:beyond}. Here let's take a closer look at the IR cutoff, which is given by the comoving size of the Universe $L_{\rm IR}$ here. This scale defines the size of the box for the system that we are interested in. For instance, at late time $t$, the physical size of the Universe  is given by $a(t) L_{\rm IR}$.  One way to interpret it is to consider a minimum Fourier mode $k_{\rm IR}$ which exits the horizon at some initial time $t_i$, such that $k_{\rm IR} = a(t_i) H \simeq L_{\rm IR}^{-1}$. This initial time may be seen as the beginning of inflation. Then all the long wavelength perturbations we are interested in are generated after $t_i$ and have $k>k_{\rm IR}$.

\vskip4pt
As both momentum integration and nested time integration are involved, in general in-in computations for loop-level correlators are notoriously difficult. For cases with IR divergences, one can perform computations from one-loop corrections  (see \cite{vanderMeulen:2007ah,Tsamis:1997za,Tsamis:2005hd,Burgess:2009bs} for previous literature and Appendix \ref{app:inin} for several examples), but it becomes much more complicated if we consider the computation of higher loops.\footnote{
Some efforts in this direction have been made in \cite{Baumgart:2019clc}.}
Meanwhile, one can also tackle the problem using the wavefunction formalism. Recently it has been realized that loop computations become simpler in the wavefunction fortmalism, and one can even derive the general form of the wavefunction coefficients with IR loops at any order in perturbation theory \cite{Gorbenko:2019rza}.
In the following we shall go one step further and show that, the computation for IR-divergent {\it correlators} can also be greatly simplified by using the wavefunction method, and the leading contributions always come from classical loops.

\paragraph{Setup for computation}
Our basic object of interest is the wavefunction coefficients $\psi_n$ and we will apply the Feynman rules derived in Section \ref{sec:wav}. For massless scalars, the bulk-to-boundary propagator is given by solving \eqref{btod}
\be
K(k,\eta) = (1-ik\eta)e^{ik\eta}~.
\ee
Then for a free theory, its two-point wavefunction coefficient can be directly read off from the boundary term in \eqref{Psi:tree}
\be \label{psi2}
\psi_2^{\rm free} ({\bf k},{\bf k}') = \frac{i k^2}{\hbar H^2\eta_0 (1-ik\eta_0)} (2\pi)^3\delta({\bf k}+{\bf k}')~,
\ee
whose real part gives ${\rm Re} \psi_2^{\prime\rm free} = -k^3/H^2\hbar$ and thus generates the standard power spectrum of massless scalars by using \eqref{equ:2pt}.
The bulk-to-bulk propagator follows from \eqref{green}
\bea
G(k,\eta,\eta') &=&  \frac{\hbar H^2}{2k^3} \[(1+ik\eta)(1-ik\eta') e^{ik(\eta'-\eta)}\Theta(\eta-\eta') + (1-ik\eta)(1+ik\eta')e^{ik(\eta-\eta')}\Theta(\eta'-\eta)\right.\nn\\
&&~~~~~~~~\left. -(1-ik\eta)(1-ik\eta')e^{ik(\eta'+\eta)} \]~.
\eea
Here as we approach the boundary with $\eta\rightarrow 0$ this propagator scales as
\be \label{latelimit}
\lim_{\eta\rightarrow 0} G(k,\eta,\eta')  \rightarrow -\frac{i}{3}H^2 \eta^3~(1-ik\eta')e^{ik\eta'} ~.
\ee
Thus, even for massless scalars the wavefunction $G$ propagator quickly decays on super-horizon scales, instead of becoming constant. This property becomes important for the analysis of IR divergences in wavefunction coefficients.

\vskip4pt
As in Section \ref{sec:wav} we have clarified the quantum and classical contributions in the calculation, now we  set $\hbar=1$ to avoid clutter.
Particularly the computation will be performed for two massless scalars $\Phi$, $\Sigma$ in the bulk, with $\phi$ and $\chi$ being their boundary fluctuations correspondingly. 
Our analysis mainly focuses on three types of bulk interactions:
\be \label{int3}
\frac{\lambda}{4!}\Phi^4~, ~~~~~~ \frac{g}{3!} \Phi^3~, ~~~~~~ \frac{\alpha}{2} \dot\Phi\Sigma^2~.
\ee
Let us give a few more words about the choice of these vertices before moving to particular computations.
The first two are commonly expected for interacting scalars and have been extensively discussed for the studies of IR divergences in dS. For inflation the scalar field $\phi$ can be the inflaton itself, and then normally the  two couplings $\lambda$ and $g$ are suppressed by slow-roll parameters. The third interaction may not be so familiar as it breaks the boost symmetry. 
Usually this is seen in many multi-field inflation models where we have a rolling background of the inflaton field, and the coupling $\alpha$ is typically associated with the field-space curvature \cite{Wang:2019gok}.
Moreover, as the inflaton here is protected by a shift symmetry, this coupling may not be necessarily slow-roll suppressed. We shall see later that the time derivative in this vertex simplifies the analysis of IR divergences, and thus this type of interaction provides a working example for us to go from perturbative computation to the non-perturbative regime.

\subsection{IR Divergences at Tree Level}
\label{sec:tree}

The tree-level wavefunction coefficients of interacting massless scalars have already been computed in the literature. Especially for the ones with IR divergences, the recent analysis using the cosmological bootstrap provides a systematic classification for their full analytical forms \cite{Wang:2022eop}, which can be derived by solving a set of interesting differential equations from {\it anomalous} conformal Ward identities (see also the corresponding analysis of CFT correlators in momentum space in \cite{Bzowski:2013sza, Bzowski:2015pba,Pajer:2016ieg, Bzowski:2018fql, Bzowski:2019kwd, Bzowski:2022rlz}). 
Here we mainly collect the result of tree-level wavefunction coefficients of interacting massless scalars, from both contact and exchange diagrams.

\begin{itemize}
\item {\bf $\Phi^4$ interaction}: the four-point contact diagram leads to
\bea \label{phi4contact}
\psi'_4 &=& i\lambda \int_{-\infty}^{\eta_0} d\eta a(\eta)^4 K(k_1,\eta) K(k_2,\eta) K(k_3,\eta) K(k_4,\eta)   \nn\\
&=& -\frac{\lambda}{H^4} \[ \frac{i}{3\eta_0^3} +  \frac{i}{2\eta_0} \sum_{a=1}^4 k_a^2 + \frac{1}{3}\(\sum_{a=1}^4 k_a^3\)\log\(-k_T \eta_0\) \]
 ~,
\eea
where $k_T=k_1+k_2+k_3+k_4$.
Then the corresponding correlation function is given by the first term in \eqref{equ:4pt}, while there is no disconnected terms from $\phi^4$ interaction. As the imaginary parts do not contribute, here the logarithmic term with secular growth gives rise to the IR divergence in the correlator.

\item {\bf $\Phi^3$ interaction}: we first have a three-point contact diagram  (see the first one in Figure \ref{fig:trees})
\bea \label{psiphi3}
\psi'_3 &=& i g \int_{-\infty}^{\eta_0} d\eta a(\eta)^4 K(k_1,\eta) K(k_2,\eta) K(k_3,\eta)   \nn\\
 &=& -\frac{g}{H^4} \[ \frac{i}{3\eta_0^3} +  \frac{i}{2\eta_0} \sum_{a=1}^3 k_a^2 + \frac{1}{3}\(\sum_{a=1}^3 k_a^3\)\log\(-k_t \eta_0\) \] ~,
\eea
where $k_t=k_1+k_2+k_3$.
The correlator follows from \eqref{equ:3pt}. Similarly the IR divergence is caused by the logarithmic secular term.
Let's also consider the exchange diagram (see the second one in Figure \ref{fig:trees}) from this interaction, which leads to the following
four-point wavefunction coefficient
\bea \label{phi3psi4}
\psi'_4 &=& - g^2 \int_{-\infty}^{\eta_0} d\eta d\eta' a(\eta)^4 K(k_1,\eta) K(k_2,\eta) G(s,\eta , \eta') K(k_3,\eta') K(k_4,\eta')  + {\rm perms.}\nn\\  
&=& - \frac{g^2}{18 H^6} \Big[(k_1^3+k_2^3+s^3)\log^2\(-E_L\eta_0\)+(k_3^3+k_4^3+s^3)\log^2\(-E_R\eta_0\)  \nn\\
&& +\text{$t$- and $u$-channels} + {\rm imaginary ~ part} \Big]~,
\eea
where ${s} =|{\bf k}_1+{\bf k}_2|$, ${t} =|{\bf k}_1+{\bf k}_3|$, ${u} =|{\bf k}_1+{\bf k}_4|$, and $E_L=k_1+k_2+s$, $E_R=k_3+k_4+s$.
In this case, to derive the corresponding correlator, in to addition the exchange $\psi_4$ above, there is also the disconnected piece in \eqref{equ:4pt} with the product of two $\psi_3$ in \eqref{psiphi3}.
Notice that here these two parts both contribute at the $\log^2$ order, thus one needs to take into account connected and disconnected terms to capture the leading IR-divergent behaviour in the correlator.

\begin{figure} [t]
   \centering
            \includegraphics[width=.5\textwidth]{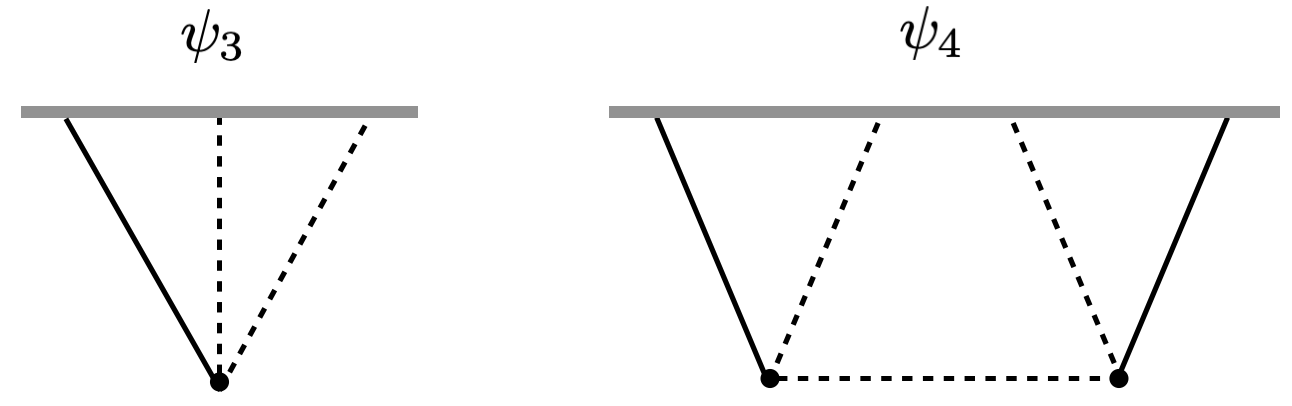}  
   \caption{Tree diagrams from the two-field interaction $\dot\Phi\Sigma^2$. The solid lines are for $\Phi$ fields, while the dashed lines represent $\Sigma$ propagators.}
  \label{fig:treea}
\end{figure}

\item {\bf $\dot\Phi\Sigma^2$ interaction}: we first have the three-point contact diagram with one $\phi$ and two $\chi$ fields (see Figure \ref{fig:treea})
\bea \label{psi3-chi}
\psi'_3 = i\alpha \int_{-\infty}^{\eta_0} d\eta a(\eta)^3 \partial_\eta K(k_1,\eta) K(k_2,\eta) K(k_3,\eta)   = \frac{\alpha}{H^3}\[ \frac{ik_1^2}{\eta_0} + k_1^3\log(-k_t \eta_0)\]~,
\eea 
which again contributes to a logarithmic IR divergent term in the corresponding correlator.
Next we also compute the four-point exchange diagram with two $\phi$ and two $\chi$ fields as shown in Figure \ref{fig:treea}
\bea \label{psi4-chi}
\psi'_4 &=& -\alpha^2 \int_{-\infty}^{\eta_0} d\eta d\eta' a(\eta)^3 a(\eta')^3 \partial_\eta K(k_1,\eta) K(k_2,\eta) G(s,\eta,\eta') \partial_\eta K(k_3,\eta) K(k_4,\eta) \nn\\
&=& -\frac{\alpha^2k_1^3 k_3^3}{2H^4s^3} \[ \log\(\frac{E_L}{k_T}\)\log\(\frac{E_R}{k_T}\)
+ {\rm Li}_2\(\frac{k_1+k_2-s}{k_T}\)+   {\rm Li}_2\(\frac{k_3+k_4-s}{k_T}\) \right.\nn\\
&& \left. ~~ + \frac{2s(s^2-k_3(k_3+k_4))}{k_3(k_3+k_4-s)E_R}\log\(\frac{E_L}{k_T}\)+ \frac{2s(s^2-k_1(k_1+k_2))}{k_1(k_1+k_2-s)E_L}\log\(\frac{E_R}{k_T}\)\right.\nn\\
&& \left. ~~  +  \frac{2s^2}{E_LE_R} \( 1 + \frac{(k_T-k_2)(k_T-k_4)s}{k_1k_3k_T}+\frac{s^2}{k_1k_3} \)-\frac{\pi^2}{6}
\]+\text{$t$- and $u$-channels}.
\eea
This wavefunction is actually IR-finite with no secular term depending on $\eta_0$.\footnote{Naively it seems there are folded singularities at $s=k_1+k_2$ and $s=k_3+k_4$, but one can easily check that they are absent in the folded limit of the above result. A simple way to derive this is to use weight-shifting operators and four-point scalar seed of massless exchange as shown in Ref. \cite{Wang:2022eop}.}
Technically, it can be understood by looking at the late-time limit of the bulk-to-bulk propagator in \eqref{latelimit}:  $G$ decays as $\eta^3$ thus the integrand in this case does not lead to significant  contribution in the late-time part of the bulk integration.
However, the IR behaviour of the four-point correlator $\langle \phi\chi\phi\chi \rangle$ is not simply related to the $\psi_4$ above. As now the disconnected part with the product of two ${\rm Re}~\psi_3$ gives $\log^2$ divergent terms, the IR-finite term from the connected part becomes sub-dominant. Thus from this example, one can see the difference between wavefunction coefficients and correlators. Especially their IR behaviours may differ when we look at exchange diagrams. See \cite{Wang:2022eop} for more discussions about the distinction. 
\end{itemize}

There are several lessons that we can learn from the tree-level computation of wavefunction coefficients. First, for interacting massless scalars, logarithmic IR divergences arise when the number of derivatives is less than 2. For higher-derivative interactions one finds the $\psi_n$s are in term of IR-finite rational polynomials \cite{Goodhew:2022ayb}.
Second, while the order of the divergence is one for contact diagrams, it would depend on the type of interactions for exchange diagrams. For the examples above, the exchange $\psi_4$ from two $\Phi^3$ vertices has $\log^2$ divergence, since the super-horizon decay of the $G$ propagator does not stop the growth of the integrand in \eqref{phi3psi4}.
Meanwhile the one from $\dot\Phi\Sigma^2$ is IR-finite because of the time derivative in the vertex.
But when we compute the corresponding correlators, both exchange diagrams are IR-divergent, as the disconnected parts always lead to $\log^2$-type contributions.
Thus, in general we expect derivative interactions lead to simpler IR-singular structure than the non-derivative ones do. Technically this is because derivatives (either spatial or time ones) increase the power of $\eta$ in the integrand and make it less likely to cause late-time singular behaviour. Later we will  show that this is also the case for loop diagrams.

\subsection{One-Loop Structure of IR-Divergent Correlators}
\label{sec:1loop}

Next we consider IR divergences in loop corrections to massless scalar correlators using the wavefunction method.
The analysis is rather technical and it is better to gain some intuition from simple working examples. 
Thus after a general discussion on loop integrals, we first focus on the one-loop diagrams, and then look into higher-order loops in Section \ref{sec:multiloop}.
Some of the loop analysis have been sketched in Ref. \cite{Gorbenko:2019rza}, while in this paper we perform a more explicit computation with concrete examples.
All in all, the key takeaway is a simplification: the infrared (IR) divergences in loop-level correlation functions are consistently subdominant  with respect dominant classical loop contributions stemming from the tree-level wavefunction coefficients. Consequently, the quantum contributions can be effectively neglected in this context.

\vskip4pt
Let's begin with a generic form of the loop wavefunction coefficients.
By using the Feynman rule we derived for quantum theories, we can quickly write down the $L$-loop correction to a $n$-point wavefunction coefficient with $m$ vertices
\bea \label{Lloop}
\psi_n^{L-\rm loop} &\sim & \int d\eta_1 ... d\eta_m a(\eta_1)^4...a(\eta_m)^4 K(k_1,\eta_1) ... K(k_n,\eta_m) \nn\\
&& ~~~~~~ \int_{\bf p_1,...,p_L}  G(p_1,\eta_a,\eta_b) ... G(p_L,\eta_c,\eta_d) G(|{\bf p_x+k_y}|,\eta_e,\eta_f)...
\eea
where ${\bf p_x}$ and ${\bf k_y}$ represent various combinations of internal and external momenta. One needs to be careful about the bulk time of each insertion and the momentum conservation conditions for particular types of interactions.
Schematically we see that, in addition to the bulk time integral for each vertex, there are also  integrations for internal momenta. These two types of integrations also arise in the in-in computation: the former generates secular logarithmic terms (as for tree-level diagrams); while the latter leads to IR divergences regularized by the cutoff $L$. In general one would expect to see both types of divergences in the final correlator.
Although the computation looks similar in the wavefunction approach, it turns out to be simpler for the above $\psi_n^{L-\rm loop}$.

\vskip4pt
To see this explicitly, let us briefly review an important observation in \cite{Gorbenko:2019rza} that, in the  computation of loop-level wavefunction coefficients momentum integrations do not become divergent at IR.
We know that  momentum integrals can be regularized by the UV cutoff $\Lambda$ and the IR cutoff $1/L_{\rm IR}$. UV divergences are usually renormalized by introducing counterterms. The IR divergences are manifested in the $L_{\rm IR}\rightarrow\infty$ limit,  and we can use the following results of the loop integrations 
\begin{align}  
\int_{\bf p} \frac{1}{p^n}= 
 \frac{1}{(2\pi)^3} \int_{1/L_{\rm IR}}^{\Lambda} 4\pi p^{2-n}dp   \xrightarrow{L_{\rm IR}\rightarrow\infty} \begin{dcases} \text{IR-finite} ~, ~~~~ & n<3\nn\\
 \frac{1}{2\pi^2} \log(k L_{\rm IR}) ~, ~~~~ & n=3~.
\end{dcases}
\end{align}
Then it is easy to see that all the momentum integrals in \eqref{Lloop} are IR-finite by noticing the soft  behaviour of the bulk-to-bulk propagator
\be 
\lim_{p\rightarrow0} G  (p,\eta,\eta') = -\frac{i}{6}H^2(\eta^3+\eta'^3) + \mathcal{O}(p)~.
\ee
Thus in the loop-level calculation of $\psi_n$ one can safely take the $L_{\rm IR}\rightarrow\infty$ limit and the momentum integrals never become singular at IR.  
As a result, the wavefunction coefficients can only have secular IR divergences $\log(-k\eta_0)$ from bulk time integration.

\vskip4pt
Next, we consider the computation of IR-divergent correlators using the corresponding wavefunction coefficients.
As explicit demonstrations, we derive the two-point correlation functions of $\phi$ at one-loop level from three different types of interactions in \eqref{int3}.

\begin{figure} [h]
   \centering
            \includegraphics[width=.8\textwidth]{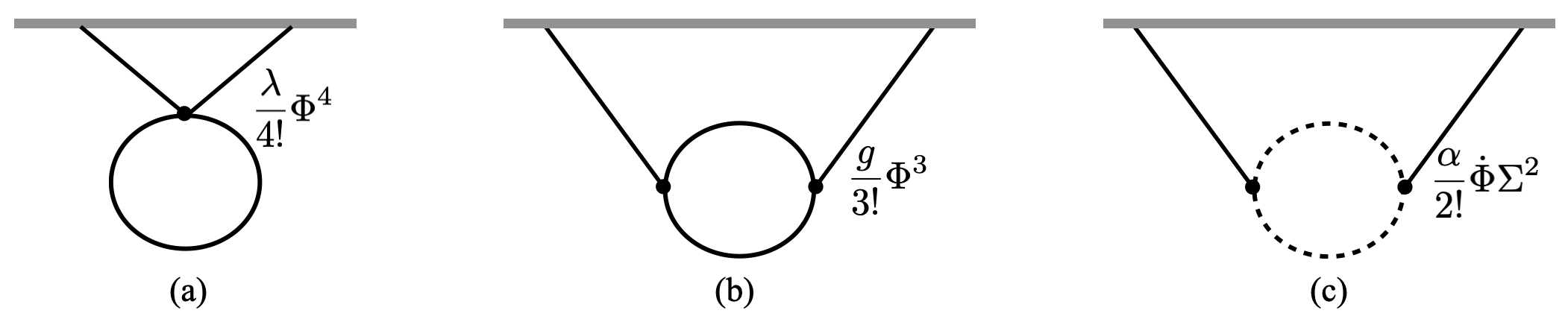}  
   \caption{One-loop diagrams of two-point functions from three types of interactions with IR divergences. The solid lines are for $\Phi$ fields, while the dashed lines represent $\Sigma$ propagators.}
  \label{fig:1loop}
\end{figure}

\begin{itemize}

\item{\bf One $ \Phi^4$ vertex}: This one-loop wavefunction coefficient has
a single internal line. 
%This correction to the two-point wavefunction coefficient can only arise in the quantum theory. 
Using the Feynman rules we derived in Section \ref{sec:wav}, we find
\begin{small}
\bea \label{phi4-qloop}
\psi_2^{\prime\rm 1-loop} &=& \frac{i\lambda}{2} \int_{-\infty}^{\eta_0} d\eta  a(\eta)^4   K (k,\eta)    K (k,\eta )   \int_{\bf p} G  (p,\eta,\eta)  \\
&=&    \frac{\lambda}{6H^2}\int_{\bf p} \[ \log(-2(k+p)\eta_0) + \frac{k^3}{p^3} \log\(\frac{k+p}{k}\) + \gamma_E -\frac{i\pi}{2} 
- \frac{12k^3+6k^2p+13kp^2+28p^3}{12p^2(k+p)}~,
%+ \frac{k(p-2k)}{2p^2}  + \frac{3k}{4(k+p)}-\frac{7}{3}
\]\nn
\eea
\end{small}where there is a $1/2$ as the symmetric factor, and in the second line we performed the time integral.
As argued above, this loop correction to $\psi_2$ has only secular divergence from bulk time integration.
However, to compute the loop correction to correlators from wavefunction coefficients, we 
also need to take into account the {\it classical loops} from tree-level wavefunction coefficients. 
For $\Phi^4$ interaction, this corresponds to the second term in \eqref{classicaloop} which requires the $\psi_4$ in \eqref{phi4contact} from the contact diagram.
Then unlike the quantum loop from $\psi_2^{\rm 1-loop}$, the momentum integral here leads to the IR divergence regularized by the cutoff $L_{\rm IR}$. As a result, at $\lambda$ order the classical loop contribution dominates over the quantum one, and the leading IR divergence in this correlator is given by
\begin{small}
\be \label{phi4-1loop}
\langle \phi_{{\bf k}}\phi_{-{\bf k}} \rangle'_{\rm 1-loop} \simeq - \frac{1}{8{\rm Re}\hskip 1pt \psi'_2 (k) \hskip 1pt {\rm Re}\hskip 1pt 
\psi'_2 (k)} \int_{\bf p} \frac{{\rm Re}\hskip 1pt\psi'_4(k,k,p,p)}{{\rm Re}\hskip 1pt\psi'_2(p)}
 = \frac{H^2}{2k^3} \frac{\lambda}{12\pi^2}\log(kL_{\rm IR})\log(-2k\eta_0) + ...~
\ee 
\end{small}
This result agrees with the in-in computation as shown  in Appendix \ref{app:inin}. 
From this example, we see for the first time that the main contribution of the IR divergence of a loop correlator is due to  the classical wavefunction coefficient.

\item{\bf Two $\Phi^3$ vertices}: The one-loop correction here corresponds to the second diagram in Figure \ref{fig:1loop}, the expression of the wavefunction coefficient is explicitly given in \eqref{psi2:phi3}. Using the massless scalar propagators and performing the time integral, we find 
\bea \label{phi3-qloop}
\psi^{\prime\rm 1-loop}_{{\bf k}{\bf k}'} = - \frac{g^2}{27H^4} \int_{\bf p} \log\(-(2k+p)\eta_0\)~,
\eea
which again only contains secular-type IR divergence, as the loop integration is IR-finite.
To compute  the two-point correlator at one-loop ($g^2$) order, we need to use the relation \eqref{classicaloop} derived by the Born rule, and all three terms there are present for the $\Phi^3$ theory.
As the first term with $\psi^{\rm 1-loop}_2$ has no loop IR divergences, the leading contributions come from the second and third terms, which are explicitly given by
\bea
\text{2nd term} &=& \left(\frac{H^2}{2k^3} \right)^2\int_{\bf p} \frac{H^2}{2p^3} {\rm Re}\hskip 1pt \psi_4^{\prime}({{\bf k},{\bf k}'},{\bf p},-{\bf p}) = 
\frac{g^2}{8k^3} \frac{-5}{18\pi^2} \log(k L_{\rm IR}) \log(-2k\eta_0)^2 ~, \nonumber\\
\\
\text{3rd term} &=& \left(\frac{H^2}{2k^3} \right)^2 \int_{\bf p} \frac{H^2}{2p^3}  \[ \frac{H^2}{2|{\bf k}+{\bf p}|^3}  2{\rm Re}\hskip 1pt \psi'_3({\bf k},{\bf p},-{\bf p}-{\bf k}){\rm Re}\hskip 1pt \psi'_3({\bf k}',-{\bf p},{\bf p}+{\bf k}) \right.\nn\\
&& \left. ~~~~~~ +   \frac{H^2}{2|{\bf k}+{\bf k}'|^3}  2{\rm Re}\hskip 1pt \psi'_3({\bf k},{\bf k}',-{\bf k}-{\bf k}'){\rm Re}\hskip 1pt \psi'_3({\bf k}+{\bf k}',{\bf p},-{\bf p})\] \nn\\
& =&  \frac{g^2}{8k ^3} \frac{1}{3\pi^2} \log(k L_{\rm IR}) \log(-2k \eta_0)^2  ~,
\eea
where $\psi_3$ and $\psi_4$ are the tree-level wavefunction coefficients \eqref{psiphi3} and \eqref{phi3psi4} from contact and exchange diagrams.
As both of them have $\log^2$ order secular divergences, the  two-point correlator in the end is a sum of two contributions
\be
\langle \phi_{{\bf k}}\phi_{-{\bf k}} \rangle'_{\rm 1-loop} \simeq  \frac{g^2}{8k^3} \frac{1}{18\pi^2} \log(k L_{\rm IR}) \log(-2k\eta_0)^2~,
\label{loop_cubic}
\ee
which again reproduces the in-in result shown in Appendix \ref{app:inin}. 
This example also shows that the leading IR divergence of the correlator is given by classical loops from tree-level wavefunction coefficients, although in this case both the contact $\psi_3$ and exchange $\psi_4$ contribute at the same order.

\item{\bf Two $\dot\Phi\Sigma^2$ vertices}: Here the one-loop process also corresponds to the third diagram in Figure \ref{fig:1loop}, though the two internal lines are for $\Sigma$ field and there is one time derivative on $\Phi$ for each vertex
\be \label{chi-qloop}
\psi^{\prime\rm 1-loop}_{{\bf k}{\bf k}'}= -\frac{\alpha^2}{2} \int d\eta  d\eta' a(\eta)^3 a(\eta')^3  \partial_\eta K (k,\eta)  \partial_{\eta'} K (k,\eta')   \int_{\bf p} G (p,\eta,\eta') G (|{\bf k}+{\bf p}|,\eta,\eta')~.
\ee
Then because of the time derivative, this wavefunction coefficient is actually IR-finite, even without secular logarithms from time integrations, as we found for the exchange $\psi_4$  in \eqref{psi4-chi}. However, since the contact $\psi_3$ in \eqref{psi3-chi} has a secular term, we still get IR-divergent contributions from the third term in \eqref{classicaloop}.
\bea
\langle \phi_{{\bf k}}\phi_{-{\bf k}} \rangle'_{\rm 1-loop} &\simeq & \frac{1}{ 8{\rm Re}\hskip 1pt\psi'_2 (k) \hskip 1pt {\rm Re}\hskip 1pt\psi'_2 (k)} 
 \int_{\bf p} \frac{ {\rm Re}\hskip 1pt \psi'_3({\bf k},{\bf p},-{\bf p}-{\bf k}) 
 {\rm Re}\hskip 1pt \psi'_3(-{\bf k},-{\bf p},{\bf p}+{\bf k}) }{ {\rm Re}\hskip 1pt\psi'_2 (p) {\rm Re}\hskip 1pt\psi'_2 (|{\bf p}+{\bf k}|)} \nn\\
 &=& \frac{H^2}{2k^3} \frac{\alpha^2}{8\pi^2}\log(kL_{\rm IR})\log(-2k\eta_0)^2 ~,
\eea 
which also reproduces the in-in result in Appendix \ref{app:inin}. Again, we find the correlator is dominated by the contribution from classical loops. This time the computation is further simplified, as the exchange wavefunction coefficient does not contribute and one only needs to consider the term from the product of two contact $\psi_3$.
\end{itemize}
 
\paragraph{Summary}
From these three examples, we see that leading IR divergences of correlators come from classical loops, which means only tree-level wavefunction coefficients are needed.
While the explicit computations above are for two-point functions,
it is straightforward to  extend the analysis to any higher-point functions with IR-divergent one-loop corrections. In general, the momentum integrations in loop-level wavefunction coefficients do not lead to IR divergences, and thus the quantum loops only contain secular divergences, as we see below \eqref{Lloop}. Meanwhile, at the same order of the couplings, 
 the classical loops contain both IR divergences from momentum integrals and also secular terms from time integrals, thus giving rise to the leading logarithms in correlators.
In general, to explicitly compute the classical loops for higher-point functions, we need wavefunction coefficients of more complicated tree-level exchange diagrams with multiple internal lines. The integration over the internal momentum $\bf p$ is IR-divergent as there is always a $\psi_2(p)\propto p^3$ in the denominator of the integrand. 
As a result,  the classical loops always give the dominant contribution to the final correlators.

\subsection{Multi-Loop Diagrams and Leading Logarithms}
\label{sec:multiloop}

\begin{figure} [t]
   \centering
            \includegraphics[width=0.95\textwidth]{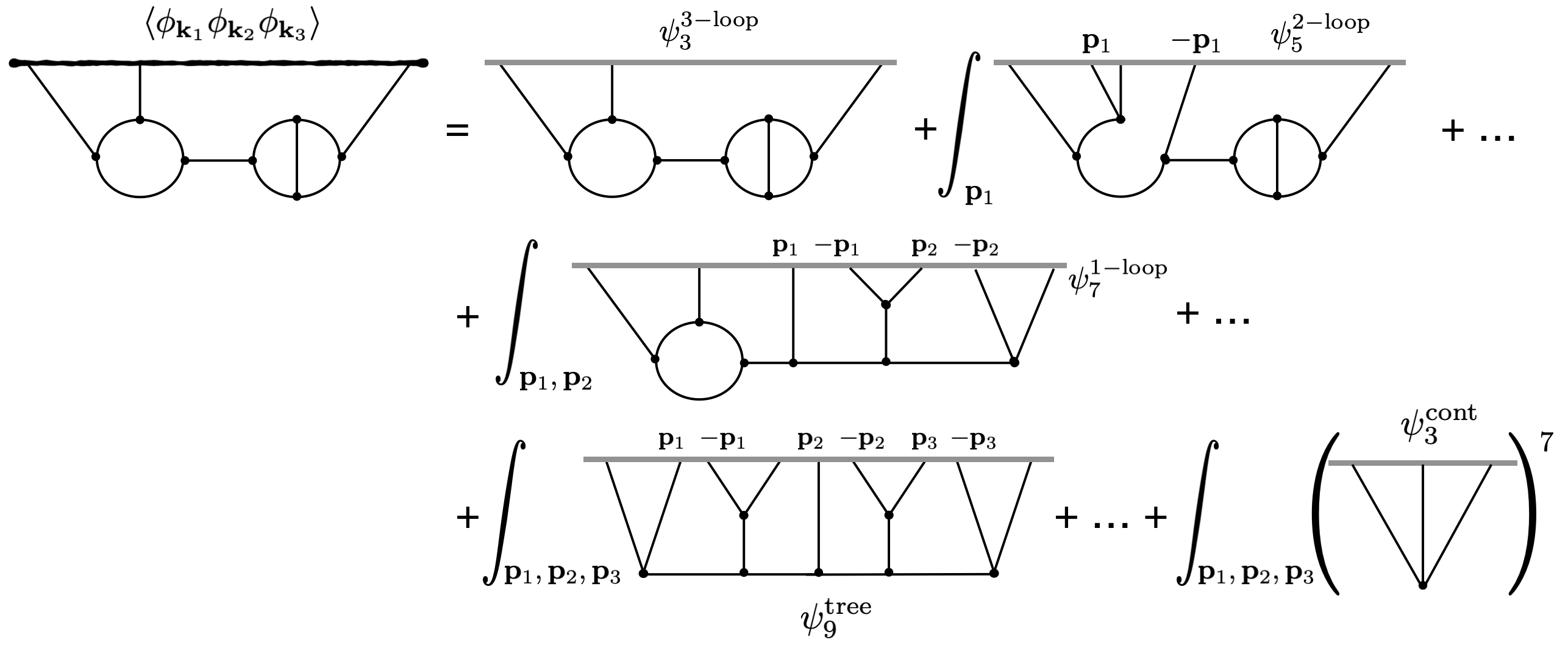}  
   \caption{A diagrammatic illustration of a multi-loop relation:  the correlator on the left (the diagram with a ragged black line as the late-time boundary) receives various contributions from  both loop-level (the first and second lines) and tree-level (the third line) wavefunction coefficients on the right (the diagrams with grey late-time boundaries). In this example one needs to incorporate all the $\psi_n$'s and their products at the $g^7$ order, as required by the Born rule \eqref{bornrule} in perturbation theory.}
  \label{fig:mloop}
\end{figure}

Although the explicit computation becomes complicated once we consider multi-loop processes, the notion of classical loops can still help us easily identify the leading logarithms in IR-divergent correlators. 
Let's take $g\Phi^3$ interaction as an explicit example.
At the $g^V$ order, the $n$-point correlator of $L$-loop diagram can be computed by using all the wavefunction coefficients and their products with $V$ vertices. Schematically, the relation is
\begin{small}
 \begin{align} \label{lloop}
\langle \phi^n \rangle_{L-\rm loop } & \sim 
\frac{1}{({\rm Re} \psi_2)^n} \Bigg[
{{\rm Re}~\psi_n^{L-\rm loop }}  +  \int_{\bf p_1} \frac{{\rm Re} \psi_{n+2}^{(L-1)-\rm loop }} {{\rm Re} \psi_2(p_1)} +  ... + \int_{\bf p_1,...,p_{L-1}} \frac{{\rm Re} \psi_{n+2(L-1)}^{1-\rm loop }} {({\rm Re} \psi_2(p_1) ... {\rm Re} \psi_2(p_{L-1}) )} \nn
\\
& +  \int_{\bf p_1,...,p_L} \frac{1}{{\rm Re} \psi_2(p_1) ... {\rm Re} \psi_2(p_{L}) }\( {\rm Re} \psi^{\rm ex}_{n+2L} + \frac{{\rm Re} \psi^{\rm ex}_{n+2L-1}{\rm Re}~\psi_3 }{{\rm Re}~\psi_2} + ...  + \frac{({\rm Re}~\psi_3 )^V}{({\rm Re}~\psi_2)^{3V-2L-n}} \) \Bigg]~,
\end{align}   
\end{small}where all the numerical coefficients are dropped. In the first line we collect all the wavefunction coefficients containing loops, and the tree-level contributions  are listed in the second line, which are the classical loops.
One example is shown in Figure \ref{fig:mloop}.
As we have argued in the one-loop case, the momentum integration in wavefunction coefficients does not lead to IR divergences, while each momentum integration explicitly shown in \eqref{lloop} contributes one $\log(k_E L_{\rm IR})$.\footnote{Here $k_E$ represents the magnitude for one combination of various external momenta. The same notation applies for the discussion in the rest of the section.} Thus the second line leads to more divergent  contributions than the ones from loop-level wavefunction coefficient.
In other words, the classical loops always dominate over the quantum ones for IR-divergent correlators.

\vskip4pt
This analysis in the end leads us to a simple conclusion: the saddle-point approximation of the wave function \eqref{saddle}, which contains all the tree-level information, captures the leading IR behaviour.
Intuitively, it is easy to understand: IR divergences are generated on {\it superhorizon scales}, which are normally due to {\it semi-classical effects}.
As we have shown in Section \ref{sec:wav}, the wavefunction of the Universe provides a clear distinction for quantum and classical effects, which helps us here to  confirm the intuitive understanding with concrete computations.
This interesting conclusion provides simplifications for the analysis of IR divergences in dS, and will play an important role for our discussion of the stochastic formalism in the following sections. 
The same results can be found for the $\lambda \Phi^4$ and $\alpha\dot\Phi\Sigma^2$ interactions.

\vskip4pt
As one example of the simplifications, now let us show that in the perturbative computation how we can directly identify the leading IR logarithms. The trick here is to simply look at the last term in the second line in \eqref{lloop}, which is the contribution from the product of $V$ three-point wavefunction coefficient. As contact diagrams are simple, the IR divergence of  ${\rm Re} \psi_3$ is known in \eqref{psiphi3}, with each of them contributing a secular term $\log(-k_\Sigma\eta_0)$. Meanwhile, each classical loop integral contains a  $\log(k_E L_{\rm IR})$. 
As a result, we find the leading logarithmic divergence for this $L$-loop $n$-point correlator is given by
\be
\langle \phi^n \rangle_{L-\rm loop } \propto g^V  \log(k_E L_{\rm IR})^L \log(-k_\Sigma\eta_0)^V~.
\ee
One can quickly check that the above  result also works for other interactions with IR divergences. In general, considering an $m$-point vertex, we have the  relation $L = 1 + \(\frac{m}{2}-1\) V - \frac{n}{2}$ for diagrams with $n$ external legs, $V$ vertices and $L$ loops. Then for the two-point functions from a given interaction, the order of leading divergences depends on the number of vertices only
\begin{align}
\langle \phi_{\bf k}\phi_{-\bf k} \rangle'_{L-\rm loop } 
\propto  \begin{cases}
    \lambda^V  \log(k_E L_{\rm IR})^V \log(-k_\Sigma\eta_0)^V ~, ~~~~ & {\rm for}~ \Phi^4 \\
     g^V  \log(k_E L_{\rm IR})^{V/2} \log(-k_\Sigma\eta_0)^V ~, ~~~~ & {\rm for}~ \Phi^3 ~{\rm and }~ \dot\Phi\Sigma^2 ~. \end{cases}
\label{2point_IRdiv}     
\end{align}
Thus the order of divergences and the power of the couplings are uniquely related for classical loops. This fact plays an important role in the following analysis about resummation.
Meanwhile the IR divergences from quantum loops for the same diagram are suppressed, as they have the same power of the coupling, but lower power of logarithmic functions.
In the next section we will show that when we go to the coordinate space, this leading-log behaviour agrees with the result in Ref. \cite{Baumgart:2019clc} by using the retarded in-in formalism.

\vskip4pt
Finally, let's comment on the differences for the various interactions. 
For $g\Phi^3$ and $\lambda\Phi^4$, the leading divergent piece may also receive classical loop contributions from wavefunction coefficients of exchange diagrams, which have the same order of $\log$s with the products of contact $\psi_n$s, as we have seen in the previous analysis of the one-loop correlator from $\Phi^3$ interaction. 
The situation for $\dot\Phi\Sigma^2$ can be further simplified, since the only IR-divergent tree-level wavefunction coefficient is the contact $\psi_3$ in \eqref{psi3-chi}. 
As a result, the last term with the product of ${\rm Re} \psi_3$s in \eqref{lloop} gives us all the contribution to the leading  logarithms of IR-divergent correlators.
This means for this special two-field theory, all the higher-order $\psi_n$s in the semi-classical wavefunction are IR-safe, and we only need the simplest tree-level term $\psi_3$  to capture its IR-singular behaviour. This simple toy model example provides an interesting playground for studying IR divergences, which even allows us to go beyond perturbation theory as we shall discuss in Appendix \ref{app:wave}.

\newpage

\section{What does the Stochastic Formalism Resum?}
\label{sec:resum}

The stochastic formalism provides another framework to study IR effects of quantum field theories in dS, which is supposed to work even in the non-perturbative regime.
In this approach, the Fokker-Planck equation is applied to study the probability distribution $P$ of long wavelength perturbations. It has been argued for a long time that by doing so, there is an equilibrium state which resums  the IR-divergences arising from all the higher loop  contributions, and thus in the end   no IR divergences remain for interacting massless scalars in dS. However, it is still unclear  how this resummation was achieved, and how this formalism connects with the perturbation theory.
Meanwhile, for any quantum system, it is well-known that the probability distribution is simply given by $P=|\Psi|^2$. Therefore the wavefunction of the Universe provides an obviously natural setup to look into this remaining issue.
With results of IR-divergent correlators from field-theoretic computations in perturbation theory, now we explore the  connection with the stochastic formalism.

\vskip4pt
In this section, we work in the perturbative regime, and 
explicitly compare the calculation 
of IR-divergent correlators from the coarse-grained wavefunction (Section \ref{sec:long}) with the results from the stochastic formalism (Section \ref{sec:stochastic_correlations}). 
Interestingly, this analysis in the end provides a simple answer to the question proposed in the title of the section: 
\begin{center}
{ \it 
The stochastic formalism is a resummation of classical loops.}
\end{center}
In addition to comparing perturbative correlators in this section, we shall provide more supportive evidence in Appendix \ref{app:wave} by computing the probability distribution directly from the wavefunction. Then in Section \ref{sec:beyond} we further justify this conclusion in the non-perturbative regime.

\subsection{Long Modes Correlations from the  Wavefunction}
\label{sec:long}

To make the comparison, there is one more step for the computation of correlators in the wavefunction approach. 
Since the stochastic formalism describes long wavelength perturbations, we need to coarse grain the system by integrating over the short wavelength modes.
Let's first define the long wavelength perturbations in coordinate space by using a window function 
\be \label{philong}
\phi_l({\bf x}) = \int_{\bf k}  \Omega_\Lambda(k)e^{i\mathbf{k}\cdot\mathbf{x}}\phi_{\bf k}~,~~~~~~\chi_l({\bf x}) = \int_{\bf k}  \Omega_\Lambda(k)e^{i\mathbf{k}\cdot\mathbf{x}}\chi_{\bf k}.
\ee
This window function $\Omega_\Lambda(k)$  vanishes for $k>\Lambda$ and approaches $1$ for $k<\Lambda$, with $\Lambda$ being the UV cutoff. 
One simple example of the window function is the Heaviside step
$\Omega_\Lambda(k) = \Theta (\Lambda -k)$, and there are also smooth versions of the window function which preserve the locality of short-scale interactions~\cite{Gorbenko:2019rza}. 
%The explicit form of $\Omega_\Lambda$ will not be used in most of our computations.
The role of the UV cutoff is to define a fixed physical length $L_{UV}=a(t)/\Lambda$ such that perturbations with shorter wavelengths are smoothed out. This means the comoving scale $\Lambda$ should be time-dependent.
Meanwhile, as argued in \cite{Creminelli:2008es}, the minimum length scale for smoothing should be larger than the Hubble radius such that the quantum nature of fields can be neglected. We choose
$ \Lambda(t) = \epsilon a(t) H $ with $\epsilon \ll 1$.
As this UV cutoff is artificial, it should not appear in physical observables at the end of the computation. 

\vskip4pt
With this window function, we will look into the coarse-grained description  in Section \ref{sec:beyond}, and derive the probability distribution of the long wavelength perturbations in Appendix \ref{app:wave}. Here in this section we compute correlators of long wavelength modes using the results of perturbative wavefunction. 
In order to compare with the stochastic formalism, we are interested in the correlators at  the coincident point ${\bf x}$, which can be expressed in terms of the following Fourier transformation
\begin{align}
\langle\phi_l({\bf x})^n\rangle \equiv\int_{{\bf k}_1,\dots,{\bf k}_n}\prod_{i=1}^n {\Omega_\Lambda(k_i)}
\langle \phi_{{\bf k}_1} \dots\phi_{{\bf k}_n} \rangle e^{i({\bf k}_1+\dots+{\bf k}_n)\cdot {\bf x}}~,
\label{smeared_correlators}
\end{align}
where $\langle \phi_{{\bf k}_1} \dots\phi_{{\bf k}_n} \rangle$ are the correlators computed in Section \ref{sec:pert}.

\vskip4pt
Let's begin with the  free theory in order to explain our choice for regularization. In this simplest case, the wavefunction is fully determined by $\psi_2$ in \eqref{psi2}, and the two-point function at ${\bf x}=0$ is just the variance of Gaussian distribution
\be \label{variance}
\sigma^2\equiv \langle\phi_l({\bf x}=0)^2\rangle = \int_{\bf k} \frac{\Omega_\Lambda(k)}{2{\rm Re}\psi_2(k)} = \frac{H^2}{4\pi^2} \int_{1/L_{\rm IR}}^\Lambda \frac{dk}{k} = \frac{H^2}{4\pi^2} \log(\Lambda L_{\rm IR})~.
\ee
Naively, this two-point correlator in real space seems to depend on both the IR and UV cutoff scales, which should not be the case for physical observables.
However, the important part of the above logarithmic function is its time-dependence, while the rest can be seen as  regularization-dependent constants. Recall that  the comoving size of the Universe can be associated with some initial time $L_{\rm IR}^{-1} = a(t_i) H$, and we are free to pick a scale factor normalization $a(t_i)=\epsilon$. As a result, we get
\be \label{variancet}
\langle\phi_l^2\rangle =  \frac{H^2}{4\pi^2} \log a(t)~,
\ee
which reproduces the famous result for the stochastic behaviour of a free massless scalar in de Sitter, $\sigma^2 = H^3t/(4\pi^2)$ by using $a(t)=e^{Ht}$.
Physically, this secular growth with time can be explained as follows: the physical size of the box is expanding with $a(t)L_{\rm IR}$ while the smallest scale for screening is fixed to be $L_{UV}=a(t)/\Lambda(t) = 1/(\epsilon H)$. Therefore, there are more and more comoving modes entering the coarse-grained system that we are looking at, while no long wavelength perturbations are leaving the box.

\vskip4pt
Higher order correlation functions will follow the same logic, \textit{ie.} when integrating over momenta we will consider the same integration limits. Notice that the presence of IR divergence translates into higher powers of $\log a $. For instance, whereas the four point function in $\lambda\phi^4$ contains a single power of $\log(-k_T\eta_0)$ the correlator $\langle\phi_l^4\rangle$ goes as $\lambda H^4\log a^4$. The reason behind this is because there are now four different momenta entering the horizon, each adding a factor of $\log a$. Notice that the number of IR divergences is crucial, as it clearly indicates the larger terms. In this section we will make use of this fact to understand how the semi-classical solution is always the dominant one.

\vskip4pt
We can make more explicit the relation between perturbative computations and the stochastic formalism. To do so let us start by considering a collection of fields $\phi^a$ where $a$ labels the field. The probability distribution functions for the long wavelength part of $\phi^a_l$ can be written as,
\begin{align}
P[\phi_l^a]=\int  \[\prod_a  \mathcal{D}\phi_{\bf k}^a\  \delta\left(\phi_l^a-\int_\mathbf{k}\Omega_\Lambda(k)\phi_{\bf k}^a\right) \]
\vert\Psi[\phi_{\bf k}^a]\vert^2 ~.
\label{long_mode_Prob:def}
\end{align}
where $\Omega_\Lambda(k)$ is the window function defined earlier. 
As has been proven for a single scalar field in \cite{Gorbenko:2019rza}, the time evolution of $P[\phi_l^a]$ is given by the Fokker-Planck equation (We show how this is extended to two fields in Appendix \ref{app:2field}). If we write the wavefunction $\Psi[\phi^a]$ on perturbation theory,  the path integral will contains all long wavelength correlation functions. To see this fact, in detail it is instructive to write the $\delta$ function in \eqref{long_mode_Prob:def} as an exponential,
\begin{align}
P[\phi_l^a,t]= \int \mathcal D\phi_k^a\int dJ^a \exp\left[i\sum_a\left(\phi_l^a  -\int_{
\mathbf{k}
}\Omega_k\phi_k^a \right)J^a\right]\vert\Psi(\phi_k^a)\vert^2 ~.
\end{align}
Upon closer examination, the path integral over the short modes can be viewed as two successive Legendre transforms. The first transform introduces a current that only has support on the long-wavelength modes, while the second transform focuses on the long-wavelength part of the fields.

\vskip4pt
Let's delve into the first transformation. By introducing a current with restricted support over the square of the wavefunction, we effectively generate all connected correlators. This process is analogous to the creation of a linear interaction term, $\phi J$, in quantum field theory. After integrating over $\phi_{\bf k}$, we obtain all correlators with an arbitrary number of external $J$ currents. The number of external legs, denoted as $J$, encompasses all possible ways of connecting these legs using the available wavefunction coefficients. This includes not only tree-level diagrams but also all loops (up to the order specified by the wavefunction coefficients). This object corresponds to a partition function for the long wavelength correlators, denoted as $Z[J]$, whose expression is left in Appendix \ref{app:wave}. As usual, we can compute correlation functions by taking functional derivatives of  $\ln Z[J]$ and setting $J$ to 0. This operation effectively eliminates all external legs, leaving us with the connected correlations. 
In this sense the partition function computes the same quantities as the perturbative solution to the Fokker-Planck equation. Moreover, one of the advantages of solving the path integral is that we can understand how the semi-classical part dominates over all quantum corrections. 

Next, we  use this approach to derive the long modes correlators from the perturbative wavefunction. Our analysis again covers three types of interactions shown in \eqref{int3}.

\begin{itemize}

\item{$\lambda \Phi^4/4!$ \bf interaction}:
Let us start by considering the quartic vertex. Either using the partition function or the relation \eqref{smeared_correlators} , we can write down the two-point function of long modes as
\begin{align}
\langle\phi^2\rangle&\equiv \left.\frac{\delta^2\log Z[J]}{\delta J^2}\right\vert_{J=0}\nonumber\\
&=\frac{1}{Z[0]}\int_\mathbf{k}\frac{\Omega(k)}{2\mathrm{Re}\psi_2(\mathbf{k})}\left[1+\int \mathcal{D}\phi_k\vert\Psi\vert^2\left(\int_{\bf p}\frac{2\mathrm{Re} \psi_4(\mathbf{k},-\mathbf{k},\mathbf{p},-\mathbf{p})}{2\mathrm{Re}\psi_2(\mathbf{p})}\right) \phi_k^2 +\dots\right]~,
\label{2point_current}
\end{align}
where dots represent  terms higher order in $\lambda$.
Here we  focus solely on the tree-level wavefunction coefficients, and the contribution from the first term alone yields the conventional two-point function in the free theory. The window function $\Omega_\Lambda(k)$ effectively eliminates all short-wavelength modes.
Then the second term with the tree-level wavefunction coefficient $\psi_4$ gives the leading one-loop correction to the correlation function, as we have found in \eqref{phi4-1loop}. Performing the $\bf k$ integration of the smeared Fourier transformation,  we find 
\begin{align}
\langle\phi_l^2({\bf x}=0)\rangle&=\int_\mathbf{k}\Omega(k)\frac{H^2}{2k^3}\left[1+\frac{\lambda}{12\pi^2}\log(k L_{\mathrm{IR}})\log(-2k\eta_0)+\mathcal{O}(g^4)\right]\nonumber\\
&=\frac{H^2}{4\pi^2}\log a-\frac{\lambda H^2}{144\pi^2}(\log a)^3+\mathcal{O}(\lambda^2(\log a)^5 )~.
\label{2point_cubic_ft}
\end{align}
The leading logarithmic divergences from two-loop and arbitrarily higher-loop terms can be estimated in a similar way, though the numerical coefficients may become very difficult to compute.
For the $\Phi^4$ theory, the $L$-loop correction to the power spectrum from classical loops is given in \eqref{2point_IRdiv}, which lead to the following correction to the two-point function in coordinate space
\be
\langle\phi_l^2\rangle_{L-{\rm loop}} \propto \lambda^V (\log a)^{2V+1}~,
\ee
where the number of vertices satisfies $V= L$ for the two-point function from quartic interactions.
This result of the leading logarithms is in agreement with the analysis in Ref.~\cite{Baumgart:2019clc}.\footnote{To make the explicit comparison, we notice that the number of internal lines for a Feynman diagram  satisfies $I=L+V-1$. In Ref.~\cite{Baumgart:2019clc}, the leading order is given by $\lambda^V \log^Pa$, with $P=n+I$ being the total number of propagators. For the two-point function in $\Phi^4$ theory, we simply have $P=2V+1$.} Basically, it shows that if we go to higher order in perturbation theory, the corrections come as a power of the combination $\lambda(\log a)^2$.

Let us now consider the corrections from quantum loops. At the linear order in $\lambda$, we find a contribution from $\psi_2^{1-\mathrm{loop}}$ in \eqref{phi4-qloop}. After integrating over the external momentum $\bf k$, we find a term proportional to $\lambda \log^2 a$. This contribution is clearly subdominant compared to the loop correction from the tree-level wavefunction coefficient. 
To isolate the semi-classical contributions, we may define an effective coupling $\tilde \lambda=\lambda(\log a)^2$ and consider the limit $\lambda\to 0$ and $\log a\to\infty$ with $\tilde \lambda$ being constant. 
Then the quantum loops are suppressed by a factor of  $\tilde{\lambda}/\log a$ at least. This analysis can be simply extended to higher order terms.

\item{$g \Phi^3/3!$ \bf interaction}: The cubic vertex  follows the same logic as the quartic case. After a straightforward computation the two-point function is given by
\begin{align}
\langle\phi_l^2\rangle&\equiv \left.\frac{\delta^2\log Z[J]}{\delta J^2}\right\vert_{J=0}\nonumber\\
&=\frac{1}{Z[0]}\int_\mathbf{k}\frac{\Omega(k)}{2\mathrm{Re}\psi_2(\mathbf{k})}\nonumber\\
&\times\left[1+\int \mathcal{D}\phi_k\vert\Psi\vert^2\left(\int_\mathbf{p}\frac{\mathrm{Re}\psi_4(\mathbf{k},\mathbf{k}',\mathbf{p},-\mathbf{p})}{2\mathrm{Re}\psi_2(\mathbf{p})}\right.\right.\nonumber\\
&\qquad\left.\left.+\int_\mathbf{p}\frac{2\mathrm{Re} \psi_3(\mathbf{k},\mathbf{p},-\mathbf{p},-\mathbf{k})\mathrm{Re} \psi_3(\mathbf{k}',\mathbf{p},-\mathbf{p},\mathbf{p+k})}{2\mathrm{Re}\psi_2(\mathbf{p})2\mathrm{Re}\psi_2(\vert \mathbf{k}+\mathbf{p}\vert)}\right) \phi_{\mathbf{k}}^2 +\dots\right]~.
\label{2point_current}
\end{align}
 Within the parentheses, the path integral in the second term corresponds to a 1-loop correction arising from the exchange diagram $\psi_4$ as well as  from the product of two $\psi_3$ functions, as derived in \eqref{loop_cubic}. Consequently, the  expression becomes
\begin{align}
\langle\phi_l^2({\bf x}=0)\rangle&=\int_\mathbf{k}\Omega(k)\frac{H^2}{2k^3}\left[1+\frac{g^2}{72\pi^2}\log(k L_{\mathrm{IR}})\log(-2k\eta_0)^2+\mathcal{O}(g^4)\right]\nonumber\\
&=\frac{H^3}{4\pi^2}\log a\left[1+\frac{g^2}{288\pi^2}(\log a)^3+\mathcal{O}(g^4(\log a)^6)\right]~.
\end{align}
Similarly, we can also include the quantum loop corrections to this correlator. At the leading ($g^2$) order, we find a contribution from $\psi_2^{1-\mathrm{loop}}$ in \eqref{phi3-qloop} which  results in a term proportional to $g^2\log a^3$ and is subdominant compared to the classical loops. To isolate the semi-classical contributions, again we may introduce the coupling $\tilde{g} = g\log a^3$ and consider the limit as $g \to 0$, $\log a \to \infty$, while $\tilde{g}$ remains finite. For instance, in the given example, the quantum correction from $\psi_2^{1-\mathrm{loop}}$ is suppressed by a factor of $\tilde{g}/\log a$.

Focusing on classical loop contributions, we can simply write down higher order corrections from $L$-loop processes  by using the leading logarithmic divergences that we identified in \eqref{2point_IRdiv}. For two-point correlators of long modes in coordinate space, we find
\be \label{phi3Lloop}
\langle\phi_l^2\rangle_{L-{\rm loop}} \propto g^{2L} (\log a)^{3L+1}~,
\ee
where we have used the relation for the number of vertices $V=2L$ for two-point functions from cubic interactions. This tells us at each order in perturbation theory, the leading contributions to IR divergences come with the combination $g^2\log^3a$. Quantum loops only lead to subdominant contributions suppressed by lower powers of $\log a$.

\item{$\alpha\dot\Phi\Sigma^2/2$ \bf interaction}:
This two-field case is actually simpler, as in the semi-classical wavefunction only the contact $\psi_3$ in \eqref{psi3-chi}  contains secular growth term. Therefore, we are allowed to neglect $\psi_{n>3}$ from exchange diagrams, and  the semi-classical wavefunction with $\psi_2$ and $\psi_3$ captures the leading IR divergences in this theory. Furthermore, we can explicitly perform the integration over the fields and obtain the following generator\footnote{We leave further details of this computation in Appendix \ref{app:wave}. There, we also present the computation for the probability distribution and confirm that the result from this simplified semi-classical wavefunction is a solution of the Fokker-Planck equation. This provides another support for our conclusion that the stochastic formalism is resumming classical loops.}
\begin{align}
Z[J_\phi,J_\chi]=\exp\left\{-\frac{\langle\phi^2\rangle}{2}J_\phi^2+\frac{\langle\chi^2\rangle}{1-i\frac{ \langle\phi\chi^2\rangle}{\langle\phi^2\rangle} J_\phi}\frac{J_\chi^2}{2}\right\}~.
\end{align}
It's worth noting that the cubic interaction introduces a non-local term, which corresponds to a resummation of diagrams with two legs of $\chi$ and an arbitrary number of legs of $\phi$. It can be readily verified that the two-point function is given by
\begin{align}
\langle\phi^2\rangle = \frac{H^2 }{4\pi^2}\log a + \frac{\alpha^2H^2}{16\pi^2}(\log a)^4~.
\label{2point_2fds_ft}
\end{align}
Notice that integrating over $J_\chi$ yields a collection of classical one-loop diagrams with an arbitrary number of external $\phi$ legs. This result is similar to the one obtained in \cite{Panagopoulos:2020sxp} using a different technique but with a similar coupling. The consequences of this will be explored elsewhere.

For this interaction, the contributions from quantum loops are IR-finite and thus absent, as we have seen in \eqref{chi-qloop}. Meanwhile, 
the leading logarithms for arbitrary higher order loop diagrams have the same behaviour as in \eqref{phi3Lloop}, as  the classical loops here  are similar with the ones from the $\Phi^3$ interaction.

\end{itemize}

\vskip4pt
This concludes our perturbative analysis using the wavefunction method. After identifying the leading IR divergences from classical loops, now we are ready to make the comparison with perturbative computations of  the stochastic formalism.

\subsection{Correlators in the Perturbative Stochastic Formalism}
\label{sec:stochastic_correlations}

Now we consider the stochastic formalism and try to establish the link with the field-theoretic computations. This approach is supposed to be an effective description for the stochastic behaviour on superhorizon scales while the short wavelength perturbations are smoothed out. In order to make the explicit comparison with the wavefunction calculation of three types of interactions in \eqref{int3}, here we look at the following two-field system for the stochastic analysis
\be \label{longL}
\mathcal{L}_l = \frac{1}{2}\dot\phi^2  +\frac{1}{2}\dot\chi^2 - f(\chi)\dot\phi - V(\phi,\chi)~.
\ee
where the gradients of fields have been neglected.
Strictly speaking, the two scalars here $\phi(t)$ and $\chi(t)$ correspond to boundary fluctuations $\phi_l$ and $\chi_l$ in the field-theoretic approach.
Then \eqref{longL} can be seen as an approximate Lagrangian around the future boundary with time $t$ where we perform the bootstrap analysis.
We  keep the time dependence of these long wavelength modes as the system may still evolve.
The three types of interactions can be achieved by choosing different forms of the potential function $V(\phi,\chi)$ and the $f(\chi)$ function for the kinetic coupling.

\vskip4pt
The starting point of the stochastic approach is the Fokker-Planck equation for the probability distribution of long wavelength perturbations. Here we consider an extended version of the Fokker-Planck equation that also works for the long-wavelength system in \eqref{longL} 
\begin{align} \label{FP-2field}
\frac{d P}{d t}=\frac{\partial}{\partial\phi}\left[\left(\frac{V_\phi}{3H}+f(\chi)\right)P\right]+\frac{\partial}{\partial\chi}\left[\left(\frac{V_\chi}{3H}\right)P\right]+\frac{H^3}{8\pi^2}\left(\frac{\partial^2P}{\partial\phi^2}+\frac{\partial^2P}{\partial\chi^2}\right)~.
\end{align}
We leave the details of its derivation in Appendix \ref{app:2field}, and here focus on its implications for correlators.
As is well-known, the Fokker-Planck equation describes the time evolution of the probability distribution $P[\phi,\chi;t]$ with two physical effects: {\it  drift}, which is given by the first two terms on the right-hand side of \eqref{FP-2field} above and comes from the background equation of motion; {\it  diffusion}, which is the last term caused by  quantum noise.
In the non-perturbative regime, these two effects are supposed to compete with each other and in the end lead to an equilibrium state  with ${dP}/{dt}=0$. 

\vskip4pt
With a given probability distribution function, correlation functions can be computed using the Born rule:
\begin{align}
\langle\phi^n\chi^m\rangle=\int \mathcal{D}\phi \mathcal{D}\chi \phi^n\chi^m P[\phi,\chi;t]~.\label{def:stochastic_corr}
\end{align}
By substituting  \eqref{def:stochastic_corr} into the Fokker-Planck equation, we  derive a system of differential equations governing the evolution behavior of the correlation functions. 
Assuming that $V(\phi, \chi)$ are polynomial in the fields, we can write these equations into the following form
\begin{align}
\frac{d}{dt}\langle\phi^n\chi^m\rangle&=\int d\phi d\chi \phi^n\chi^m \frac{P(\phi,\chi,t)}{dt}\nonumber\\
&=-\frac{n}{3H}\left\langle\phi^{n-1}V_\phi\chi^m\right\rangle-n\langle\phi^{n-1}\chi^m f(\chi)\rangle-\frac{m}{3H}\left\langle\phi^{n}\chi^{m-1}V_\chi\right\rangle\nonumber\\
&+\frac{n(n-1)H^3}{8\pi^2}\langle\phi^{n-2}\chi^m\rangle+\frac{m(m-1)H^3}{8\pi^2}\langle\phi^{n}\chi^{m-2}\rangle~,
\end{align}
where the second line comes from the drift and the third line is due to the diffusion.
This is an infinite set of differential equations for all the correlators in the theory.
For a free theory, the equations are decoupled and only the diffusion terms remain on the right-hand side. Then it is easy to get $\langle\phi^2\rangle =\langle\chi^2\rangle = H^3t/(2\pi)^2$, which is just the Gaussian variance we computed in \eqref{variancet}. 
However, for interesting theories, this set of equations are not solvable in general. One needs to know all of the equations even for computing lower-point correlation functions like $\langle\phi^2\rangle$.  
Next, let's try to make progress by treating the drift terms perturbatively, and explicitly show the results for the three types of interactions.

\begin{itemize}
    \item {\bf $V={\lambda\phi^4}/{4!}$ and $f(\chi)=0$}: 
This simply corresponds to the $\Phi^4$ interaction in the field-theoretic computation. 
For the single scalar, the equations  for  non-vanishing correlation functions of $\phi$ are given by
\begin{align}
\frac{d}{dt}\langle\phi^2\rangle&=-\frac{1\lambda}{9H}\langle\phi^4\rangle+\frac{H^3}{4\pi^2}\nonumber\\
\frac{d}{dt}\langle\phi^4\rangle&=-\frac{2\lambda}{9H}\langle\phi^6\rangle+\frac{3H^3}{4\pi^2}\langle\phi^2\rangle\nonumber\\
&\qquad\qquad\vdots\nonumber\\
\frac{d}{dt}\langle\phi^n\rangle&=-\frac{n\lambda}{18H}\langle\phi^{n+2}\rangle+\frac{n(n-1)H^3}{8\pi^2}\langle\phi^{n-2}\rangle~.
\end{align}
Now we solve for the lower-point correlation functions order by order in $\lambda$. 
The key point to note is that if we set $\lambda=0$, then the solution for the correlation functions is approximately given by $\langle\phi^{n}\rangle\sim H^{n}(\log a)^{n/2}$. Turning on $\lambda$ introduces corrections of $\lambda (\log a)^{n/2+1}$ to each $n$-point correlator at the leading order. Then iteratively, the lower-point functions get corrections from higher-point ones. 
For instance, truncating at $\lambda^3$ is equivalent to setting $\lambda=0$ in the equation for $\phi^6$. Doing so, we find that the two-point function becomes:
\begin{align}
\langle\phi^2\rangle=\frac{H^2}{4\pi^2}\log a-\frac{\lambda H^2}{144\pi^4}(\log a)^3 +\frac{\lambda^2H^2}{2880\pi^6} (\log a)^3 + \mathcal{O}(\lambda^3(\log a)^7  a)~.
\end{align}
One can show that for corrections to the variance  the expansion parameter is $\lambda\log^2 a$, and the $m$th order correction is proportional to $\lambda^{m} (\log a)^{2m+1} $. This exactly agrees with what we have found in field-theoretic computation for leading logarithms from classical loops. 
Explicitly, we can check the result at one-loop level as it was obtained in Eq.\eqref{2point_cubic_ft}.

\vskip4pt
This calculation confirms that the contributions from classical loops are incorporated within the stochastic formalism.  One can further check that the subleading terms from quantum loops, e.g. $\psi_2^{\mathrm{1-loop}}$, can be associated with the corrections to the Fokker-Planck equation identified in Ref. \cite{Gorbenko:2019rza}.  
It is easy to see that these terms are subdominant in perturbative expansions, as at the same order in $\lambda$ the quantum loops always contribute fewer powers of logarithmic functions. We explicitly check this in Appendix \ref{App:corrections} where we perturbatively solve the noise term in the Fokker-Planck equation. We get a contribution to the two point function that scales as $\lambda(\log a)^2$ in accordance with the result obtained using the perturbative wavefunction.

\item {\bf $V={g\phi^3}/{3!}$ and $f(\chi)=0$}: This example is supposed to be related to the $\Phi^3$ interaction. However, the potential is unbounded from below, and the stochastic computation cannot be consistently applied. Thus we do not make the explicit comparison with the field-theoretic computation from $\Phi^3$ interaction here. 

    \item {\bf $f(\chi) = \alpha\chi^2/2$ and $V=0$}.
This corresponds to the two-field example with the derivative interaction $\dot\Phi\Sigma^2$.   The differential equations for correlators are given by
\begin{align}
\frac{d}{dt}\langle\phi^2\rangle&=-\alpha\langle\phi\chi^2\rangle+\frac{H^3}{4\pi^2}\nonumber\\
\frac{d}{dt}\langle\chi^2\rangle&=\frac{H^3}{4\pi^2}\nonumber\\
\frac{d}{dt}\langle\phi\chi^2\rangle&=-\frac{\alpha}{2}\langle\chi^4\rangle+\frac{H^3}{4\pi^2}\langle\phi\rangle\nonumber\\
\frac{d}{dt}\langle\chi^4\rangle&=\frac{3H^3}{2\pi^2}\langle\chi^2\rangle\nonumber\\
\frac{d}{dt}\langle\phi\rangle&=-\frac{\alpha}{2}\langle\chi^2\rangle~.\nonumber
\end{align}
Solving these equations, we find the first-order correction to the two-point function 
\begin{align}
\langle\phi^2\rangle=\frac{H^2 }{4\pi^2}\log a+\frac{\alpha^2 H^2}{128\pi^4}(\log a)^4~,
\end{align}
which again matches the result from classical loops obtained using the wavefunction method as in Eq.~\eqref{2point_2fds_ft}.
This case is actually simpler than the $\phi^4$ and $\phi^3$ interactions. 
One way to see it is to notice that there is a conserved current for interactions with one time derivative
\begin{align}
 j^0=\dot\phi-\frac{1}{2}f(\chi)=\frac{\pi_\phi }{a^3}~,
 \label{conserved:current}
\end{align}
which is a generalisation of the usual momentum conservation in the single field case.  So in the long wavelength limit $\dot\phi$ is no longer conserved, but the more general combination given by  $j_0$ is required to be constant. Thus intuitively the massless scalar $\chi$ on the superhorizon scales sources the growth of $\phi$, with $\phi\sim \alpha\chi^2 \log a $, as is normally seen in multi-field inflation models.

In the above solution, there seems to be no corrections at higher order in $\alpha$.  However, we should notice that when constructing the Hamiltonian due to an interaction term like $\dot\phi f(\chi)$, additional terms involving self-interactions of $\chi$ will inevitably appear. This is a consequence of $\chi$ depending on momentum, as given by $a^{-3}\pi_\phi=\dot\phi-\frac{1}{2}f(\chi)$. Consequently, the interacting Hamiltonian will contain a term proportional to $f(\chi)^2/4$, which can be conceptualized as an effective potential term, denoted as $V_{\rm eff}(\phi)=f(\chi)^2/4$. The influence of this term is  subdominant, at most manifesting at the next-to-leading order of the couplings represented by $f(\chi)$. 
Including this term perturbs the Fokker-Planck equation in a manner akin to the situation when a potential like $V(\phi)=\lambda\phi^4$ is considered. In the aforementioned example, the presence of the self-interactions of $\chi$ alters the equation for $\langle\chi^2\rangle$, consequently introducing corrections that begin at order $\alpha^4$ to the two-point function $\langle\phi^2\rangle$. In this way we could derive the higher order corrections systematically. It will be possible to show that  the result will scale as  $(\alpha^2\log^3a)^m$ for the $m$th order, which is in agreement with a similar estimate for the wavefunction computation of classical loops from Eq.\eqref{2point_IRdiv}.
\end{itemize}

\vskip4pt
Now we are in the position to establish the connection between the wavefunction approach and the  stochastic formalism in perturbation theory. From the above examples, we have seen that the correlation functions computed through perturbation theory are the same as those computed by solving the Fokker-Planck equation perturbatively. 
While we  explicitly checked the matches of results for the variance, the same analysis can be extended to higher-point functions.

\vskip4pt
In general, at each order in perturbation theory, the stochastic computation always leads to the same power of $\log a$ as that found from classical loops in the wavefunction method. Meanwhile, the quantum loop contributions are suppressed because they have less powers of logarithmic functions. 
This observation provides a simple answer about what the stochastic formalism is resumming: classical loops from tree-level wavefunction coefficients.
It confirms our intuitive understanding that IR divergences in dS are governed by semi-classical physics on superhorizon scales.

Interestingly, it also explains why the Fokker-Planck equation may not be a good description for IR-finite theories, such as the ones with higher-derivative interaction: there the quantum loops are not always suppressed and can be equally important as the classical loop contributions, while they are beyond the scope of a stochastic description.

\vskip4pt
In the end, we would like to comment on the differences with Ref. \cite{Baumgart:2019clc} where similar results about leading IR divergences at any loop order were derived. There the authors applied the retarded in-in formalism  to identify the leading logarithms in perturbaton theory, and in the end ``the resummation of time" reproduces  results from the Fokker-Planck equation.
Here with the wavefunction method, we further point out that the these leading logarithms have an origin as classical loops.  
%\vskip4pt
This new understanding helps us to simplify the perturbative computation for IR divergences at the loop level. Meanwhile, it indicates that the leading IR effects should be captured by the saddle-point approximation of the wavefunction \eqref{saddle}, as all the tree-level information has been included there. With this insight, we shall go beyond the perturbation theory in the next section and directly study the implications of the semi-classical wavefunction for IR divergences in dS.

\section{Beyond Perturbation Theory}
\label{sec:beyond}

In this section, we go beyond the perturbative regime  of IR divergences in dS. We will show that, for the semi-classical piece of the wavefunction, the Fokker-Planck equation  follows as a consequence of the Schr\"{o}dinger equation and Polchinski's exact renormalization group (RG) equation from a boundary perspective. {We note that our major focus here is to explore the link between the stochastic formalism and the semi-classical wavefunction in the non-perturbative regime. While we apply the exact RG techniques to perform the computation, the precise analogy of Polchinski's equation in cosmology is beyond the scope of the current work. }

\vskip4pt
As we noticed in the previous sections, when we have IR divergences in de Sitter, 
the semi-classical wavefunction always provides the dominant contributions to correlators in perturbation theory, which match the perturbative computation in the stochastic formalism. This interesting observation suggests that the stochastic formalism, which is also expected to be valid in the non-perturbative regime, actually resums the classical loops from all the tree-level wavefunction coefficients. 
In this section we shall show that indeed the Fokker-Planck equation can be derived from
 the saddle-point approximation of the wavefunction.

\vskip4pt
Our focus lies in the coarse-grained theory at the late-time boundary, where we can effectively ``integrate out" the short-wavelength modes. This process results in an effective description for the long-wavelength modes. 
Explicitly, we  regularize the theory by requiring that only momentum
modes up to  the UV cutoff scale $\Lambda$ are allowed.
Then in analogy with \eqref{Psi:coeff}, 
here we start with the coarse-grained 
probability distribution of field fluctuations $\phi({\bf x})$  at the late-time boundary of time $t$\footnote{{We choose to consider the coarse-grained probability distribution instead of the wavefunction because of a theoretical subtlety. In general the coarse-grained wavefunction is not a well-defined object, as its time evolution may no longer be described by the Schr\"{o}dinger equation. More precisely, when we integrate out short modes, we find a density matrix (or probability distribution)  for long modes. Since $P_\Lambda$  is the object of interest here, we shall directly look at it  in the following analysis. We thank the anonymous referee for pointing this out.}}
\be \label{PLambda}
P_\Lambda[\phi, t] = |\Psi_\Lambda[\phi, t]|^2 =  e^{ W _0[\phi] + W_I[\phi] } ~,
\ee
with the free and interacting parts given by
\bea
W_0 &=& \int_{{\bf k}} 
{\rm Re}~\psi_2(k) \Omega_\Lambda^{-1}(k) ~\phi_{{\bf k}}\phi_{-{\bf k}}  ~,   \nn\\
W_I &=& \sum_{n = 3}^{\infty}\frac{1}{n!}\int_{{\bf k}_1,...,{\bf k}_n}
\, 2{\rm Re}~\psi_n^\Lambda({\bf k}_1,...,{\bf k}_n,t)  ~\phi_{{\bf k}_1}\cdot\cdot\cdot \phi_{{\bf k}_n}\, .
\eea
%version of the wavefunction can be expressed 
% %as a functional of the boundary field 
% \begin{align}
% \Psi_\Lambda[\phi,t] =& \exp \[  \frac{1}{2} \int_{{\bf k}} 
% \psi_2(k) \Omega_\Lambda^{-1}(k) ~\phi_{{\bf k}}\phi_{-{\bf k}} + \sum_{n = 3}^{\infty}\frac{1}{n!}\int_{{\bf k}_1,...,{\bf k}_n}
% \, \psi_n^\Lambda({\bf k}_1,...,{\bf k}_n,t)  ~\phi_{{\bf k}_1}\cdot\cdot\cdot \phi_{{\bf k}_n}\,
% \]\, .
%  \label{Psi:Lambda}
% \end{align}
In the free theory part $\psi_2$, we have used the window function $\Omega_\Lambda(k) $ introduced below \eqref{philong} for the blocking. 
Now we see this window function fixes the $\Lambda$-dependence of the quadratic term in $\Psi_\Lambda$, while the concrete forms of
 $\psi_{n\geq 3}^\Lambda$ remain unclear, especially when we go beyond perturbation theory. 
 This is an important question that will be analyzed in the following derivation using the RG flow techniques.
In the end we recall that  this cutoff scale depends on the time of the boundary slice $ \Lambda(t) = \epsilon a(t) H $ with $\epsilon \ll 1$.

% \vskip4pt
% Next, we consider the probability distribution of field fluctuations $\phi({\bf x})$  at the late-time boundary $t$. 
% By using the coarse-grained wavefunction, we simply have
% \be \label{PLambda}
% P_\Lambda[\phi, t] = |\Psi_\Lambda[\phi, t]|^2 =  e^{ W _0[\phi] + W_I[\phi] } ~,
% \ee
% with the free and interacting parts given by
% \bea
% W_0 &=& \int_{{\bf k}} 
% {\rm Re}~\psi_2(k) \Omega_\Lambda^{-1}(k) ~\phi_{{\bf k}}\phi_{-{\bf k}}  ~,   \nn\\
% W_I &=& \sum_{n = 3}^{\infty}\frac{1}{n!}\int_{{\bf k}_1,...,{\bf k}_n}
% \, 2{\rm Re}~\psi_n^\Lambda({\bf k}_1,...,{\bf k}_n,t)  ~\phi_{{\bf k}_1}\cdot\cdot\cdot \phi_{{\bf k}_n}\, .
% \eea
To derive the Fokker-Planck equation, we notice that the time dependence of $P_\Lambda$ comes from two different terms. 
The first one is from the wavefunction itself: as we have seen in the perturbative computation with IR divergences, the wavefunction coefficients in general contain secular terms with logarithmic  growth. The second one comes from the time dependence of the cutoff $\Lambda(t)$. Thus we find
\be \label{FP}
\frac{d}{dt} P_\Lambda[\phi,t] = \frac{\partial}{\partial t} P_\Lambda[\phi,t] + \dot\Lambda \frac{\partial}{\partial \Lambda} P_\Lambda[\phi,t]~.
\ee
In the rest of the section, we shall look into these two contributions and focus on their physical origins.
Before going into the details of the derivation, the upshot is that the first term gives the classical drift from the Schr\"{o}dinger equation, while the second term comes as a RG flow on the boundary which in the end leads to the diffusion.
Some of our derivation share similarities with the one presented in Ref.~\cite{Gorbenko:2019rza}, especially for the derivation of the drift term. 
The novelty in our approach is to notice that in the semi-classical regime the diffusion term comes as a consequence of Polchinski's equation of RG flow on the boundary.
 We shall comment on the differences with Ref.~\cite{Gorbenko:2019rza} in more detail at the end of this section.

\subsection{Diffusion as the RG Flow on the Boundary}

Let's first briefly review Polchinski's derivation of the exact renormalization equation for a general quantum field theory in $d$-dimensional space.
The basic idea of the exact RG flow starts from the requirement that physics should be independent of the artificial cutoff scale that we imposed in the theory, thus the partition function remains invariant with respect to the change of $\Lambda$
\be \label{RG-eq}
\Lambda \frac{d}{d\Lambda} Z_\Lambda = 0~, ~~~~~~ {\rm with} ~~ Z_\Lambda \equiv \int \mathcal{D} \varphi ~e^{-S_{\rm eff}^\Lambda}~,
\ee
where $S_{\rm eff}^\Lambda$ is the effective action below the scale $\Lambda$ with UV degrees of freedom being integrated out.
Then as we change the UV cutoff, the overall partition function should be unaffected,
which basically means that, the low-energy effective action would vary correspondingly to account
for the change in the short-wavelength modes ($k>\Lambda$) that are being integrated out.
Specifically, Polchinski's idea is to cut off the theory with a smooth window function $\Omega_\Lambda$ as the one introduced in \eqref{PLambda}. Then the effective action can be written as the free theory part and the interacting part $S_{\rm eff}^\Lambda = S_{0}^\Lambda+ S_{\rm int}^\Lambda$, with
\be
S_{0}^\Lambda = \frac{1}{2}\int \frac{d^d k}{(2\pi)^d} \frac{1}{\Omega_\Lambda} \varphi_{\bf k}G_k^{-1} \varphi_{-\bf k}~.
\ee
Here $G_k$ is the propagator in momentum space, and in the above expression we see that it has been regularized by the smooth window function $\Omega_\Lambda(k)$.
Then the condition \eqref{RG-eq} in the end leads to a non-perturbative functional differential equation for the $\Lambda$-dependence of $S_{\rm int}^\Lambda$
\be \label{Polchinski}
\Lambda \frac{d}{d\Lambda} e^{-S_{\rm int}^\Lambda} =  -\frac{1}{2} \int \frac{d^d k}{(2\pi)^d} \frac{d \Omega_\Lambda}{d\ln \Lambda} {G_k} 
 \frac{\delta^2}{\delta\varphi_{\bf k}\delta\varphi_{-\bf k}} e^{-S_{\rm int}} ~.
\ee
By substituting \eqref{Polchinski} into \eqref{RG-eq} and performing integration by parts twice, one can check that the $\Lambda$-dependence of $S_{\rm int}^\Lambda$ cancels out the $\Lambda$-dependence of $S_{0}^\Lambda$, and thus the partition function remains independent of $\Lambda$. 
Equation \eqref{Polchinski} is
known as Polchinski's exact renormalization group  \cite{Polchinski:1983gv}, which provides a general and non-perturbative description for the running of couplings in the low-energy effective theory. See~\cite{Bagnuls:2000ae,Berges:2000ew,Pawlowski:2005xe,Rosten:2010vm,Cotler:2022fze} for recent reviews and discussions.
It is worth mentioning that, Polchinski's equation \eqref{Polchinski} is not the only option to satisfy the condition in \eqref{RG-eq}. As integrations by parts are included, we are left with the freedom to add boundary terms in \eqref{Polchinski} without affecting the $\Lambda$-independence of $Z_\Lambda $. These boundary terms correspond to different choices of RG schemes.

\vskip4pt
Similarly, for theories with IR divergences in dS, an effective description for long wavelength modes is given by the coarse-grained probablity distribution \eqref{PLambda}. 
For now let us forget about the time evolution in the bulk, and consider this 3-dimensional Euclidean field theory on the boundary.
As the object of interest here is the probability distribution \eqref{PLambda}, we start with the requirement that the conservation of probability $\int \mathcal{D}\phi~P_\Lambda=1$ should be unaffected by changing the cutoff scale
\be \label{Pcons}
\frac{d}{d\ln \Lambda} \int \mathcal{D}\phi~ P_\Lambda[\phi,t] = \int \mathcal{D}\phi ~\[ 
\frac{d W_0}{d\ln \Lambda} e^{W_I} + \frac{d e^{W_I}}{d\ln \Lambda}  \] e^{W_0} =0~.
\ee
This means that the $\Lambda$-dependence of the free theory part should be cancelled out by the $\Lambda$-dependence of the interacting part. 
For $W_0$, it depends on the cutoff scale through the window function
\be \label{W0L}
\frac{d W_0}{d\ln \Lambda} = \int_{{\bf k}} 
{\rm Re}~\psi_2(k)  ~\phi_{{\bf k}}\phi_{-{\bf k}}  \frac{d}{d\ln \Lambda}\Omega_\Lambda^{-1}(k) ~.
\ee
For the $\Lambda$-dependence of $W_I$, there are also various choices that can ensure the probability conservation with varying cutoff scales.
In analogy with \eqref{Polchinski}, we propose the following version of Polchinski's equation
\begin{eBox}
\be \label{modi-polchin}
e^{W_0} \frac{d e^{W_I}}{d\ln \Lambda} = \frac{1}{4} \int_{\bf k} \frac{d \Omega_\Lambda}{d\ln \Lambda} \frac{1}{{\rm Re} \psi_2} 
\[ \(\frac{\delta^2}{\delta\phi_{\bf k}\delta\phi_{-\bf k}} e^{W_I} \) e^{W_0} - 2 \frac{\delta}{\delta\phi_{\bf k}} \( e^{W_0} \frac{\delta e^{W_I}}{\delta\phi_{-\bf k}}\)\]~,
\ee
\end{eBox}
which has an extra boundary term.
This is supposed to be the exact RG equation on the late-time boundary for the semi-classical wavefunction.
Let us justify it step by step.
First of all, it is straightforward to check that
with this ansatz indeed \eqref{Pcons} is satisfied. By substituting \eqref{modi-polchin} into \eqref{Pcons}, we find 
\bea
\frac{d}{d\ln \Lambda} \int \mathcal{D}\phi ~ P_\Lambda[\phi,t] &= &  \frac{1}{4} \int_{\bf k} \frac{1}{{\rm Re} \psi_2}  \frac{d \Omega_\Lambda}{d\ln \Lambda} \int D \phi \[ 
-\frac{\delta^2 e^{W_0}}{\delta\phi_{\bf k}\delta\phi_{-\bf k}} e^{W_I} \right.\nn\\
&& ~~~~~~~~~~~~  \left. + \(\frac{\delta^2}{\delta\phi_{\bf k}\delta\phi_{-\bf k}} e^{W_I} \) e^{W_0} - 2 \frac{\delta}{\delta\phi_{\bf k}} \( e^{W_0} \frac{\delta e^{W_I}}{\delta\phi_{-\bf k}}\) \] 
~,
\eea
where the first term comes from rewriting \eqref{W0L}. The last term is a total functional derivative and thus can be neglected.
Then doing integration by parts twice for the second term in the bracket, we move the $\frac{\delta}{\delta\phi}$ derivatives on $e^{W_I}$ onto $e^{W_0}$. Neglecting the boundary terms, we see that the remaining piece just cancels out the first term from the $\Lambda$-dependence of $W_0$.

\vskip4pt
Next, we need to justify the appearance of the last term in \eqref{modi-polchin}.
In general, we have the freedom to add arbitrary boundary terms in Polchinski's equation, which correspond to different  RG schemes. 
The choice in \eqref{modi-polchin} corresponds to the semi-classical scheme that will be spelled out below.
As we shall see  soon after, this particular form of the boundary term  in \eqref{modi-polchin} becomes important for deriving the diffusion term in the Fokker-Planck equation.

\paragraph{The semi-classical scheme}
So far we have kept our analysis fully non-perturbative, while the RG flow equation \eqref{modi-polchin} is also expected to be valid in the perturbative regime.
Equivalently, we can write it as the following flow equation satisfied by the interacting part of the wavefunction 
\be \label{flow}
\frac{dW_I}{d\ln \Lambda} = - \frac{1}{4} \int_{\bf k} \frac{1}{{\rm Re} \psi_2} \frac{d \Omega_\Lambda}{d\ln \Lambda} \[ \frac{\delta^2 {W_I}}{\delta\phi_{\bf k}\delta\phi_{-\bf k}} + \frac{\delta {W_I}}{\delta\phi_{\bf k}}\frac{\delta {W_I}}{\delta\phi_{-\bf k}}+2 \frac{\delta {W_0}}{\delta\phi_{\bf k}}\frac{\delta {W_I}}{\delta\phi_{-\bf k}} \]~,
\ee
which tells us the running of wavefunction coefficients with the cutoff scale  in interacting theories.\footnote{It is an analogue to the running of couplings in the standard quantum field theory. Here we choose the couplings, like $g$ and $\lambda$, to be constant, and the $\Lambda$-dependence comes from the logarithmic terms in the coarse-grained wavefunction coefficients. Therefore, one  way to think about this running is to consider the effective couplings that absorb the logarithmic $\Lambda$-dependence in $\psi_n$.}
Now we explicitly show that the flow equation has to be the one above in order to match the perturbation theory computation with the semi-classical wavefunction.

\vskip4pt
For concreteness, we take the $g\Phi^3$ interaction as an example. At the $g$ and $g^2$ orders, we have the following components in the semi-classical wavefunction
\be
W_I^{(1)} = \frac{1}{3} \int_{\bf k_1,k_2,k_3} {\rm Re} \psi_3^{\Lambda}~ \phi_{\bf k_1}\phi_{\bf k_2}\phi_{\bf k_3} ~,~~~~~~
W_I^{(2)} = \frac{1}{12} \int_{\bf k_1,k_2,k_3} {\rm Re} \psi_4^{\Lambda}~ \phi_{\bf k_1}\phi_{\bf k_2}\phi_{\bf k_3}\phi_{\bf k_4}~,
\ee
which correspond to the three-point contact and four-point single-exchange diagrams.
Meanwhile, there is also a quantum wavefunction coefficient $\psi_2^{\rm 1-loop}$ at the $g^2$ order
\be \label{W1loop}
\Tilde{W}^{(2)}_I  =  \int_{\bf k_1,k_2}  {\rm Re} \psi_2^{\rm 1-loop} ~ \phi_{\bf k_1}\phi_{\bf k_2} ~.
\ee
Let's first consider the semi-classical wavefunction from tree-level processes.
In perturbation theory, their $\Lambda$-dependence can be derived by requiring that the corresponding correlators are independent of the cutoff scales. As correlators of long wavelength modes are physical observables, {at the leading order of the coupling constant,} the final results should not be affected by the artificial UV cutoff scale that we choose for smoothing out the short wavelength modes.\footnote{{This argument also applies for tree-level processes only. Beyond the leading order,  it is the correlator of the full field $\phi=\phi_l+\phi_s$ that is cutoff independent, but individually $\langle\phi_l^n\rangle$s are still cutoff dependent \cite{Gorbenko:2019rza}.}}
At the linear order in $g$, the three-point correlator is given by
\be
\langle \phi_l^3 \rangle = \int D\phi \phi^3 P_\Lambda = -  \int_{\bf k_1,k_2,k_3} \frac{\Omega_\Lambda(k_1)\Omega_\Lambda(k_2)\Omega_\Lambda(k_3)}{8 {\rm Re} \psi_2 (k_1){\rm Re} \psi_2 (k_2){\rm Re} \psi_2 (k_3)} 2{\rm Re} \psi_3^{\Lambda}~.
\ee
Then up to the $g$ order, its $\Lambda$-independence leads to
\be \label{psi3Lambda}
\frac{d}{d\ln \Lambda} \langle \phi_l^3 \rangle =0 ~~ \Rightarrow ~~ \frac{d{\rm Re} \psi_3^{\Lambda}}{d\ln \Lambda}  = -\frac{d }{d\ln \Lambda} \big(\ln\[\Omega_\Lambda(k_1)  \Omega_\Lambda(k_2)\Omega_\Lambda(k_3)  \]\big) {\rm Re} \psi_3^{\Lambda}+ \mathcal{O}(g^2)~,
\ee
which can be written as
\be  
\frac{dW_I^{(1)}}{d\ln \Lambda} = - \frac{1}{2} \int_{\bf k} \frac{1}{{\rm Re} \psi_2} \frac{d \Omega_\Lambda}{d\ln \Lambda}  \frac{\delta {W_0}}{\delta\phi_{\bf k}}\frac{\delta {W_I}^{(1)}}{\delta\phi_{-\bf k}} + \mathcal{O}(g^2)~.
\ee
This is the last term in \eqref{flow}. At this order, the first two terms in the flow equation do not contribute, thus \eqref{flow} is satisfied.
Likewise, we can check the $g^2$ order result using both the contact $\psi_3^\Lambda$ and the single exchange $\psi_4^\Lambda$.
Now by requiring that the four-point function of long wavelength perturbations order $g^2$ is invariant with respect to the change of the cutoff scale, we find
\bea
\frac{d}{d\ln \Lambda} \langle \phi_l^4 \rangle =0 ~~ \Rightarrow ~~ \frac{d{\rm Re} \psi_4^{\Lambda}}{d\ln \Lambda} & =& -\frac{d }{d\ln \Lambda} \bigg(\ln\[\Omega_\Lambda(k_1)  \Omega_\Lambda(k_2)\Omega_\Lambda(k_3) \Omega_\Lambda(k_4)  \]\bigg) {\rm Re} \psi_4^{\Lambda} \nn\\
&&- \frac{d\Omega_{\Lambda}(s)}{d\ln \Lambda} \frac{{\rm Re}\psi_3^\Lambda~{\rm Re}\psi_3^\Lambda}{{\rm Re}\psi_2(s)} + {\rm perms}. ~
\eea
where we have used the result in \eqref{psi3Lambda} for $\psi_3^\Lambda$'s cutoff dependence.
 Thus at this order, we find
\be
\frac{dW_I^{(2)}}{d\ln \Lambda} =- \frac{1}{4} \int_{\bf k} \frac{1}{{\rm Re} \psi_2} \frac{d \Omega_\Lambda}{d\ln \Lambda} \[  \frac{\delta {W_I^{(1)}}}{\delta\phi_{\bf k}}\frac{\delta {W_I^{(1)}}}{\delta\phi_{-\bf k}}+2 \frac{\delta {W_0}}{\delta\phi_{\bf k}}\frac{\delta {W_I^{(2)}}}{\delta\phi_{-\bf k}} \]
 + \mathcal{O}(g^3)~,
\ee
which is the last two terms in \eqref{flow}, while the first term there comes at higher order in $g$.
We can continue with this computation to check that the higher-point tree-level wavefunction coefficients also satisfy this flow equation, and thus   the extra boundary term in \eqref{modi-polchin} is justified.

\vskip4pt
However, the flow equation does {\it not} hold for loop-level wavefunction coefficients. Let's take a look at $\Tilde{W}_I^{(2)}$ in \eqref{W1loop} for instance. Its $\Lambda$-dependence can be derived from the requirement that $\langle \phi_l^2 \rangle_{\rm 1-loop}$ is independent of the cutoff scale
\bea
\frac{d}{d\ln \Lambda} \langle \phi_l^2 \rangle =0 ~~ \Rightarrow ~~ \frac{d}{d\ln \Lambda} {\rm Re} \psi_2^{\rm 1-loop} &=& -2 \frac{d\ln\Omega_\Lambda}{d\ln \Lambda} {\rm Re} \psi_2^{\rm 1-loop} - \frac{1}{2} \int_{\bf p} \frac{d \Omega_\Lambda(p)}{d \ln \Lambda} \frac{1}{{\rm Re}\psi_2(p)}  \[ {\rm Re} \psi_4^\Lambda \right.\nn\\
&& \left. -
\Omega_\Lambda \frac{{\rm Re} \psi_3^\Lambda {\rm Re} \psi_3^\Lambda}{{\rm Re} \psi_2^\Lambda(|{\bf p}+{\bf k}_1|)} -{\rm perm.} \]
\eea
which differs from the result of the flow equation
 at the $g^2$ order
\be
\frac{d\Tilde{W}^{(2)}_I}{d\ln \Lambda} = - \frac{1}{4} \int_{\bf k} \frac{1}{{\rm Re} \psi_2} \frac{d \Omega_\Lambda}{d\ln \Lambda} \[ \frac{\delta^2 {W_I^{(2)}}}{\delta\phi_{\bf k}\delta\phi_{-\bf k}} +2 \frac{\delta {W_0}}{\delta\phi_{\bf k}}\frac{\delta {\Tilde{W}^{(2)}_I}}{\delta\phi_{-\bf k}} \]~,
%-2 \frac{d\ln\Omega_\Lambda}{d\ln \Lambda} {\rm Re} \psi_2^{\rm 1-loop} - \frac{1}{2} \int_{\bf p} \frac{d \Omega_\Lambda(p)}{d \ln \Lambda} \frac{1}{{\rm Re}\psi_2(p)}   {\rm Re} \psi_4^\Lambda 
\ee
Notice that the second term in \eqref{flow} does not contribute because it contains higher powers of $\phi$ at the $g^2$ order.
This suggests that the modified version of the Polchinski equation in \eqref{modi-polchin} works for the semi-classical wavefunction, but cannot capture the cutoff-dependence of loop-level $\psi_n$.
It remains interesting to see if there is a universal form of the boundary term that works for loop-level wavefunction coefficients.
In the end, we notice that the above computation for $g\Phi^3$ can be easily extended to other types of interactions, like $\lambda\Phi^4$.

\vskip4pt
Let's summarize the logic of justifying the extra boundary term in \eqref{modi-polchin}. 
Going back to the perturbation theory, we  require that correlators of long wavelength perturbations $\langle \phi_l^n \rangle$ should be independent of the cutoff scale, which allows us to fix the $\Lambda$-dependence of $\psi_n^\Lambda$ order by order.
Then by explicitly comparing with \eqref{flow}, we see that tree-level wavefunction coefficients satisfy the flow equation, while loop-level ones do not.
Thus in Polchinski's equation  \eqref{modi-polchin}, our choice of the extra boundary term reflects the cutoff-dependence of the semi-classical piece of the wavefunction. Accordingly we dub it  the  {\it semi-classical scheme}. 

\vskip4pt
Therefore, corrections to \eqref{modi-polchin} are expected to be higher order in $\hbar$ from quantum loops. As we have seen, for theories with IR divergences in dS, the quantum loop effects are always subdominant, and thus \eqref{modi-polchin} gives us the leading $\Lambda$-dependence of $W_I$. However, this may not be the case for IR-finite theories where the quantum wavefunction can become equally important at the loop order.
It remains unclear if there is a flow equation that works for loop-level wavefunction coefficients.
Meanwhile, although the current computation using the leading order perturbation theory is complete on its own, {the precise analogy to the Polchinski's equation in cosmology is not fully developed.} It would be more satisfying to find a systematic derivation that takes care of all the corrections and to understand better the physics of different schemes.  We leave these interesting open questions for future exploration.

\paragraph{Diffusion from RG flow}
Now we are ready to derive the diffusion term in the Fokker-Planck equation. 
Using the exact RG equation \eqref{modi-polchin}, we find the running of the probability distribution with the cutoff scale
\bea \label{diffusion1}
\frac{d}{d\ln \Lambda} P_\Lambda 
&=& \frac{d e^{W_0}}{d\ln \Lambda} e^{W_I} + \frac{d e^{W_I}}{d\ln \Lambda}    e^{W_0}   
%&=& - \frac{1}{4} \int_{\bf k} \frac{1}{{\rm Re} \psi_2} \frac{d \Omega_\Lambda}{d\ln \Lambda} \[  \frac{\delta^2e^{W_0} }{\delta\phi_{\bf k}\delta\phi_{-\bf k}}  e^{W_I} - \frac{\delta^2e^{W_I}}{\delta\phi_{\bf k}\delta\phi_{-\bf k}}    e^{W_0}  + 2 \frac{\delta}{\delta\phi_{\bf k}}  \( e^{W_0} \frac{\delta e^{W_I}}{\delta\phi_{-\bf k}} \) \]\nn\\
= - \frac{1}{4} \int_{\bf k} \frac{1}{{\rm Re} \psi_2} \frac{d \Omega_\Lambda}{d\ln \Lambda} \frac{\delta^2}{\delta\phi_{\bf k}\delta\phi_{-\bf k}} P_{\Lambda}~.
\eea
This already has the form of the diffusion equation, though it is in the Fourier space and the functional derivatives on the right hand side are for the Fourier mode $\phi_{\bf k}$.
To make the connection with the stochastic formalism more explicitly,
we need to use the long wavelength perturbation in coordinate space $\phi_l$ introduced in \eqref{philong}.
Then the Fourier transformation of the functional derivatives is given by
\be \label{longFourier}
\frac{\delta}{\delta\phi_{\bf k}} = \int d^3x e^{i{\bf k}\cdot{\bf x}} \Omega_\Lambda \frac{\delta}{\delta\phi_l(\bf x)}~.
\ee
By doing so, we rewrite \eqref{diffusion1} in the coordinate space
\be
\frac{d}{d\ln \Lambda} P_\Lambda = \int d^3x_1\int d^3x_2   \(- \frac{1}{4} \int_{\bf k}e^{i{\bf k}\cdot({\bf x}_1-{\bf x}_2)} \frac{1}{{\rm Re} \psi_2} \frac{d \Omega_\Lambda}{d\ln \Lambda} \) \frac{\delta^2}{\delta\phi_l({\bf x}_1)\delta\phi_l({\bf x}_2)} P_{\Lambda}~,
\label{diffusion2}
\ee
where $k$ is left in the bracket.
%Notice that the Fourier transform does not act $P_\Lambda$.  
By using $\mathrm{Re}\psi_2=-k^3/H^2$ and $ {d \Omega_\Lambda}/{d\ln \Lambda} \simeq \Lambda \delta(k-\Lambda) $, we find this momentum integration part becomes 
\begin{align}
- \frac{1}{4} \int_{\bf k}e^{i{\bf k}\cdot({\bf x}_1-{\bf x}_2)} \frac{1}{{\rm Re} \psi_2} \frac{d \Omega_\Lambda}{d\ln \Lambda}
%= \frac{1}{4}\int \frac{dk}{2\pi^2} \frac{\sin(\vert {\bf k}\cdot({\bf x}_1-{\bf x}_2)\vert)}{\vert{\bf k}\cdot({\bf x}_1-{\bf x}_2)\vert}k^2 \frac{H^2}{k^3}\frac{d \Omega_\Lambda}{d\ln \Lambda} 
%=\frac{H^3}{8\pi^2}\frac{\sin(\Lambda\vert{\bf x}_1-{\bf x}_2\vert)}{\Lambda\vert{\bf x}_1-{\bf x}_2\vert}
=\frac{H^2}{8\pi^2}j_0(\Lambda\vert{\bf x}_1-{\bf x}_2\vert)~,
\end{align}
with $j_0(x)=\sin(x)/x$.
Finally we restore the time-dependence of the cutoff scale via
$
d\Lambda = \epsilon aH^2 dt =\Lambda H dt 
$.
Then the second term in \eqref{FP} becomes
\be
 \dot\Lambda \frac{\partial}{\partial \Lambda} P_\Lambda[\phi,t]
 = H \frac{dP_\Lambda}{d\ln \Lambda} = 
\int d^3 x_1\int d^3 x_2\  \frac{H^3}{8\pi^2}j_0(\Lambda\vert{\bf x}_1-{\bf x}_2\vert)\frac{\delta^2}{\delta\phi_l({\bf x}_1)\delta\phi_l({\bf x}_2)}P_\Lambda~.
\ee
Notice that $j_0(x)$ has a maximum at the origin and decays for $x>1$. Thus only points with short separations $\Lambda\vert{\bf x}_1-{\bf x}_2\vert<1$ contribute to the spatial integrals on the right hand side. This concludes our derivation of the diffusion term from the boundary RG flow. 
Interestingly, with no input from the bulk, we find that the Polchinski equation for a boundary quantum field theory at a fixed time $t$ is enough to deduce the diffusion effects of long wavelength fluctuations. This derivation works for quantum field theories in de Sitter  beyond the perturbative regime, while the only nontrivial assumption is that the semi-classical piece dominates the wavefunction. 

\subsection{Fokker-Planck = Schr\"{o}dinger + Polchinski}

With the above derivation for the diffusion, now we look into the drift  which is supposed to be given by the first term in \eqref{FP}.  The explicit time derivative of the probability distribution is due to the bulk time flow of the wavefunction. Using the Schr\"{o}dinger equation, it is   straightforward to show that in the semi-classical regime we get the drift term in the Fokker-Planck equation  \cite{Gorbenko:2019rza}. Here we  recap the derivation.

\vskip4pt
We start with the Schr\"{o}dinger equation at the late-time boundary with cosmic time $t$
\begin{align}
i\frac{\partial}{\partial t}\Psi_\Lambda[\phi({\bf x}), t]= \mathcal{H} \Psi_\Lambda[\phi({\bf x}), t]~.
\end{align}
The form of the Hamiltonian $\mathcal{H}$ depends on the theory. In this section, we shall just focus on the simplest single field case with interactions in the potential $V(\Phi)$. The extension to the two-field scenario with derivative interactions is presented in Appendix \ref{app:2field}. More precisely, here we use
\begin{align}
\mathcal{H}=\int d^3 x\[ \frac{1}{2a(t)^3} \Pi_\phi^2 +\frac{a(t)}{2}(\partial_i\phi)^2+\frac{a(t)^3}{2}V(\phi) \]~,
\end{align}
where the  momentum operator is given as $\Pi_\phi({\bf x}) = -i\frac{\delta}{\delta\phi({\bf x})}$.
Notice that everything is evaluated at the boundary time $t$, and we have used $\Phi({\bf x},t)=\phi({\bf x})$.
Since the Hamiltonian is Hermitian, {i.e.} unitary, we can use the Schr\"{o}dinger equation to show that the probability distribution satisfies a continuity equation 
\begin{align} \label{continuity}
\frac{\partial P_\Lambda[\phi,t]}{\partial t}
&=-\frac{i}{2a^3}\int_{\bf k}\left[\frac{\delta}{\delta\phi_{-\bf k}}\left(\Psi_\Lambda^*\frac{\delta}{\delta\phi_{\bf k}}\Psi_\Lambda-\Psi_\Lambda\frac{\delta}{\delta\phi_{\bf k}}\Psi_\Lambda^*\right)\right]
%\\&=-\int d^3 x  \frac{\delta}{\delta\phi(\bf{x})}\left(\frac{1}{a^3}\Pi_\phi P_\Lambda[\phi,t]\right)\label{continuity_eq}
\end{align}
where we have performed the Fourier transformation on the right hand side.
Then to derive a closed equation for $P_\Lambda$, we notice that the functional derivative of the wavefunction can be explicitly computed using the saddle-point approximation \eqref{saddle}
\be
-i\frac{\delta}{\delta\phi_{\bf k}}\Psi_\Lambda \simeq  
\frac{\delta S_\Lambda[\Phi_{\rm cl}]}{\delta\phi_{\bf k}}\Psi_\Lambda
=\int d^3x e^{i{\bf k}\cdot {\bf x}}  \Omega_\Lambda(k)
\frac{\delta S_\Lambda[\Phi_{\rm cl}]}{\delta\phi_l({\bf x})}\Psi_\Lambda
= -\int d^3x e^{i{\bf k}\cdot {\bf x}}  \Omega_\Lambda(k) \Pi_{l} ({\bf x},t) \Psi_\Lambda
\ee
where we have used the Fourier transformation of functional derivatives in \eqref{longFourier} to introduce the derivative with respect to $\phi_l$ in coordinate space. 
This simply shows that we can express $\delta\Psi/\delta\phi_l$ using the corresponding field momentum at the boundary $\Pi_{l}  \equiv a^3 \dot\Phi_{l}$, and then \eqref{continuity} becomes an equation for $P_\Lambda$ only.
After performing the Fourier transformation for another functional derivative, \eqref{continuity} becomes
\begin{align}  
\frac{\partial P_\Lambda}{\partial t}
&= -\frac{1}{a^3}\int d^3x d^3x' \int_{\bf k}  e^{i{\bf k}\cdot ({\bf x}-{\bf x'})}  \Omega_\Lambda(k)  \frac{\delta}{\delta\phi_l({\bf x})} \Pi_{l} ({\bf x},t) P_\Lambda = -\int d^3x   \frac{\delta}{\delta\phi_l({\bf x})}\( \dot\Phi_l({\bf x},t) P_\Lambda\) ~,
\end{align}
which is the drift term. Up to this point, again we have only used the saddle-point approximation of the wavefunction. If we neglect the gradient terms, the field momentum can be further expressed as $\dot\Phi_l \simeq V'(\phi_l)/(3H)$ by using the approximated equation of motion for perturbations on superhorizon scales.
Then the drift term becomes the one that we normally see in the stochastic formalism.

\vskip4pt
Now we can join two pieces of derivation together and   describe the full time evolution of the probability distribution.   Then \eqref{FP} becomes the Fokker-Plank equation 
\begin{eBox}
\begin{align} \label{FP-general}
\frac{d P_\Lambda}{d t}=-\int d^3 x_1\frac{\delta}{\delta\phi(\mathbf{x}_1)}\left(\dot\Phi_l\Lambda P_\Lambda\right)+
\int d^3 x_1\int d^3 x_2\  \frac{H^3}{8\pi^2}j_0(\Lambda\vert{\bf x}_1-{\bf x}_2\vert)  \frac{\delta^2}{\delta\phi_l({\bf x}_1)\delta\phi_l({\bf x}_2)}P_\Lambda~.
\end{align}
\end{eBox}In summary, here the drift term corresponds to the bulk time flow given by the Schr\"odinger equation, while the diffusion comes from the boundary RG flow described by the Polchinski equation.
This aligns with the equations derived in previous works~\cite{Starobinsky:1994bd, Gorbenko:2019rza}, although we have arrived at this equation exclusively through the consideration of boundary terms. 
It's worth noting that in interacting theories characterized by shift symmetries, the momentum effectively becomes zero at the boundary, and the time evolution is exclusively dictated by the diffusion term. In such cases, higher-order terms in the action may become directly linked to higher-order derivatives in the  (RG) equation. 
If we are interested in the probability distribution of $\phi_l$ at one spatial location $P_\Lambda[\phi_l({\bf x})]$, we get $\delta$-functions from  functional derivatives: ${\delta\phi_l({\bf x})}/{\delta\phi_l({\bf x}_1)} = \delta^{(3)}({\bf x}_1-{\bf x})$, which remove the spatial integrals. In the end, the above equation simplifies to the usual Fokker-Planck equation
\begin{align}
\frac{d P_\Lambda}{d t}=\frac{\partial}{\partial\phi_l}\left(\frac{V'}{3H}P_\Lambda\right)+\frac{H^3}{8\pi^2}\frac{\partial^2P_\Lambda}{\partial \phi_l^2}
\end{align}
where we have used that $\lim_{x\to 0}j_0(x)=1$.

\vskip4pt
We close the non-perturbative analysis with some concluding remarks.
First of all, the semi-classical wavefunction plays an important role in the above derivation of the stochastic formalism. For the diffusion term, we used the tree-level computation in perturbation theory to justify a modified version of the Polchinski equation for the boundary field theory.
For the drift term, we simply applied the saddle-point approximation in the derivation. 
The final equation \eqref{FP-general} describes the evolution of the probability distribution that comes from the semi-classical wavefunction, and thus can be seen as the resummation of classical loops. 
This simplification works for theories with IR divergences where the quantum corrections are always subdominant. 
However, we may expect this semi-classical approximation breaks down for IR-finte theories.

\vskip4pt
As we mentioned in the beginning of this section, our derivation of the Fokker-Planck equation shares  similarities with the approach in \cite{Gorbenko:2019rza}. Now let us clarify the differences.
{The starting point of Ref.~\cite{Gorbenko:2019rza} is the probability distribution of the long modes defined in \eqref{long_mode_Prob:def}, while for us it is the coarse-grained version in \eqref{PLambda}. 
They are supposed to be equivalent, and do not lead to much difference in the derivation of the drift term. But for the diffusion, \eqref{PLambda} allows us to apply the techniques of the non-perturbative RG equation. 
This new attempt provides an interesting perspective for the effective description of the superhorizon physics and the  origin of the diffusion effects, though the precise analogy to Polchinski's equation is still missing.}
In addition, Ref.~\cite{Gorbenko:2019rza} presented a quite systematical derivation, with all the corrections carefully taken into account. Our approach is more based on the insight from perturbation theory that the semi-classical wavefunction and the Fokker-Planck equation are equivalent to each other. 
Therefore, we used the saddle-point approximation as our starting point and focus on showing the stochastic formalism as a consequence of that. It will be interesting to understand how corrections to the Fokker-Planck equation arise from first order corrections to the leading saddle-point.

\vskip4pt
Finally, we would like to briefly discuss possible connections with the holographic RG in AdS/CFT, {as a speculative comment}. There it has been shown that the RG flow of the boundary CFT is dual to the Hamilton-Jacobi equation in the bulk, and the  cutoff scale on the boundary is suspected to be related to the bulk radial coordinate \cite{deBoer:1999tgo, deBoer:2000cz, Skenderis:1999mm, Heemskerk:2010hk}. 
Our computation suggests that something different happened in dS. Here the ``radial" direction becomes time, and its evolution is governed by the Schr\"odinger equation, or equivalently the Hamilton-Jacobi equation in the semi-classical regime. However, as we have shown, the bulk time flow and the boundary RG flow lead to two distinct physical effects for the probability distribution: the former is responsible for the drift, and the later generates the diffusion.  This distinction may have nontrivial implications for the holographic description of de Sitter space, which may deserve future exploration. 

%=======================================
%=======================================
\section{Summary and Outlook}
%=======================================
\label{sec:concl}
Understanding the structure of the IR divergences in de Sitter  not only has conceptual interests in theoretical considerations  but also can lead to new phenomenological consequences of cosmic inflation. In many previous works, the problem was investigated through two approaches: the conventional cosmological perturbation theory, which demonstrate the IR-singular behaviour of tree-level and loop-level correlators, and the stochastic formalism that describes an equilibrium state in the non-perturbative regime. 
{Recently, the connection between the two is being established, and in particular it has been shown that the stochastic formalism is  a resummation of IR divergences from higher order loops \cite{Gorbenko:2019rza,Baumgart:2019clc}, though it remained less clear which part of the field-theoretical computation is being resummed. }

\vskip4pt
{The main focus of this paper is to find an answer to the above question. Built on previous studies, } we have built a  precise link between the semi-classical piece of the wavefunction and the stochastic formalism, by exploiting the wavefunction of the Universe and recent developments of the cosmological bootstrap.
Our analysis first made the explicit comparison between two approaches in the perturbative regime, and then presented a non-perturbative derivation of the Fokker-Planck equation with minimal assumptions.
More precisely, the main results are summarized as follows:
\begin{itemize}
    \item Within perturbation theory, we look into the loop-level cosmological correlators with IR divergences using the wavefunction method and the stochastic formalism. 
    \begin{itemize}
        \item Our starting point is to identify that in the perturbative computation of wavefunction coefficients, while tree-level processes can be produced in semi-classical approximations, loop diagrams there have a truly quantum origin. As both of them contribute to the cosmological correlators, in general the loop-level correlators are given by combinations of classical loops from tree-level wavefunction coefficients, and quantum ones that cannot be generated in classical theories. 
        \item The notion of classical loops  becomes especially important for IR-divergences in dS. We find loop-level correlators there are always dominated by the classical loop contributions.  This observation helps us to  significantly simplify the computation of leading IR-divergences in loop diagrams at any order. Furthermore, it indicates that the leading IR behaviour of the  system  is captured by the semi-classical wavefunction which contains all the tree-level information.
        \item Then we made the comparison with cosmological correlators from the stochastic formalism. By solving the Fokker-Planck equation perturbatively, we identify that at each order of coupling constants, the late-time divergent behaviour of correlators matches the results from classical loops. Quantum loop contributions are absent in the original stochastic computation, but can be incorporated by considering the next-to-leading order corrections to the Fokker-Planck equation. This simply shows that the stochastic formalism is resumming classical loops from the perturbative wavefunction.
    \end{itemize}
    
  \item Beyond perturbation theory, we explore the direct consequence of the semi-classical approximation. In particular, we look into the coarse-grained probability distribution and study its time evolution. We show that the Fokker-Planck equation can be understood as a combination of two distinct components: {\it i}) a drift term from the bulk time evolution that is governed by the Schr\"odinger equation of the semi-classical wavefunction; {\it ii}) a diffusion term that is associated with the boundary RG flow and given by a revised version of the Polchinski equation.
  As our analysis is based on the minimal assumption of the saddle-point approximation, this non-perturbative derivation further demonstrates the  relation between the semi-classical wavefunction and the stochastic formalism.

\end{itemize}

This work  also initiates many  future directions for explorations. Here we list several obvious  questions that would be interesting for further considerations.

\vskip4pt
First, it will be interesting to explore the phenomenological consequences of IR divergences in cosmological correlators. 
For theories of cosmic inflation, it has been shown that the perturbative treatment of IR divergences leads to local-type non-Gaussianities from multi-field models \cite{Wang:2022eop}.
Meanwhile, in past several years, there has been a significant interest in extending the analysis of inflationary fluctuations to the non-perturbative regime, which may generate novel behaviour at the tail of the probability distribution \cite{Panagopoulos:2019ail,Achucarro:2021pdh,Ezquiaga:2019ftu,Pattison:2021oen,Mishra:2023lhe}. 
In particular, Ref.~\cite{Celoria:2021vjw} showed that rare events on the tail can also be captured by the semi-classical wavefunction in IR-finite theories. 
Possible connections with our approach would be appealing, which might lead to a systematical new understanding about non-Gaussian tails.

\vskip4pt
Second, the current analysis has been mainly focused on  the semi-classical wavefunction and the resulting classical loops, which capture the leading IR divergences, while  quantum corrections may become important at the next-to-leading order, or in IR-finite theories. It will be interesting to compute these  corrections to the Fokker-Planck equation from our approach, and check if a resummation of quantum loops is possible. 
Another possible correction comes from gravitational interactions.
The current analysis has been performed in a fixed dS background, and ideally we also wish to consider a more realistic setup with  dynamical gravity.

\vskip4pt
Third, the renormalization group analysis for quantum field theories in de Sitter is another new topic that needs to be better understood, {and the analogy to Polchinski's equation is far from complete}. In this work, we have just focused on the semi-classical wavefunction and performed the  derivation for the Polchinski equation on the boundary, which can be seen as the RG flow of an Euclidean classical field theory. In our current derivation, there is an extra boundary term in addition to Polchinski's original equation, which we justify by returning to the leading order perturbation theory. It would be more satisfying to perform a  systematic analysis and understand better the physics of this particular choice.
Meanwhile, there could be other derivations of the RG flow from a bulk perspective, and it remains curious to see how the bulk version and the boundary one are related to each other. Last but not least, it is tempting to explore  possible connections with the holographic RG in AdS/CFT. Excitingly, new insights along this direction may shed light on the holographic description of  de Sitter spacetime.

\vskip32pt
%=======================================
\paragraph{Acknowledgements} 
%=======================================

We  acknowledge many inspiring discussions with Santiago Agui Salcedo, Carlos Duaso Pueyo, Scott Melville, Enrico Pajer, Guilherme Pimentel, Arttu Rajantie and Kostas Skenderis. SC is supported in  by the STFC Consolidated Grants ST/T000791/1 and ST/X000575/1. DGW is supported by a Rubicon Postdoctoral Fellowship awarded by the Netherlands Organisation for Scientific Research (NWO), and partially by the  VIDI grant with Project No. 680-47-535 from NWO and the STFC Consolidated Grants ST/T000694/1 and ST/X000664/1. ACD acknowledges partial support from STFC Consolidated Grant ST/T000694/1.

\newpage

\appendix

\section{In-In Computation of One-Loop Corrections}\label{app:inin}

In this section we apply the in-in formalism to compute the one-loop corrections to the IR-divergent two-point correlators. The results here confirm our computation in Section \ref{sec:pert} using the wavefunction method. As we can see, the classical and quantum contributions are mixed up in the in-in computation. 
Although one can derive the leading IR divergences at one-loop level, it is hard to identify their classical origin.

For the in-in computation of  massless scalar correlators,  we introduce the  
 bulk-to-boundary propagators    as
\be \label{Kpm}
K_+(k,\eta) =  \phi_{k}(\eta_0) \phi^*_{k}(\eta) , ~~~~~~~~
K_-(k,\eta) =  \phi^*_{k}(\eta_0) \phi_{k}(\eta) ,
\ee
and the
 bulk-to-bulk propagators as
\bea \label{b2b}
G_{++} (k,\eta, \eta') &=& \phi_{k}(\eta) \phi^*_{k}(\eta') \theta(\eta - \eta') +
\phi^*_{k}(\eta) \phi_{k}(\eta') \theta(\eta' - \eta) \nn\\ 
G_{+-} (k,\eta, \eta') &=&  
\phi^*_{k}(\eta) \phi_{k}(\eta')  \nn\\ 
G_{-+} (k,\eta, \eta') &=&  
\phi_{k}(\eta) \phi^*_{k}(\eta') \nn\\ 
G_{--} (k,\eta, \eta') &=& \phi_k(\eta)\phi_k^*(\eta')\theta(\eta'-\eta) +
\phi^*_{k}(\eta) \phi_{k}(\eta') \theta(\eta - \eta')~, 
\eea
where the mode function of massless scalars is given by
\bea \label{phik}
\phi_{k}(\eta) = \s_{k}(\eta)= \frac{H}{\sqrt{2k^3}}(1+i k\eta)e^{-ik\eta}~.
\eea
In the following we perform  the in-in computation for the three types of interactions that were considered in Section \ref{sec:pert}.

\paragraph{One-loop two-point function with one $\Phi^4$ vertex}
We start with the one-loop correction to the two-point correlation function $\langle \phi \phi \rangle$ from the quartic self-interaction $\lambda{\Phi^4}/{4!}$. This is a well-studied example of the in-in computation for IR divergent correlators. 
\bea
\langle \phi_{\bf k} \phi_{-\bf k} \rangle'_{\rm 1-loop} 
&=& - \frac{i\lambda}{2}\int d \eta  a(\eta)^4 \int_{\bf p} \Big[ 
  K_+(k,\eta)  K_+(k,\eta) G_{++} (p,\eta, \eta)   - c.c.
\Big] 
  \nn ~,
\eea
where we have a symmetry factor $2$ in the denominator of the overall coefficient.
Next, we put the momentum integrals outside and perform the time integrals first and then the IR-divergent part of the correlator is given by
\be
\langle \phi_{\bf k} \phi_{-\bf k} \rangle'_{\rm 1-loop}  
=   \frac{\lambda H^2}{12k^3} \int_{\bf p}  \frac{1}{p^3} \log(-2k\eta_0) =  \frac{H^2}{2k^3} \frac{\lambda}{12\pi^2}\log(\Lambda L)\log(-2k\eta_0) ~.
\ee

\paragraph{One-loop two-point function with two $\Phi^3$ vertices}
Next, similarly let's compute the one-loop correction to the  $\langle \phi \phi \rangle$ correlator from the cubic vertex $g  \Phi^3/3! $
\bea
\langle \phi_{\bf k} \phi_{-\bf k} \rangle'_{\rm 1-loop}  
&=& - \frac{g^2}{2}\int d \eta d\eta' a(\eta)^4 a(\eta')^4 \int_{\bf p,q} \Big[ 
  K_+(k,\eta)K_+(k,\eta') G_{++} (p,\eta, \eta') G_{++} (q,\eta, \eta')  \nn\\
&& ~~~~~~~~~~~ -  K_+(k,\eta) K_-(k,\eta') G_{+-} (p,\eta, \eta') G_{+-} (q,\eta, \eta')  \\
&& ~~~~~~~~~~~ -   K_-(k,\eta)  K_+(k,\eta') G_{-+} (p,\eta, \eta') G_{-+} (q,\eta, \eta') \nn\\
&& ~~~~~~~~~~~  +   K_-(k,\eta)  K_-(k,\eta') G_{--} (p,\eta, \eta') G_{--} (q,\eta, \eta') 
\Big] 
(2\pi)^3 \delta({\bf p+q-k}) \nn ~.
\eea
Again, we put the momentum integrals outside and perform the time integrals first
\be
\langle \phi_{\bf k} \phi_{-\bf k} \rangle'_{\rm 1-loop}  
= - \frac{g^2}{2} \int_{\bf p,q} (2\pi)^3 \delta({\bf p+q-k}) \mathcal{I}~.
\ee
As we are mainly interested in the IR-divergent part,  the secular growth piece from bulk time integration is given by
\be
\lim_{\eta_0\rightarrow 0}   \mathcal{I} = - \frac{1}{36 p^3q^3 }  \Big(\log\(-(k+p+q)\eta_0 \)\Big)^2 + ... 
\ee
Performing the loop integration, we find the IR-divergent correction to the two-point function 
\be
\langle \phi_{\bf k} \phi_{-\bf k} \rangle'_{\rm 1-loop} 
= \frac{g^2}{2} \int_{\bf p} \frac{H^2}{36 p^3 k^3 }  \Big(\log\(-(2k+p)\eta_0 \)\Big)^2 
= \frac{H^2}{2k^3} \frac{g^2}{72\pi^2H^2} \log(kL)\log(-2k\eta_0)^2 ~.
\ee

\paragraph{One-loop two-point function with two $\dot\Phi\Sigma^2$ vertices}
Now we compute the one-loop correction to   $\langle \phi \phi \rangle$ from the cubic vertex $\alpha \dot\Phi \Sigma^2/2 $
\bea
\langle \phi_{\bf k} \phi_{-\bf k} \rangle'_{\rm 1-loop} 
&=& - \frac{\alpha^2}{2}\int d \eta d\eta' a(\eta)^3 a(\eta')^3 \int_{\bf p,q} \Big[ 
\partial_\eta K_+(k,\eta) \partial_{\eta'}K_+(k,\eta') G_{++} (p,\eta, \eta') G_{++} (q,\eta, \eta')  \nn\\
&& ~~~~~~~~~~~ - \partial_\eta K_+(k,\eta) \partial_{\eta'}K_-(k,\eta') G_{+-} (p,\eta, \eta') G_{+-} (q,\eta, \eta')  \\
&& ~~~~~~~~~~~ - \partial_\eta K_-(k,\eta) \partial_{\eta'}K_+(k,\eta') G_{-+} (p,\eta, \eta') G_{-+} (q,\eta, \eta') \nn\\
&& ~~~~~~~~~~~  + \partial_\eta K_-(k,\eta) \partial_{\eta'}K_-(k,\eta') G_{--} (p,\eta, \eta') G_{--} (q,\eta, \eta') 
\Big] 
(2\pi)^3 \delta({\bf p+q-k}) \nn ~.
\eea
Next, we put the momentum integrals outside and perform the time integrals first
\be
\langle \phi_{\bf k} \phi_{-\bf k} \rangle'_{\rm 1-loop}  
= - \frac{\alpha^2}{2} \int_{\bf p,q} (2\pi)^3 \delta({\bf p+q-k}) I~,
 %\big[ I_{\rm TO} - I_{\rm NTO} \big]~,
\ee
with
\be
\lim_{\eta_0\rightarrow 0}  I
%\big[ I_{\rm TO} - I_{\rm NTO} \big] 
= - \frac{H^2}{4 p^3q^3 }  \Big(\log\(-(k+p+q)\eta_0 \)\Big)^2  ... ~.
\ee
Then at one-loop level, the IR-divergent correction to the two-point correlator becomes
\be
\langle \phi_{\bf k} \phi_{-\bf k} \rangle'_{\rm 1-loop}  
= \frac{g^2}{2} \int_{\bf p} \frac{H^2}{4 p^3k^3 }  \Big(\log\(-2k\eta_0 \)\Big)^2 = \frac{H^2}{2k^3} \frac{\alpha^2}{8\pi^2} \log(kL)\log(-2k\eta_0)^2 ~.
\ee

While we can still perform the one-loop computation using  in-in formalism here, it becomes more complicated once we go beyond the lowest order for loop corrections. 
Meanwhile, the IR-singular piece of the correlators in the end look quite simple, with only two types of logarithmic functions: one from the time integration; the other from the momentum integration. This indicates that some simplification may be achieved if one is mainly interested in the leading IR-divergent contributions. 
As we discussed in Section \ref{sec:multiloop}, the wavefunction method provides a more intuitive understanding that identifies the origin of dominant IR divergences as classical loops. Meanwhile, by looking at one particular contribution from the product of contact wavefunction coefficients, we can easily derive the leading logarithmic singularities.

\section{Probability Distribution from the Wavefunction}
\label{app:wave}

In this appendix, we compute the probability distribution of long wavelength fluctuations from the coarse-grained wavefunction in perturbation theory.
We show that contributions from the saddle-point approximation satisfy the Fokker-Planck equation, while the ones from loop-level wavefunction coefficients are subdominant.
This provides another justification for our major conclusion that the stochastic formalism is resuming the classical loops and thus is equivalently described by the semi-classical wavefunction.

This method,  with a particular choice of $\Omega_k$ is similar to the path integral used in \cite{Gorbenko:2019rza} to derive the the Fokker-Planck.  Our approach will be different: instead of finding an equation for $P[\phi_l]$ we will do the path integral show how the long wavelength physics appears and which diagrams are considered.   To be able to  do  this we will rely on knowing the perturbative wavefunction $\Psi$,  hence our results will be valid only while perturbation theory is valid.  Still these computations will be instructive when comparing with the solutions of the Fokker-Planck equation.

We start by defining the probability for the long wavelength part of a scalar fields as  a functional integral
\begin{align}
P[\phi_l(\mathbf{x})]=\int \mathcal D\phi\ \delta\left(\phi_l(\mathbf{x})-\int_{\bf k}\Omega_ke^{i\mathbf{k}\cdot\mathbf{x}}\phi_{\bf k}\right)\vert\Psi[\phi_{\bf k}]\vert^2~.
\label{long_mode_Prob:def}
\end{align}
The idea is to use the already found wavefunction $\Psi[\phi_{\bf k}]$ and then compute the path integral at leading order in a given coupling. The path integral can be thought of as two consecutive Legendre transform. The first is analogous to plugging a current with support only for the long modes. In the end it computes a partition function composed only of connected correlators of the long modes of the fields. The second Legendre transform rewrites the partition function as a function  of the long mode fields $\phi_l$ similarly as the wavefunction.

\paragraph{Free Theory} To be more concrete let us start our discussion with an example.  If the field is free then we can write the probability as
\begin{align}
P[\phi_k]=\mathcal{N}\sqrt{\mathrm{Re\psi_2}({\bf k})}\exp\left(-\frac{1}{2}\int_{\bf k}2\mathrm{Re}\psi_2({\bf k}) \phi_{\bf k}^2\right),
\end{align}
where $\mathcal{N}$ is a real number such that the probability is normalized.   We want to now compute the probability of finding the long wavelength part of the field $\phi_l$.   As discussed previously this  can be done through the path integral,
\begin{align}
P[\phi_l({\bf x})]&=\mathcal{N}\sqrt{\mathrm{Re\psi_2(\mathbf{k})}}\int D\phi_{\bf k}\ \delta\left(\phi_l(\mathbf{x})-\int_{\bf k}e^{i\mathbf{k}\cdot\mathbf{x}}\Omega(k)\phi_{\bf k}\right)\exp\left(-\frac{1}{2}\int_{\bf k}2\mathrm{Re}\psi_2({\bf k})\phi_{\bf k}^2\right).\nonumber\\
\end{align}
Let us consider the case when we pick a point ${\bf x}$, by momentum invariance the field $\phi_l$ has to have the same value on all space so it can depends only on time $\phi_l(\mathbf{x},t)=\phi_l(t)$.  Secondly,  we can see that the role of the $\delta$ function is to integrate over all modes such that the average of the field $\phi_{\bf k}$ corresponds to a fixed value $\phi_l$.  The path integral can be done analytically by writing the $\delta$ function as an integral over a current $J$
\begin{align}
P(\phi_l)&=\mathcal{N}\sqrt{\mathrm{Re\psi_2}({\bf k})}\int dJ\int D\phi_{\bf k} \exp\left(+i J\phi_l-\int_{\bf k}\left(\mathrm{Re}\psi_2({\bf k})\phi_{\bf k}^2-i\Omega(k)J\phi_{\bf k} \right)\right)\nonumber\\
&=\int dJ \exp\left(i J \phi_l-\frac{1}{2}\int_{\bf k}\frac{\Omega_k}{2\mathrm{Re \psi_2}({\bf k})}J^2\right)\nonumber\\
&=\sqrt{\frac{2\pi}{\int _{\bf k}\frac{\Omega_k}{2\mathrm{Re \psi_2}({\bf k})}}}\exp\left(-\frac{1}{2}\frac{\phi_l^2}{\int _{\bf k}\frac{\Omega_k}{2\mathrm{Re \psi_2}({\bf k})}}\right).
\end{align}
Now using that $1/2\mathrm{Re}\psi_2({\bf k})=\langle\phi_{\bf k}^2\rangle'$ we notice that the integral over the two point function is nothing more than the variance of the field
\begin{align}
\sigma_\phi^2\equiv\int_{\bf k}\frac{\Omega_k}{2\mathrm{Re \psi_2}({\bf k})}=\int_{\bf k}\Theta(k-a H)\langle\phi_k^2\rangle=\frac{H^2}{4\pi^2}\log a,
\end{align}
and so  we can write that
\begin{align}
P(\phi_l)=\sqrt{\frac{2\pi}{\sigma_\phi^2}}\exp\left(-\frac{\phi_l^2}{2\sigma_\phi^2}\right),
\end{align}
which is the usual result obtained by solving the Fokker-Planck equation for a free field.   It is possible to extend this results to higher order by using the Born approximation.  

For interacting theories, before doing this in detail let us understand what is the role of the path integral.  Given an action $S[\phi]$ for a interacting scalar field $\phi$, let us call $S_\mathrm{cl}[\phi]$ the semi-classical action,  on-shell $S_{\mathrm{cl}}[\phi]$  is a polynomial in $\phi_{\bf k}$ that contains all possible tree-level interactions.  Then we can write the generator of connected correlators as \cite{Weinberg:2005vy}
 \begin{align}
 Z[J]=e^{W[J]}=\int D\phi e^{-i \int_{{\bf k}}\phi J \Omega_k-2i\mathrm{Im} S_\mathrm{cl}[\phi_k]}.
 \end{align}
With this preparation, now let us take a look at some examples.

\paragraph{$\lambda\Phi^4/4!$ \bf interaction}
Let us generalise the free theory to the case with a quartic interaction.  Up to first order in $\lambda$ there are only two terms in the tree-level  wavefunction
\begin{align}
\Psi(\phi)\sim\exp\left[\frac{1}{2}\int_{\bf k}\psi_2({\bf k}) \phi_{\bf k}^2+\frac{1}{4!}\int_{{\bf k}_1,{\bf k}_2,{\bf k}_3,{\bf k}_4}\psi_4({\bf k}_1,{\bf k}_2,{\mathbf{k}}_3,{\mathbf{k}}_4)\  \phi_{{\bf k}_1}\phi_{{\bf k}_2}\phi_{{\bf k}_3}\phi_{{\bf k}_4} \right],
\end{align}
where $\psi_2$ and $\psi_4$ are derived in Section \ref{sec:pert}.  The generating functional for the connected correlators of long modes is
\begin{small}
\begin{align}
e^{W[J]_l}=&\frac{1}{Z[0]}\int D\phi \exp\left[-i \int_{\bf k}\Omega_k \phi_{\bf k} J-\frac{1}{2}\int_{\bf k}2\mathrm{Re}\psi_2({\bf k})\phi_{\bf k}^2-\frac{1}{12}\int_{\mathbf{k}_1\dots\mathbf{k}_4}2\mathrm{Re}\psi_4(\mathbf{k}_1,\dots,\mathbf{k}_4)\phi_{\mathbf{k}_1}\dots\phi_{\mathbf{k}_4}\right].
\end{align}
\end{small}We can remove the linear term in $\phi_k$ by shifting the field  $\phi_k\to\phi_k-i\frac{\Omega_k}{2\mathrm{Re\psi_2}}J$.  Doing this the measure does not change and we get,
\begin{align}
e^{W[J]_l}&=\frac{1}{Z[0]}\exp\left(-\frac{\sigma_\phi^2}{2}J^2-\frac{1}{16}\int_{\mathbf{k}_1\dots\mathbf{k}_4}\prod_{i=1}^4\frac{\Omega(k_i)}{\mathrm{\Re} \psi_2({\bf k}_i)}\mathrm{Re}\psi_4(\mathbf{k}_1,\dots,\mathbf{k}_4)J^4\right)\nonumber\\
&\times\int D\phi_{\bf k}\exp\left(-\frac{1}{2}\int_{\bf k}2\mathrm{Re}\psi_2({\bf k})\phi_{\bf k}^2-\frac{1}{12}\int_{\mathbf{k}_1\dots\mathbf{k}_4}2\mathrm{Re}\psi_4(\mathbf{k}_1,\dots,\mathbf{k}_4)\varphi_{\mathbf{k}_1}\dots\varphi_{\mathbf{k}_4}\right.\nonumber\\
&\left.\hspace{2.2cm}+\frac{1}{4}\int_{\mathbf{k}_1\dots\mathbf{k}_2}\mathrm{Re}\psi_4(\mathbf{k}_1,\dots,\mathbf{k}_4)\frac{\Omega(k_3)\Omega(k_4)}{\mathrm{Re}\psi_2({\bf k}_3)\mathrm{Re}\psi_2({\bf k}_4)}\phi_{\mathbf{k}_1}\phi_{\mathbf{k}_2}J^2+5\ \mathrm{perm.}\right).
\end{align}
Notice that the second line corresponds to $Z[0]$  so the path integral in the second lines computes the correlation function of the term in the exponential. Since the integral is still Gaussian in the fields it can be computed at first order in $\lambda$ yielding,
\begin{small}
\begin{align}
\left\langle e^{\frac{1}{4}\int_{\mathbf{k}_1\dots\mathbf{k}_4}\frac{\Omega(k_3)\Omega(k_4)\mathrm{Re}\psi_4(\mathbf{k}_1,\dots\mathbf{k}_4)}{\mathrm{Re}\psi_2({\bf k}_3)\mathrm{Re}\psi_2({\bf k}_4)}\phi_{\mathbf{k}_1}\phi_{\mathbf{k}_2}J^2+5\ \mathrm{perm.}}\right\rangle=\left(1-\frac{J^2}{8}\int_{\mathbf{k},\mathbf{p},\mathbf{q}}\frac{\mathrm{Re}\psi_2(\mathbf{k})\mathrm{Re}\psi_4(\mathbf{k},\mathbf{p},\mathbf{q},-\mathbf{k})\Omega(p)\Omega(q)}{\mathrm{Re}\psi_2(\mathbf{p})\mathrm{Re}\psi_2(\mathbf{q})}\right)^{-1/2}.
\end{align}
\end{small}
It is easy to see that is related to the loop correction we computed on the first part. Indeed it appears as a loop correction to the determinant, which is expected since we are truncating at first order in $\lambda$.

Using the notation from Eq.~\eqref{smeared_correlators} we can write that
\begin{align}
\int_{\mathbf{k}_1\dots\mathbf{k}_4}\prod_{i=1}^4\frac{\Omega(k_i)}{\mathrm{\Re} \psi_2(\mathbf{k}_i)}\mathrm{Re}\psi_4({\bf k}_1,{\bf k}_2,{\bf k}_3,{\bf k}_4)=\langle\phi_l^4\rangle~.
\end{align}
We then have that,
\begin{align}
P[\phi_l]=\int dJ \frac{\exp\left(i J\phi_l+\frac{\langle\phi_l^2\rangle_0}{2}J^2-\frac{\langle\phi_l^4	\rangle_0}{16}J^4\right)}{\sqrt{1+J^2\langle\phi^2\rangle_{1-\mathrm{loop}}}}.
\end{align}
and so we have checked explicitly that $\exp(W[J]_l)$ is the generator of connected correlator at long wavelengths. We can compute correlation functions by taking functional derivatives of  $\exp(W[J]_l)$ and setting the current to $0$ at the end of the computation. 
For the two-point function we get
\begin{align}
\langle\phi^2\rangle=\left.\frac{\delta^2 W[J]}{\delta J^2}\right\vert_{J=0}=\langle\phi_l^2\rangle_0+\langle\phi_l^2\rangle_{\mathrm{1-loop}}.
\end{align}

\paragraph{Two-field model}
The two-field interaction $\alpha \dot\Phi\Sigma^2/2$ can lead to further simplifications in the computation of the probability distribution from the perturbative wavefunction.  Explicitly, we have
\begin{small}
\begin{align}
P[\phi_l,\chi_l, t]&=\frac{1}{Z[0]}\int D\phi D\chi \ \delta\left(\phi_l({\bf x})-\int_{{\bf k}}\Omega_{\Lambda}e^{i{\bf k}\cdot{\bf x}}\phi_{\bf k}\right)\delta\left(\chi_l({\bf x})-\int_{\bf k}\Omega_{\Lambda}e^{i{\bf k}\cdot{\bf x}}\chi_{\bf k}\right)P[\phi, \chi, t]\nonumber\\
&=\frac{1}{Z[0]}\int dJ_\phi dJ_\chi\int D\phi_k D\chi_k \exp\left(i J_\phi \phi_l+iJ_\chi \chi_l\right)\nonumber\\
&\times\exp\left[-\int_{\bf k}\left(\mathrm{Re}\psi_2^{\phi}({\bf k})\phi_{\bf k}^2-ie^{i{\bf k}\cdot{\bf x}}\Omega(k)J_\phi\phi_{\bf k}\right)-\int_{\bf k}\left(\mathrm{Re}\psi_2^{\chi}({\bf k})\chi_{\bf k}^2-ie^{i{\bf k}\cdot{\bf x}}\Omega(k)J_\chi\chi_k\right)\right.\nonumber\\
&\qquad\qquad\left.-\frac{1}{2}\int_{{\bf k}_1,{\bf k}_2,{\bf k}_3}\mathrm{Re}\psi_3({\bf k}_1,{\bf k}_2,{\bf k}_3)\phi_{{\bf k}_1}\chi_{{\bf k}_2}\chi_{{\bf k}_3}\right]\nonumber\\
&=\frac{1}{Z[0]\sqrt{\mathrm{Re}\psi_2^\phi({\bf k})}}\int dJ_\phi dJ_\chi\int D\phi_k  \exp\left[+i J_\phi \phi_l+iJ_\chi \chi_l-\frac{\langle\phi^2\rangle}{2}J_\phi^2+i\int_{{\bf k},{\bf s}}\frac{\mathrm{Re}\psi_3({\bf k},-{\bf k},{\bf s})}{2\mathrm{Re}\psi_2^\phi({\bf{s}})}\chi_{{\bf k}}^2 J_\phi\right.\nonumber\\
&\left.\qquad+i\int_{\bf k} \mathrm{Re}\psi_2^{\chi}({\bf k}) \chi_{\bf k}^2+\frac{1}{4}\int_{{\bf k_1}\dots{\bf k}_4}\frac{\mathrm{Re}\psi_3({\bf k}_1,{\bf k}_2,{\bf s})\mathrm{Re}\psi_2(\mathbf{k}_3,\mathbf{k}_4,\mathbf{s})}{\mathrm{Re}\psi_2(\mathbf{s})} \chi_{{\bf k}_1}\chi_{{\bf k}_2}\chi_{{\bf k}_3}\chi_{\bf{k}_4}+\mathrm{perms}.\right]\nonumber\\
&=\int dJ_\phi dJ_\chi \exp\left[-\frac{\langle\phi^2\rangle}{2}J_\phi^2+\frac{\langle\chi^2\rangle}{1-i\frac{ \langle\phi\chi^2\rangle}{\langle\phi^2\rangle} J\phi}\frac{J_\chi^2}{2}+\mathcal{O}(J_\chi^4)\right]~,
\end{align}\end{small}.To get the last line we have neglected the quartic term in $\chi$. 
We also know the long wavelength limit of the three-point function
\begin{align}
\langle\phi\chi^2\rangle=-\int_{{\bf k}_1,{\bf k}_2,{\bf k}_3}\frac{\mathrm{Re}\psi_3(\mathbf{k}_1,\mathbf{k}_2,\mathbf{k}_3)\Omega(k_1)\Omega(k_2)\Omega(k_3)}{\mathrm{Re}\psi_2^\phi(\mathbf{k}_1)\mathrm{Re}\mathbf{k}_2)\mathrm{Re}\psi_2^\chi(\mathbf{k}_3)}+\mathrm{perm}~.
\end{align}
Notice that $W[J_\phi,J_\chi]$ is non-local in $J_\phi$.  This effect comes from integrating out $J_\chi$ and it will be suppressed when $\chi$ has a large mass.  Expanding in small $\alpha$ there will be a series of terms quadratic in $J_\chi$ accounting for all possible exchange diagrams with two external $\chi$ legs connecting internal $\chi$ propagators with external $\phi$ legs.  For the purpose of computing the probability we can  expand in $\alpha$  and solve the integrals over $J_\chi$ and $J_\phi$.  We get
\begin{align}
P[\phi_l,\chi_l]&=\int dJ_\phi dJ_\chi \exp\left[i J_\phi \phi_l+iJ_\chi \chi_l-\frac{\langle\phi^2\rangle}{2}J_\phi^2+\frac{\langle\chi^2\rangle}{2}J_\chi^2-i\frac{ \langle\phi\chi^2\rangle }{2} J_\phi J_\chi^2\right]\nonumber\\
&=\int dJ_\phi  \frac{ \exp\left[i J_\phi \phi_l-\frac{\langle\phi^2\rangle}{2}J_\phi^2-\frac{1}{1+i\frac{\langle\phi\chi^2\rangle}{\langle\chi^2\rangle} J_\phi}\frac{\chi_l^2}{2\langle\chi^2\rangle}\right]}{\sqrt{\langle\chi^2\rangle+i\frac{\langle\phi\chi^2\rangle}{\langle\phi^2\rangle}J_\phi}}\nonumber\\
&\approx\sqrt{\frac{2\pi}{\langle\chi^2\rangle\langle\phi^2\rangle}}\exp\left[-\frac{\chi_l^2}{2\langle\chi^2\rangle}-\frac{1}{2\langle\phi^2\rangle}\left(\phi_l-\frac{\langle\phi\chi^2\rangle}{2\langle\chi^2\rangle^2}\chi_l^2\right)^2\right]~,
\end{align}
where, in the second line, we have expanded in $\alpha$.  This is the probability distribution which was obtained  in \cite{Achucarro:2021pdh} as a solution of the Fokker-Planck equation. The two-point functions $\langle \phi^2 \rangle$ and $\langle \chi^2 \rangle$ are given by the variance in \eqref{variancet}.
For the three-point function, we have
\begin{align}
\langle\phi\chi^2\rangle&=-\lambda H^3\int_{{\bf k}_1,{\bf k}_2,{\bf k}_3}\frac{\log(-(k_1+k_2+k_3)\eta_0)}{k_1^3k_2^3k_3^3}\Omega(k_1)\Omega(k_2)\Omega(k_3)(2\pi)^3\delta^{(3)}({\bf k}_1+{\bf k}_2+{\bf k}_3)\nonumber\\
&=-\lambda H^3\int_{{\bf k}_1,{\bf k}_2} \frac{\log(-2(k_1+k_2)\eta_0)}{k_1^3k_2^3(k_1+k_2)^3}\Omega(k_1)\Omega(k_2)~.\nonumber\\
&=-\frac{\lambda H^3}{2(4\pi^2)^2}\log(a)^3~,
\end{align}
where we have used the delta function in the first line to eliminate the $k_3$ integral.  We can compute correlation functions by taking derivatives from $W_l[J_\phi,J_\chi]$ and setting the currents to zero.  Let us check the 1-loop corrections.  First,  integrating over $J_\chi$, we obtain
\begin{align}
W_l[J_\phi]=-\frac{\langle\phi^2\rangle}{2}J_\phi^2-\frac{1}{2}\log\left(\frac{1}{\langle\chi^2\rangle}+\frac{i\langle\phi\chi^2\rangle}{\langle\chi^2\rangle}J_\phi\right)~.
\end{align}
Then the  1-loop correction to the two-point function is given by
\begin{align}
\langle\phi^2\rangle_{1-\mathrm{loop}}&=\langle\phi^2\rangle-\frac{\langle\phi^2\chi\rangle^2}{2\langle\chi^2\rangle^2}=\frac{H^2}{4\pi^2}\log a+\frac{\lambda^2H^2}{128\pi^4}\log(a)^4~,
\end{align}
which matches the perturbative computation we did in Section \ref{sec:pert}.

\section{Stochastic Formalism with Two Fields}
\label{app:2field}

In this Appendix, we present another version of the derivation for the Fokker-Planck equation, especially focusing on the situation with two interacting massless scalars. 

To proceed let us first study the long wavelength dynamics.  One of the advantages of using wavefunctions is that it allow us to define a probability through the relation $P[\phi,\chi;t]\sim \vert\Psi\vert^2$.  The dynamics of this probability distribution is given by the Schrodinger equation,
\begin{align}
\frac{d}{dt}\Psi=iH\Psi~,
\end{align}
which is strictly valid in the decoupling limit of the EFT (otherwise the Hamiltonian is a constraint).  Of course the equation for the probability distribution is just the usual continuity equation from quantum mechanics.  A difference between our derivation and the analysis of Ref.~\cite{Gorbenko:2019rza} is that we will include derivative interactions.  Their main effect will on modifying the classical dynamics.  For this reason let us keep discussing the two field scenario and let us consider the following interaction Lagrangian,
\begin{align}
S_{\mathrm{int}}=\int d^4 x\sqrt{-g}\left(V(\chi)+\dot\phi f(\chi)\right)~,
\end{align} where $V(\phi)$ and $f(\chi)$ are polynomial in $\chi$.  The effect of the derivative interactions is twofold.  Besides the usual interaction it modifies the momenta which is now $a^{-3}\pi_\phi=\dot\phi-\frac{1}{2}f(\chi)$ which in turns implies an extra term in the interaction Hamiltonian $f(\chi)^2/8$.  
The Hamiltonian is given by,
\begin{align}
H=\frac{\pi_\phi}{2a^3}+\frac{\pi_{\chi}}{2a^3}+\frac{a^2}{2}((\nabla\chi^2)-m^2a^2\chi^2)+\frac{a^2}{2}(\nabla\phi)^2-\frac{1}{2}\pi_\phi f(\chi)+\frac{a^3}{8}(f(\chi)^2-4V(\chi))~.
\end{align}
Notice that by Noether theorem the shift symmetry of $\phi$ implies a conserved current in the continuity equation, 
\begin{align}
 j^0=\dot\phi-\frac{1}{2}f(\chi)=\frac{\pi_\phi }{a^3}~,
 \label{conserved:current}
\end{align}
which is just a generalisation of the usual momentum conservation in models with a single field.  All this equation is saying is that in the long wavelength limit $\dot\phi$ is not conserved,  as in single field inflation, but the more general combination given by  $j_0$. 

 Assuming unitarity we can write a continuity equation for the probability density by taking the time derivative of $P$ and using the  Schrodinger equation.
Using this we can write the continuity equation as,
\begin{align}
\frac{\partial P[\phi,\chi,t]}{\partial t}&=[H,P]\nonumber\\
&=-\int d^3x \left[\frac{i}{2a^3}\frac{\delta}{\delta\phi({\bf x})}\left(\Psi^*\frac{\delta}{\delta\phi({\bf x})}\Psi-\Psi\frac{\delta}{\delta\phi({\bf x})}\Psi^*\right)+\frac{i}{2a^3}\frac{\delta}{\delta\chi({\bf x})}\left(\Psi^*\frac{\delta}{\delta\chi({\bf x})}\Psi-\Psi\frac{\delta}{\delta\phi({\bf x})}\Psi^*\right)\right.\nonumber\\
&\hspace{2cm}\left.-f(\chi)\frac{\delta}{\delta\chi({\bf x})}\Psi\Psi_*\right]\nonumber\\
&=-\int d^3 x  \frac{\delta}{\delta\phi({\bf x})}\left(\left(\Pi_\phi +f(\chi)\right)P[\phi,\chi,t]\right)+ \frac{\delta}{\delta\chi({\bf x})}(\Pi_\chi P[\phi,\chi, t])~,\label{continuity_eq}
\end{align}
where we have used that $\pi_\phi\to-i\frac{\delta}{\delta\phi}$.  If we apply this definition over the wavefunction coefficients we can define a classical momentum as,
\begin{align}
\Pi_{\phi^i}[\phi,{\bf x},t]=- \frac{\delta}{\delta\phi^i({\bf x})}\mathrm{Im} \log \psi=\frac{\delta S(\zeta,t)}{\delta\phi^i({\bf x})}~,
\label{momenta_def}
\end{align} 
where by $S$ we mean that it is evaluated over the classical part of the wavefunction.  In this sense this definition can be thought of as the on-shell long wavelength limit of the momentum.
By using this we can write the continuity equation as,
\begin{align}
\frac{\partial P[\phi,\chi,t]}{\partial t}&=-\int d^3 x  \frac{\delta}{\delta\phi({\bf x})}\left(\left(\frac{\Pi_\phi}{a^3} +f(\chi)\right)P[\phi,\chi,t]\right)+ \frac{\delta}{\delta\chi({\bf x})}\left(\frac{\Pi_\chi}{a^3} P[\phi,\chi, t]\right)~.
\end{align} 
The advantage  of writing the equation in this form  relies on the fact that we can make use of the long wavelength limit of the equations of motion.  In our case,  the shift symmetry of $\phi$ implies that the classical momentum vanishes at late times.
For the second field we can use the equations of motion to get that  $\Pi_\chi\to -(V'(\chi)-f'(\chi)/8)/3H$. 
\\
 In order to incorporate the effect of short wavelengths we first coarse grain the fields into long and short wavelength modes.  This is achieved by writing,
\begin{align}
\phi({\bf x})=\int_{0}^{\Lambda}\frac{d^3k}{(2\pi)^3}e^{i{\bf k}\cdot{\bf x}}\phi({{\bf x}})+\int_{\Lambda(t)}^{\infty} \frac{d^3k}{(2\pi)^3}e^{i{\bf k}\cdot{\bf x}}\phi({{\bf x}})\equiv \phi_L({\bf k})+\phi_S({\bf k})~,
\end{align}
where $\Lambda(t)$ is a time dependent  cut-off  given by  $\Lambda=\epsilon a(t)H$ with $\epsilon$ a small number (not to be confused with the slow roll parameter).  In order to circumvent problems with the fact that the cut-off is sharp let us define the smoothed window function,
\begin{align}
\Omega_{\Lambda}(k)=\begin{cases}
1\qquad \mathrm{for} \ k\leq \Lambda(t)\\
0\qquad\mathrm{for} \ k\geq (1+\delta)\Lambda(t)~.
\end{cases}
\label{cutt-off}
\end{align}
such that in the limit $\delta\to 0$, the window function becomes a Heavyside step-function,  $\Omega_{\Lambda}=\Theta(k-\Lambda(t))$.   An example of such function is given in~\cite{Polchinski:1983gv}.
With the help of the window function~\eqref{cutt-off} we can define the smeared field as,
\begin{align}
\phi_{\Omega}({\bf x})=\int_\mathbf{k}\Omega_{\Lambda}(k)e^{i{\bf k}\cdot{\bf x}}\phi({\bf k})~,
\end{align}
and similarly for $\chi$.
Notice that  at leading order in $\delta$,  $\phi_{\Omega}$ is just the long wavelength part of the curvature modes.  In other words, expanding in $\delta$ we find that,  $\phi_{\Omega}=\zeta_l(1+\mathcal{O}(\delta))$.  At the same time  we can define the PDF for the long wavelength modes as a path integral using the window function, 
\begin{align}
P_{\Omega}[\phi_{\Omega},\chi_\Omega, t]&=\int \mathcal{D}\phi\mathcal{D}\chi \ \delta\left(\phi_\Omega({\bf x})-\int\frac{d^3 k}{(2\pi)^3}\Omega_{\Lambda}e^{i{\bf k}\cdot{\bf x}}\phi({\bf k})\right)\nonumber\\
&\qquad\times\delta\left(\chi_\Omega({\bf x})-\int\frac{d^3 k}{(2\pi)^3}\Omega_{\Lambda}e^{i{\bf k}\cdot{\bf x}}\chi({\bf k})\right)P[\phi, \chi, t]~.
\label{path_integral}
\end{align}
This expression generalises the notion of the long wavelength limit probability we have discussed before.  Notice that because the window function also has a time dependence there will be another term when taking the derivative of $P_{\Omega}$. One will be given by the classical solution whereas the new one will take into account the short wavelength fluctuations.  Schematically we can write,
\begin{align}
\frac{\partial P_{\Omega}}{\partial t}=\mathrm{Drift}\ +\mathrm{Diff.}
\end{align} 
In analogy with the usual Fokker-Planck equation.  Let us unpack this.  For the drift term we  have already found an expression for the derivative of $P$ in \eqref{continuity_eq}.  Replacing it we find,
\begin{align}
\mathrm{Drift}&=\int \mathcal{D}\zeta\mathcal{D}\chi\ \delta[\dots]\left(-\int d^3 x\frac{\partial}{\partial\phi({\bf x})}\left(\Pi_\phi({\bf x})-f(\chi)\right)P[\phi,\chi,t]\right.\nonumber\\
&\qquad\qquad\qquad\left.-\int d^3 x\frac{\partial}{\partial\chi({\bf x})}(\Pi_\chi({\bf x}) P[\phi,\chi,t]\right)~.
\end{align}
Integrating by parts we find  at leading order in $\delta$ that,
\begin{align}
\mathrm{Drift}&=\int d^3 x\int _0^{\Lambda(t)}\frac{d^3 k'}{(2\pi)^3}\times\left(\frac{\partial}{\partial\phi({\bf x})}\int \mathcal{D}\phi\mathcal{D}\chi\delta[\dots]\left(\Pi_\phi({\bf k})+f(\chi)\right)P[\phi,\chi, t]\right.\nonumber\\
&\qquad\qquad\left.+\frac{\partial}{\partial\chi({\bf x})}\int \mathcal{D}\phi\mathcal{D}\chi\delta[\dots]\Pi_\chi({\bf k})P[\phi,\chi, t]\right)(1+\delta)~,
\label{drift:zeta}
\end{align}
where $\Pi_{\phi,\chi}({\bf k})$ are the Fourier transform of the momenta~\eqref{momenta_def}.  After swapping the momentum integral inside the functional derivative we see that  we are  basically taking the Fourier transform of the momenta until $\Lambda(t)$.  At leading order this,  of course, corresponds to considering only the long wavelength modes.  Moreover the path integral considers only the long wavelength probability.  Hence we can write,
\begin{align}
\mathrm{Drift}=\int d^3 x\left( \frac{\partial}{\partial\phi({\bf x})}\left(\left (\Pi_\phi({\bf x})+f(\chi)\right)_{\Lambda(t)}P_{\Omega}(t)\right)+\frac{\partial}{\partial\chi({\bf x})}(\Pi_\chi({\bf x})_{\Lambda(t)}P_\Omega(t))\right)~,
\end{align}
where the subscript in the momentum indicates that we are considering the long  wavelength part field. 

The computation for the diffusion follows the same steps as in \cite{Gorbenko:2019rza} which we review in the appendix.  After a straightforward computation we arrive at the following,
\begin{align}
\mathrm{Diff}=\frac{1}{2}\frac{\partial^2}{\partial\phi^2}\left(\langle\dot\Delta\phi\Delta\phi\rangle P\right)+\frac{1}{2}\frac{\partial^2}{\partial\chi^2}\left(\langle\dot\Delta\chi\Delta\chi\rangle P\right)+\dots~,
\end{align}
where the $\Delta\phi$ is defined as,
\begin{align}
\Delta\phi({\bf k})=\int_{\Lambda(t)}^{(1+\delta)\Lambda(t))}\frac{d^3k}{(2\pi)^3}e^{i{\bf k}{\bf x}}\Omega_{\Lambda}(k)\phi({\bf k})
\end{align}
 such that we integrate over a thin layer of width $\delta$.   The cross term $\frac{\partial^2}{\partial\phi\partial\chi}\left(\langle\dot\Delta\phi\Delta\chi\rangle P\right)$ gives a subdominant contribution. There will be higher order correlations contributing to the diffusion terms. 
  Notice that the time derivative over $\Delta\phi$ only acts over the window function $\Omega_\Lambda$.  Our task then reduces to compute a regulated Fourier transform of a two point function.  Using the results from the previous section we find
\begin{align}
\langle\dot\Delta\zeta\Delta\zeta\rangle&=\frac{H^3}{4\pi^2}\nonumber\\
\langle\dot\Delta\chi\Delta\chi\rangle&=\frac{H^3}{4\pi^2}~,
\label{regularised_corr_func}
\end{align}
Finally,  the quadratic Fokker-Planck equation is,
\begin{align}
\frac{\partial P}{\partial t}&=\frac{\partial}{\partial\phi}\left(\left(f(\chi) \right)P\right)+\frac{H^3}{8\pi^2}\frac{\partial^2P }{\partial\phi^2}-\frac{\partial}{\partial\chi}\left(\frac{m^2}{3H}\chi P\right)+\frac{H^3}{8\pi^2}\frac{\partial^2}{\partial\chi^2}P~.
\label{FPeq}
\end{align}
 Notice that this equation generalises the results found in \cite{Achucarro:2021pdh}, where the Fokker-Planck equation was obtained from a set of Langevin equations. In this case with a minimal set of assumptions we have arrived to the same equation.
\section{Corrections to the Fokker-Planck equation}
\label{App:corrections}
In this appendix we show how to compute some corrections to the Fokker-Planck equation
\eqref{FPeq}. 
%\subsection*{Diffusion terms}
The effect from a quartic interaction can generate a mass term on long distances $\lambda\phi_l^2$. To compute how this changes the diffusion term let us consider the wavefunction coefficient of a massive scalar field
\begin{align}
    \psi_2(k)=\frac{H}{(-H \eta)^d}\left(-\frac{d-2\nu}{2}+\frac{k\eta H_{\nu-1}(-k\eta)}{H_{\nu}(-k\eta)}\right)~,
\end{align}
where $\nu=\sqrt{d^2/4-m^2/H^2}$. If we expand for small $m^2$ and take the real part we get that,
\begin{align}
    2\mathrm{Re}\psi_2(k)=\frac{2k^3}{H^2}\left(1-(\log k-\psi(3/2))\frac{2m^2}{3H^2}\right)~,
\end{align}
where $\psi(x)$ is the digamma function. As we mentioned, the mass term is generated by the long modes and so we identify $m^2=\lambda\phi_l^2$ 
If we plug this result back into the formula for the diffusion term Eq.~\eqref{diffusion2} we find the coefficient gets shifted to,
\begin{align}
\mathrm{Diff}=\frac{H^3}{8\pi^2}\frac{\partial^2}{\partial\phi^2}\left(\left(1+\frac{2\lambda}{H^2}(\log\Lambda-\psi(3/2))\phi^2\right)P(\phi)\right)~,
\end{align}
which coincides with other results \cite{Gorbenko:2019rza,Mirbabayi:2020vyt,Cohen:2021fzf}. This new terms adds a subleading correction to the correlation function computed using the Fokker-Planck equation. Indeed we have that the recursion relations are now,
\begin{align}
    \frac{d\langle\phi^n\rangle}{dt}=-\frac{n}{3H}\langle\phi^{n-1}V_\phi\rangle+\frac{n(n-1)H^2}{8\pi^2}\left(\langle \phi^{n-2}\rangle+\frac{2\lambda}{H^2}a\langle\phi^n\rangle\right)~,
\end{align}
where $a=(\log\epsilon-\psi(3/2))$. If we solve for the variance up to first order in $\lambda$ we get
\begin{align}
\langle\phi^2\rangle=\frac{H^2}{4\pi^2}\log a-\frac{ \lambda H^2}{144\pi^4}(\log a)^3+\frac{a \lambda H^2}{32\pi^4}(\log a)^2~.
\label{corrected_variance}
\end{align}
Notice that the third term is a subleading loop contribution. Since it contains less IR divergences we can trace it back to  the 1-loop correction to the wave function coefficient,
\begin{align}
    \mathrm{Re}\psi_2^{\mathrm{1-loop}}(k)\sim\frac{\lambda k^3}{H^2}\log(-k\eta)~,
\end{align}
which leads to a similar contribution to the two point function as the third term of Eq.\eqref{corrected_variance}. 
Notice that the precise coefficient will depend on the regularisation scheme. It will be interesting to explore this further. 
\bibliographystyle{utphys}
\bibliography{references.bib}
\end{document}